\documentclass[8pt,a4paper]{article}
\usepackage{amsfonts,amssymb,amsmath,epsfig,theorem}
\usepackage[latin1]{inputenc}
\usepackage{fullpage}

 \renewcommand{\S}{\mathrm{S}}
\newcommand{\DD}{\mathrm{D}}

 \newcommand{\field}[1]{\mathbb{#1}}
\newcommand{\R}{\field{R}}

\newcommand{\red}{\mathrm{peq}}

\newcommand{\Gly}{\mathrm{Gly}}
\newcommand{\Syn}{\mathrm{Syn}}
\newcommand{\Deg}{\mathrm{Deg}}

\newcommand{\Oxiun}{\mathrm{Oxi}1}

\newcommand{\Oxideux}{\mathrm{Oxi}2}

\newcommand{\Krebs}{\mathrm{Krebs}}
\newcommand{\Kout}{\mathrm{Kout}}

\newcommand{\Finun}{\mathrm{Fin}1}

\newcommand{\Findeux}{\mathrm{Fin}2}

\newcommand{\G}{\mathrm{G}}
\newcommand{\K}{\mathrm{K}}
\newcommand{\Oun}{\mathrm{O}1}
\newcommand{\Odeux}{\mathrm{O}2}
\renewcommand{\P}{\mathrm{P}}
\newcommand{\A}{\mathrm{A}}
\newcommand{\F}{\mathrm{F}}
\newcommand{\Fun}{\mathrm{F}_1}
\newcommand{\Fdeux}{\mathrm{F}_2}
\newcommand{\T}{\mathrm{T}}
\newcommand{\PP}{\mathrm{PP}}
\renewcommand{\L}{\mathrm{L}}
\newcommand{\V}{\mathrm{V}}
\newcommand{\Eun}{{\mathrm{E}_1}}
\newcommand{\Edeux}{{\mathrm{E}_2}}
\newcommand{\Etrois}{{\mathrm{E}_3}}
\newcommand{\Equatre}{{\mathrm{E}_4}}
\newcommand{\fr}[2]{{\frac{{#1}}{{#2}}}}

\newcommand{\gq}{genetically non-regulated }
\newcommand{\gqty}{fixed genetic variables }

\newcommand{\rodeux}{\ensuremath{R_{\Fdeux}^{\Oxideux}}}

\newcommand{\rosyn}{R_{\Fdeux}^{\Syn}}
\newcommand{\roun}{R_{\Fdeux}^{\Oxiun}}
\newcommand{\rk}{R_{\Fdeux}^{\Kout}}
\newcommand{\tinun}{R_{\T}^{\Finun}}
\newcommand{\tindeux}{R_{\T}^{\Findeux}}

\newcommand{\tsoyn}{R_{\T}^{\Syn}}
\newcommand{\toun}{R_{\T}^{\Oxiun}}
\newcommand{\todeux}{R_{\T}^{\Oxideux}}
\newcommand{\tkr}{R_\T^{\Krebs}}
\newcommand{\tgly}{R_{\T}^{\Gly}}

\newcommand{\chiauntot}{\chi_{\Fun}^{tot}  }
\newcommand{\chiadeuxtot}{\chi_{\Fdeux}^{tot}  }
\newcommand{\chiaunoxi}{ \chi_{\Fun}^{\Oxiun}  }
\newcommand{\chimtot}{\chi_{\A}^{tot}  }
\newcommand{\chimsyn}{  \chi_{\A}^{\Syn}}
\newcommand{\chimkr}{  \chi_{\A}^{\Krebs}}
\newcommand{\rhoaunoxi}{\rho_{\Fun}^{\Oxiun}}
\newcommand{\rhoadeuxoxi}{\rho_{\Fdeux}^{\Oxideux}}
\newcommand{\rhomsyn}{\rho_{\A}^{\Syn}}

\newcommand{\norm}[1]{\left| #1 \right|}
\newcommand{\vect}[1]{{\bf #1}}

\newcommand{\DP}[2]{ \ensuremath{ \frac{\partial #1 }{\partial #2 } } }
\newcommand{\D}[2]{ \ensuremath{ \frac{d #1 }{d #2 } }}

\def\cqfd{\nobreak\nopagebreak\rule{0pt}{0pt}\nobreak\hfill\nobreak\rule{.5em}{.5em}}

\theoremstyle{break}\newtheorem{condition}{Condition}
\theoremstyle{break}\newtheorem{prediction}{Biological prediction}
\theoremstyle{break}\newtheorem{proposition}{Proposition}
\theoremstyle{break}\newtheorem{theorem}{Theorem}
\theoremstyle{break}\newtheorem{lemma}{Lemma}
\theoremstyle{break}\newtheorem{corollary}{Corollary}

\numberwithin{equation}{section} \numberwithin{figure}{section}

\numberwithin{table}{section} \numberwithin{proposition}{section}
\numberwithin{lemma}{section} \numberwithin{theorem}{section}
\numberwithin{corollary}{section}

\author{O.Radulescu$^{1,2}$, A.Siegel$^2$, E.P\'ecou$^{3,4}$,  and S.Lagarrigue$^5$}

\title{A model for regulated fatty acid  metabolism in liver; equilibria and their changes}

\date{}
\begin{document}

\maketitle

\noindent $^1$Institut de Recherche
Math\'ematique de Rennes (UMR CNRS, Rennes, France). \\
$^2$Projet Symbiose.  Institut de Recherche en Informatique et Syst\`emes Al\'eatoires (UMR CNRS, INRIA, Rennes, France) \\
$^3$Institut de Math\'ematiques de Bourgogne, (UMR CNRS, Dijon, France)\\
$^4$Centro de modelamiento mat\'ematico (UMI CNRS, Santiago, Chili)\\
$^5$UMR de G\'en\'etique Animale (INRA-Agrocampus, Rennes, France) \\

 \begin{abstract}
We build a model for the hepatic fatty acid metabolism and its
metabolic and genetic regulations. The model has two functioning
modes: synthesis and oxidation of fatty acids. We provide a
sufficient condition (the strong lipolytic condition) for the
uniqueness of its equilibrium. Under this condition, modifications
of the glucose input produce equilibrium shifts, which are gradual
changes from one  functioning mode to the other. We also discuss the
concentration variations of various metabolites during equilibrium
shifts. The model can explain a certain amount of experimental
observations, assess the role of poly-unsaturated fatty acids in
genetic regulation, and predict the behavior of mutants. The
analysis of the model is based on block elimination of variables and
uses a modular decomposition of the system dictated by mathematical
global univalence conditions.
 \end{abstract}

\section*{Introduction}

\noindent{{\bf Metabolic and genetic model of fatty acid
metabolism.}} Recent advances in genetics and in physiology show the
necessity for melting together several types of cultures in order to
understand animal and human nutrition  and solve important health
and economic problems. Metabolic analysis approaches study dynamics
of biochemical pathways and employs detailed knowledge of
biochemical reactions mechanisms
\cite{Mendes,meta-control,meta-control2,cornish,mazat,Metabolism,Price}.
Genetic functional studies mostly concern gene networks, recently
integrating some metabolites \cite{dejong2,biocham2,Langley}. Gene
network dynamics is modeled by various methods: systems of
differential equations  \cite{Tyson}, boolean or multivalued logical
automata \cite{kaufmann,sanchez}, Petri nets \cite{CRRT04,matsuno}.
Although virtual cell models are planned by many, present studies
deal with simple cell functions. At a higher level of complexity,
integrative approaches were applied to modeling various organs, the
heart being one of the best studied \cite{Noble}. The goal of these
studies is to explain physiology from molecular basis.
The two
main obstacles against this goal are the sparseness of the
biological knowledge and the mathematical difficulty of analyzing
large complex systems.

Main pathways of carbohydrate metabolism are considered to be the
cornerstones of metabolic modeling
\cite{Eisenthal,chassagnole,Teusink}. However, models of metabolism
in eukaryotes are scarce and dedicated to specific metabolic
pathways and organs \cite{melissa}. Furthermore, although recent
experiments pointed out the genetical changes induced by diet in
various organisms \cite{Carsten,Ashrafi,lee04,barnouin04}, metabolic
dynamical modeling is rarely considering the genetic context.
Concerning mathematical complexity, one can reduce the number of
numerical parameters in the models by building so-called "minimal
models" \cite{bergman,vicini,Castellanos,toffolo}. Even for such
minimal models inferring parameters from data posses non-trivial
problems \cite{magni}.

In this paper we present a mixed metabolic and genetic model of
regulated fatty acids metabolism in liver, representing a reasonable
compromise among these various cultures. Parallel work by
\cite{belovitch} on liver focus on transport processes and do not
discuss genetic regulation. The importance of genetic regulation in
fatty acid metabolism  was recently emphasized
\cite{cle03,peg03,dup04,jump04}. Here we trade off the complexity of
spatial effects against the complexity of regulations and the
possibility of analytical reasonings. Our analysis of the model do
not use numerical parameters. It puts forward a mathematical
qualitative approach that can be used more generally for the
analysis of equilibria of complex biological systems.



\medskip

\noindent{{\bf Questions raised by mixed models.}} Melting cultures
generates many, conceptually diverse, questions. Let us point out
three such questions related to multistability, timescales and the
role of genetic and metabolic regulations.

\medskip

Mixed genetical and metabolic systems are often multistable. As a
well known illustration,  the functioning of the E.Coli lactose
operon is based on bistability  \cite{jacmon,yildirim}. Thus, the
change in food (lactose) induces an equilibrium switch which
represents a jump from one attractor to another. Equilibrium switch
is efficient in saving resources (enzymes), because these are
produced only on demand. As a counterpart it is less flexible and
tuning is not possible (the response is of the binary type).

The alternative way to adapt to external changes is via equilibrium
shifts. Then, during an equilibrium shift, an equilibrium uniqueness
condition is fulfilled and there are no jumps between attractors.
The jumps are replaced by smooth, gradual changes.

Fatty acid metabolism in hepatocytes has two antagonistic
functioning modes which could suggest bistability: synthesis
produces fat reserves; lipolysis burns fats and produces energy.
This motivates the first  question we wish to answer: in higher
organisms, does the whole of regulations produce multistationarity
or a unique equilibrium of fatty acid metabolism? We shall argue
that during a change in food, a unique equilibrium shifts smoothly
between the two functioning modes and that there is no bistability.

\medskip

The second question we wish to answer is about timescales. Genes
coding for enzymes need relatively long times to change expression
levels and enzymes concentrations. On short time scales enzyme
concentrations can be considered to be constants. This suggests that
changes of nutritional conditions induces processes with various
timescales. We want to know whether there are any physiological
consequences of the multiple timescales, and to find simple ways to
take this into account in our modeling. For instance, fasting
demands a shift from synthesis to lipolysis functioning modes. This
can be done rapidly by metabolic control. Genetic regulation brings
slower changes that push the shift further. Other slow processes
(for instance diffusion-controlled lipid transport within different
organs) could be responsible for other long time scales. We shall
limit our analysis to only two timescales: a metabolic, fast one,
and a genetic, slow one.

\medskip

Among the fatty acids, many work describe the interference between genes and a special class of fatty acids (polyunsaturated fatty acids denoted by PUFA)  \cite{jump04}. PUFA are
synthetized from essential fatty acids, that are taken from the
diet. To the contrary, saturated and mono-unsaturated fatty acids
(denoted by S/MU-FA) are synthesized de novo in liver. PUFA control
their own oxidation as well as the synthesis and oxidation of the
other fatty acids. The control is due to formation of complexes that
activate or inhibit the active forms of nuclear receptors regulating
the transcription of genes coding for enzymes involved in the
corresponding pathways.

The third question we wish to answer is about the effects of the
PUFA interaction with genes, within normal and mutant genotypes. We
shall argue that, when genetic regulations are absent (for instance,
in PPAR-knocked out cells), the increase of PUFA concentration
during fasting is stronger than in wild type cells.

\medskip

\noindent{{\bf Main points of the paper.}}
The paper is structured as follows.
\begin{itemize}
\item{\bf {\em Definition of several classes of partial equilibria
associated with a mixed model.}} In Section 1, we introduce mixed
models, that is, a differential system of equations where
genetic/slow variables and metabolic/fast variables are
distinguished. Associated to this model, we derive two types of
partial equilibria: first, the well-known {\em quasi-stationnarity}
or {\em non-genetically regulated equilibrium}, where genetic
variables are supposed to be constant. Second, {\em genetic partial
equilibrium} where genetic  variables are supposed to be at
equilibrium. Partial equilibria result from a reduction method
consisting in successive elimination of variables of the model.

\item {\bf {\em Construction of a mixed differential model.}} In
Section 2 we build a differential model for the regulated fatty acid
metabolism in liver. In order to reduce complexity, we deliberately
choose a simplified description of the main metabolic pathways and
of the genetic regulations. Although caricatural, the  models
exhibits  the regulatory function of PUFA and agrees with recent
experiments.
Our model is quite general since we do not use explicit forms of the
functions relating metabolic fluxes to genetic or metabolic
variables. We just take into account the signs of the variations of
elementary fluxes with respect to the variables. Equilibrium
equations allow to extract implicit relations between variables.
This approach is close to metabolic control
\cite{cornish,meta-control,meta-control2} and also to classical
equilibrium thermodynamics \cite{Callen}.

\item {\bf {\em Study of equilibria.}} In Section 3 the steady states
of the differential model are studied.  We find a sufficient
condition for the uniqueness of equilibrium. This condition is
expressed mathematically as an inequality involving partial
derivatives of the fluxes with respect to metabolites
(elasticities). We discuss a biological interpretation of this
condition.

\item {\bf {\em Qualitative validation, prediction and illustration.}} In Section 4
we develop the qualitative analysis of the model. The behavior of
the model is coherent with known experimental data. Moreover, the
model has several predictions concerning the effect of suppressing
the genetic regulation of the important nuclear factor
PPAR-$\alpha$, also on the role of genetic regulation for energy
recovering at fasting.

 Using standard regulation functions, we propose an explicit set of differential equations
which represents the dynamics of hepatic fatty acid metabolism and
its regulations. We give numerical simulations
which illustrate the main results of the paper.

\item {\bf {\em Discussion.}} Section 5 is devoted to a discussion
of the results and of possible extensions.

\item  {\bf {\em Mathematical method.}} Section 6 provides
details about the mathematical method. This method uses a
decomposition of the system into boxes or modules that are chosen in
order to fulfil the conditions of a global univalence theorem
(Gale-Nikaido).
\end{itemize}

\section{Mathematical framework: mixed model; associated partial equilibria states}\label{mathfram}




\noindent{\bf Mixed metabolic-genetic, slow-fast decomposition.}
Main carbohydrate metabolic pathways are rather well documented
\cite{salway}. Nevertheless, actual models of these metabolic
pathways do not take into account genetic regulation. In these models
\cite{chassagnole} enzyme concentrations are parameters, rather than
dynamical variables.  We introduce here a mixed metabolic/genetic
system that contains metabolites, proteins (products of genes),
especially transcription factors and enzymes, as dynamical
variables.

Our mixed model for genetically regulated metabolism is
represented by a system of differential equations:
\begin{equation} ({\mathcal S}):
\left\{ \begin{array}{l}
\D{\vect{X}}{t}  ={\vect{\Phi}}(\vect{X},\vect{Y},\vect{p}) \\
\D{\vect{Y}}{t}  = {\vect{\Psi}}(\vect{X},\vect{Y},\vect{p}).
\end{array}
\right.
\label{gedyn}
\end{equation}
where $\vect{p} \in \Delta$, $\Delta$ being a compact subset of
$\R^q$, stands for a set of external parameters. The dynamical
variables are partitioned into two groups. Concentrations of
metabolites involved in biochemical reactions are metabolic
variables, represented by the vector $\vect{X} \in \R_+^n$.
Concentrations of proteins  (basically enzymes
and transcription factors) are genetic variables, represented by the
vector $\vect{Y} \in \R_+^m$.

Most of the genetic variables vary on timescales generally much
longer than any of the metabolic variables: genetic
variables including concentration of products of genes (enzymes, transcription
factors) have significant variations on long (genetic) time scales
$\tau_G$. On short, metabolic, time scales $\tau_M << \tau_G$ these
variables can be considered to be fixed. Thus, our partition of
the variables corresponds to the well known slow-fast decomposition
of dynamical systems \cite{murray03}.

In Eq.\eqref{gedyn}, ${\vect{\Phi}}$ is the time derivative of fast
(metabolic) variables and  ${\vect{\Psi}}$ is the time derivative of
slow (genetic) variables. Beyond timescales, there is a physical
difference between these functions. ${\vect{\Phi}}$ is a combination
of generally conservative, metabolic fluxes. ${\vect{\Psi}}$ have no
reason to be conservative. The consequences of this difference will
show up in the construction of the model (in particular when
identifying the relations among fluxes).

\medskip
\noindent{{\bf Equilibrium and  quasi-stationary states.}}
Equilibria are  defined mathematically as fixed points of a system
of differential equations and biologically as {\em stationary
states} in which measurable macroscopic quantities stop changing.


Given the parameter $\vect{p}$, recall that an {\em equilibrium
state} of the system (\ref{gedyn}) is defined by the following
equations:
\begin{equation}
 \left\{ \begin{array}{l}
\D{\vect{X}}{t} = {\vect{\Phi}}(\vect{X},\vect{Y}, \vect{p}) = 0 \\
\D{\vect{Y}}{t} = {\vect{\Psi}}(\vect{X},\vect{Y}, \vect{p}) = 0.
\end{array}
\right.
\label{geneq}
\end{equation}

Given a decomposition of variables and a parameter $\vect{p}$,
 we can define another notion of equilibrium,  called
{\em  quasi-stationary state}, which is the equilibrium state of the
subsystem associated to the variable $\vect{X}$, constrained by
fixing $\vect{Y}$. A quasi-stationary state satisfies:
\begin{equation}
\left\{ \begin{array}{l}
{\vect{\Phi}}(\vect{X},\vect{Y}, \vect{p}) = 0 \\
\vect{Y} = \vect{Y}_0  (= \text{const.)} .
\end{array}
\right.
\label{quasieq}
\end{equation}

As detailed before, mixed metabolic-genetic systems are slow-fast systems.
On  metabolic, time scales $\tau_M << \tau_G$ genetic variables
can be considered to be fixed. Therefore, within metabolic
timescale $\tau_M$ the system reaches only quasi-stationarity. After
a longer time of the order $\tau_G$, it reaches equilibrium.

\medskip
\noindent{{\bf Slow-fast dynamics;  \gq trajectory.}}
The slow manifold of a slow-fast system is defined by the
equations ${\vect{\Phi}}(\vect{X},\vect{Y}, \vect{p}) = 0$.

Although the definition  of a quasi-stationary state
(Eq.\eqref{quasieq}) can always be used, this state has a dynamical
meaning only within some conditions which are those required for the
applicability of the Tikhonov theorem \cite{tikhonov,wasow65} or of
the geometrical theory of Fenichel of singular perturbations
\cite{fenichel79}. Indeed, the slow manifold should be
hyperbolically stable with respect to the constrained dynamics
$\D{\vect{X}}{t} ={\vect{\Phi}}(\vect{X},\vect{Y}_0,\vect{p})$ (see
Section 5 for details).

Under this condition, trajectories of slow-fast systems
are made of two parts:   a rapid part finishing close to the slow
manifold and along which $\vect{Y}$ is practically constant  and a
slow part practically included in the slow manifold
\cite{wasow65,fenichel79,murray03}.

\begin{figure}[!ht]
\begin{center}
\epsfig{file=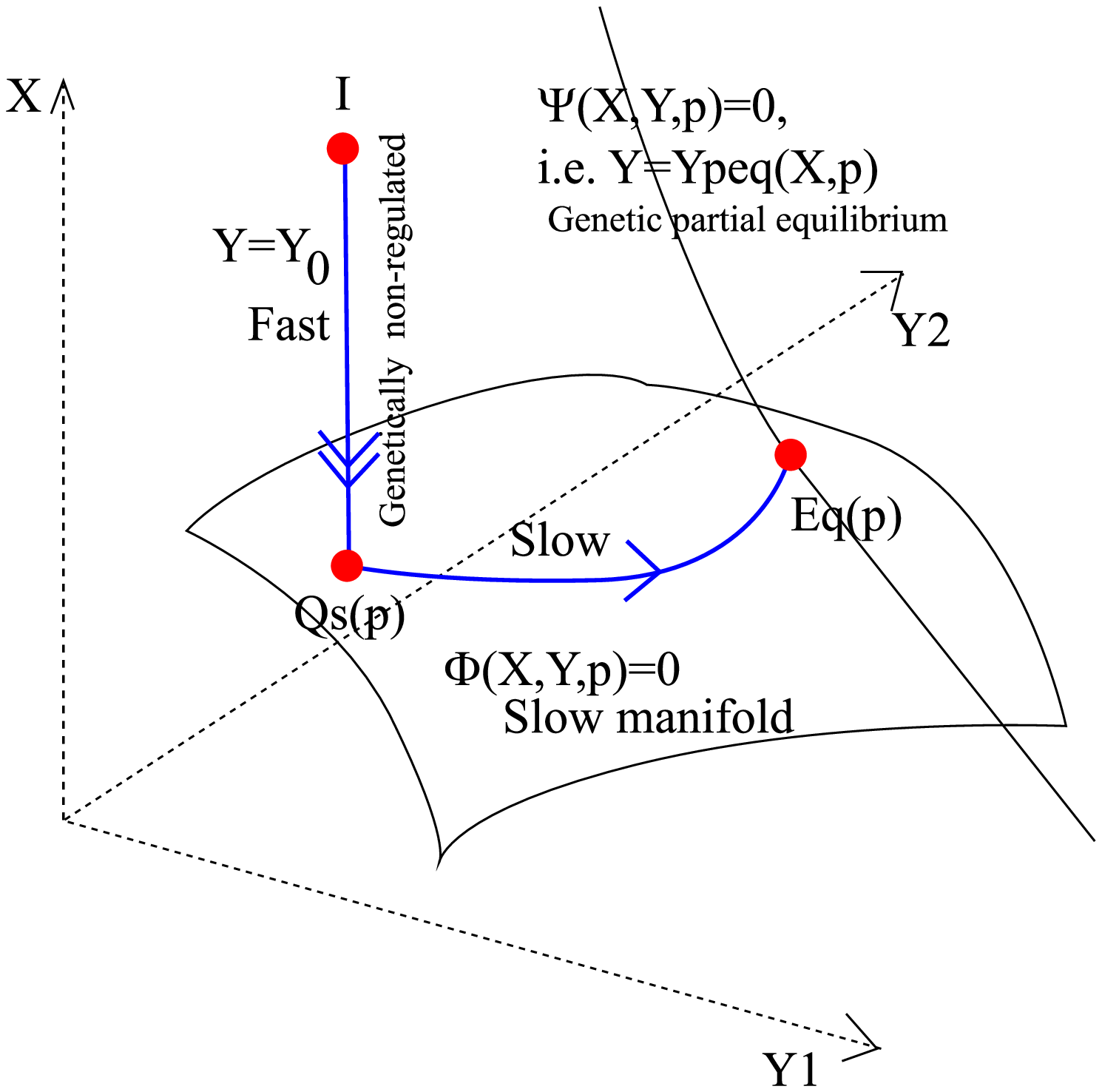,height=8cm}

\begin{tabular}{|p{2.8cm}|p{4cm}|l|p{3.8cm}|}
\hline
Model [Variables]  & Differential equations & Equilibrium equations &
  Relation with the original model \\
\hline
Mixed differential model $[\vect{X},\vect{Y}]$ &
$\begin{array}{l}
\D{\vect{X}}{t}  ={\vect{\Phi}}(\vect{X},\vect{Y},\vect{p}) \\
\D{\vect{Y}}{t}  = {\vect{\Psi}}(\vect{X},\vect{Y},\vect{p}).
\end{array}
$ &
$ \begin{array}{l}
{\vect{\Phi}}(\vect{X},\vect{Y}, \vect{p}) = 0 \\
{\vect{\Psi}}(\vect{X},\vect{Y}, \vect{p}) = 0.
\end{array}

$ &   \\
\hline
\gq model $[\vect{X}]$  &
$ \begin{array}{ll}
\D{\vect{X}}{t} & = \vect{\Phi}_{gnr}(\vect{X},\vect{p}) \\
 & = \vect{\Phi}(\vect{X},\vect{Y}_0,\vect{p})
\end{array}
$
&  $\vect{\Phi}_{gnr}(\vect{X},\vect{p})=0$
&  Fast dynamics. Equilibria correspond to quasi-stationnarity of the mixed model.
\\
\hline
 genetic partial equilibrium model $[\vect{X}]$   &
$ \begin{array}{ll}
\D{\vect{X}}{t} & = \vect{\Phi}_{\red}(\vect{X},\vect{p}) \\
& =\vect{\Phi}(\vect{X},\vect{Y}_{\red}(\vect{X},\vect{p})
,\vect{p}) \\
 \end{array}$

${\vect{\Psi}}(\vect{X},\vect{Y}_{\red}(\vect{X},\vect{p}),\vect{p})=0
$
&
$\vect{\Phi}_{\red}(\vect{X},\vect{p})=0$ & No dynamic approximation of the full system.
Equilibria correspond to equilibrium states of the mixed model.
\\
\hline
\end{tabular}

 \caption{Geometry of the mixed decomposition.
 For illustration we considered one metabolic
 variable $\vect{X}=(X)$ and two genetic variables $\vect{Y}=(Y_1,Y_2)$. The  slow
 manifold and the genetic null-cline have the equations
 $\vect{\Phi}(\vect{X},\vect{Y},\vect{p})=0$, and
 $\vect{\Psi}(\vect{X},\vect{Y},\vect{p})=0$, respectively.
The slow manifold is considered hyperbolically stable with respect
to the constrained dynamics  $\D{\vect{X}}{t}
={\vect{\Phi}}(\vect{X},\vect{Y}_0,\vect{p})$. Trajectories starting
from the initial state $I$ have a rapid part on which genetic
variables are constant $\vect{Y}=\vect{Y}_0$ and a slow part in the
slow manifold. The quasi-stationary state $Qs(p)$ is the
intersection of the slow manifold with the line
$\vect{Y}=\vect{Y}_0$ of \gq states. The equilibrium state $Eq(p)$
is the intersection of the slow manifold with the genetic null-cline
$\vect{Y}=\vect{Y}_{peq}(\vect{X},\vect{p})$ which is the line of
genetic partial equilibrium states. } \label{figstates}
\end{center}
\end{figure}

Let us call {\em \gq trajectory} the set of points approximating the
rapid part of the slow-fast trajectory, that is, following the
dynamics $\D{\vect{X}}{t}
={\vect{\Phi}}(\vect{X},\vect{Y}_0,\vect{p})$ and
$\vect{Y}=\vect{Y}_0$. The quasi-stationary state is the
intersection of the \gq trajectory and the slow manifold (see figure
\ref{figstates}).  It represents the natural intermediate stage on
the way towards equilibrium.

\medskip

\noindent{{\bf Genetically non-regulated model.}} By definition, metabolic variables on the \gq trajectory  are governed by
 a differential dynamical system for metabolic variables that we call {\em \gq model}:
\begin{equation}
\D{\vect{X}}{t} = \vect{\Phi}_{gnr}(\vect{X},\vect{p})
 \label{qsmodel}
\end{equation}
where $\vect{\Phi}_{gnr}(\vect{X},\vect{p})=\vect{\Phi}(\vect{X},\vect{Y}_0,\vect{p})$ are called {\em reduced fluxes}.

Moreover, on the  \gq trajectory, the quasi-stationary equation
\eqref{quasieq} reduces to the following equations for the metabolic
variables, that are called {\em \gq state equations}:
\begin{equation}
 \vect{\Phi}_{gnr}(\vect{X},\vect{p}) = 0
 \label{qsstateeq}
\end{equation}

Let us note that the \gq model is the playground for classical
metabolic analysis. Indeed, the vast majority of metabolic analysis
studies do not take into account genetic regulations and consider that the
concentrations of enzymes are fixed.

\medskip

\medskip
\noindent{{\bf Genetic partial equilibria.}}
According to the classical singular perturbation theory, the constraint ${\vect{\Phi}}(\vect{X},\vect{Y}, \vect{p}) = 0$ allows
expressing the fast (metabolic) variables $\vect{X}$ as functions of
the slow (genetic) variables $\vect{Y}$. In this paper we perform something different.
We have already introduced the quasi-stationarity states and the equilibrium states.
Let us introduce another type of intermediate
states obtained by equilibrating only the genetic variables.
Let us call these states {\em genetic partial equilibria}.
Contrary to quasi-stationay states (that  are natural intermediate stages on the way towards equilibrium),
genetic partial equilibria are not dynamically reachable (because genetic variables equilibrate
after and not before the metabolic ones). However, they represent
 mathematical constructions useful for the study of equilibria.

More precisely,  we intend to express  genetic variables as functions
 of the metabolic variables thanks to the genetic partial equilibrium equation
${\vect{\Psi}}(\vect{X},\vect{Y}, \vect{p}) = 0$.  The following condition ensures that
 these equations have solutions in variables $\vect{Y}$.

\begin{condition}[Unique genetic partial equilibrium]
The genetic partial equilibrium equations
${\vect{\Psi}}(\vect{X},\vect{Y}, \vect{p}) = 0$ have a unique
solution in $\vect{Y}$ which is a smooth function of $\vect{X}$ for
all $\vect{X} \in \R_+^n, p \in \Delta$. In other words, genetic
null-clines are smooth and have unique intersections with the
hyperplanes $\vect{X} = const.$. \label{uniquepeq}
\end{condition}

If Condition \ref{uniquepeq} is satisfied,  we denote the
corresponding implicit functions
$\vect{Y}=\vect{Y}_{\red}(\vect{X},\vect{p})$. This equation defines
the {\em genetic null-cline} (see Fig.\ref{figstates}), that is, the
trajectory of the dynamics $\D{\vect{X}}{t} =
\vect{\Phi}_{\red}(\vect{X},\vect{p})$,
$\vect{Y}=\vect{Y}_{\red}(\vect{X},\vect{p})$ where
$\vect{\Phi}_{\red}(\vect{X},\vect{p})=\vect{\Phi}(\vect{X},\vect{Y}_{\red}(\vect{X},\vect{p})
,\vect{p})$ are called {\em reduced fluxes}.

On the genetic null-cline,  we can simply express the equilibrium
equations \eqref{geneq} as a set of constraints for the metabolic
variables only. In analogy to classical thermodynamics \cite{Callen}
we call these reduced constraints {\em genetic partial equilibrium
state equations} for metabolic variables:
\begin{equation}
 \vect{\Phi}_{\red}(\vect{X},\vect{p}) = 0
 \label{peqstateeq}
\end{equation}

The vector field $\vect{\Phi}_{\red}(\vect{X},\vect{p})$ defines
itself a differential dynamical system governing the metabolic variables on the genetic nullcline. We call this system the {\em genetic
partial equilibrium model}:
\begin{equation}
\D{\vect{X}}{t} = \vect{\Phi}_{\red}(\vect{X},\vect{p}).
 \label{peqmodel}
\end{equation}

\medskip

\noindent{\bf Equilibria and equilibria shifts.} Our purpose is to discuss equilibria and equilibria shifts of the
full system. For this purpose, the partial equilibrium model is
fully suitable, as follows from:

\begin{proposition} \label{equilibria}
The equilibria of the partial equilibrium model
(Eq.\eqref{peqmodel}), i.e. the solutions of the genetic partial
equilibrium state equations \eqref{peqstateeq} are the equilibria of
the full system \eqref{gedyn}.

The equilibria of the \gq model (Eq.\eqref{qsmodel}), i.e. the
solutions of the \gq state equations \eqref{qsstateeq} are the
quasi-stationary states of the full system \eqref{gedyn}.
\end{proposition}

State equations help us to assess  the uniqueness or the multiplicity
of equilibria, or to estimate the variations of the metabolites when
the external parameters change. Furthermore, they allow to extend
notions from the control theory of metabolism such as elasticities \cite{cornish}
to the case when genetic regulation is present. The essential of
metabolic control is expressed by the following formula which is a
consequence of the implicit function theorem:
\begin{equation}
 \D{\vect{X}}{\vect{p}} =  - \left[ \D{\vect{\Phi}_{red}}{\vect{X}} \right]^{-1}
 \D{\vect{\Phi}_{red }}{  \vect{p} }
\label{response}
\end{equation}

Eq.\eqref{response} says that the response of a metabolic system to
changes of the parameters can be calculated from the state equation.
The choice for the function $\vect{\Phi}_{red}$ should be either
$\vect{\Phi}_{gnr}$ or $\vect{\Phi}_{peq}$ depending on the
timescale of the changes. If changes are monitored on short,
metabolic timescales then one should consider \gq state equations.
If the changes are monitored on long, genetic timescales then one
should consider partial equilibrium state equations. In all cases,
it is important to compute the derivative matrix
$\D{\vect{\Phi}_{red}}{\vect{X}}$. This matrix describes the
resistance of the metabolic variables to forcings and is analogous
to the matrix of elastic constants in elasticity theory
\cite{cornish}. $\D{\vect{\Phi}_{gnr}}{\vect{X}}$ gives the
instantaneous elastic constants, while
 $\D{\vect{\Phi}_{peq}}{\vect{X}}$  gives the
static elastic constants (resistance to slow, adiabatic changes,
taking into account genetic readjustment).

\medskip

\noindent {\bf Comments.}

\begin{itemize}
\item If Condition~\ref{uniquepeq} is not satisfied then there are several
equilibrium state equations in the metabolic state variables. We
shall not discuss this situation in this paper.
\item
This particular type of reduction implies a certain emphasis on
metabolic variables. It is justified when genetic variables are not
measured or when we are mainly interested in variations of
metabolites.
\item
The genetic partial equilibrium model represents a bad approximation
for the dynamics of the full system. On the contrary, the \gq model
represents a good approximation for the rapid part of the
trajectories of the full dynamical system Eq.\eqref{gedyn} (see
Fig.\ref{figstates}). The correct approximation of the slow parts of
the trajectories is given by the slow/fast decomposition, more
precisely  by the reduced system $\D{\vect{Y}}{t} =
\vect{\vect{\Psi}}(\vect{X}_{red}(\vect{Y},\vect{p}),
\vect{Y},\vect{p}), \, \vect{X} = \vect{X}_{red}(\vect{Y},\vect{p})$
where $\vect{\Phi}(\vect{X}_{red}(\vect{Y},\vect{p}),
\vect{Y},\vect{p})=0$.
\end{itemize}

\section{Mixed model for genetically regulated fatty acid metabolism}

We provide here the details of our model and we derive the
associated genetically non regulated and genetic partial equilibrium
models.

In different species such as chicken, rodents and humans, hepatocyte
(liver) cells have the specificity to ensure both lipogenesis and
$\beta$-oxidation. To set ideas, all the variables of the model
pertain to an "abstract" hepatocyte, capable of the two different
functioning modes.

\subsection{Variables and fluxes for the mixed model of regulated fatty acid metabolism}


\medskip
\noindent{{\bf Metabolic variables.}}

We have selected the  most important metabolites implied in the
fatty acid metabolism in liver as follows. Corresponding symbols for
these variables are given in Table \ref{table:symbols}.

\begin{itemize}
\item {\bf {\em Acetyl-CoA}}  generated in mitocondria is the first brick  for
building fatty acids. It is consumed in lipogenesis
in hepatocytes, produced in oxidation.
\item  {\bf {\em Saturated and monounsaturated fatty
acids}} (denoted by {\bf {\em S/MU-FA}}) can be produced by the organisms from Acetyl-CoA.
They can also enter the metabolism as part of the diet.
\item {\bf {\em Exogenous polyunsaturated fatty acids (PUFA)}}  are implied in genetic
regulations. They can be manufactured from essential fatty acids
which can not be produced by animals. As a simplification, we write
that PUFA can only enter the metabolism as part of the diet.
 \item {\bf {\em  Energy (ATP)}}  expresses the energy that the cell
has at its disposal.
\end{itemize}

Notice two fundamental points: first, we have introduced the level
of energy of the cell as a variable.
Second, we have divided fatty acids into two parts: the ones that
are implied in the genetic regulation, that is PUFA, and the ones
that are not. PUFA are synthetized for essential fatty acids provided
by the diet; hence we consider that this class can not be
produced by the cell. Even if simplified, this
distinction will allow us to model better the regulations of
the metabolism.

\medskip
\noindent{{\bf Parameter.}} The system is driven by the glucose
concentration, representing food. Different nutritional states such
as normal feeding or fasting are modeled by different values of this
parameter.

\medskip
\noindent{{\bf Primitive metabolic fluxes.}} Main metabolic
processes are modeled here as {\em primitive fluxes}. They are
represented as unstructured reactions, whose detailed mechanisms are
not described. The corresponding symbols are given in Table
\ref{table:symbols}.

\begin{itemize}
\item  {\bf {\em Glycolysis}} (in which we include the  Pyruvate
dehydrogenation reaction) produces Acetyl-CoA from glucose.
Glycolysis can be considered reversible. Nevertheless, we shall not
study glucose dynamics ($\G$ is a constant); reversibility will be
neither used nor rejected.
\item {\bf {\em Krebs cycle}} produces energy for
cellular needs from Acetyl-CoA.
\item {\bf {\em Ketone bodies exit}} allows the cell to
to  transfer the energy stored in Acetyl-CoA  to the outside; it
represents an important source of survival during fasting or
starving.
\item  {\bf {\em Lipogenesis}} transforms Acetyl-CoA first into citrate, then
into saturated and monounsaturated fatty  acids S/MU-FA.
\item A {\bf {\em outtake flux }} allow S/MU-FA  to exit liver
and go to storing tissues (adipocytes). Conversely, the intake flux
is fed partially from diet, partially from lipolysed adipocytes. The
intake flux is conventionally considered positive.
\item Similar {\bf {\em intake/outtake { flux of PUFA}}} allows PUFA
to enter or exit the cell. Above diet and lipolysis, the intake flux
of PUFA also includes a synthetic pathway consisting of desaturation
and elongation of essential fatty acids.
\item {\bf {\em $\beta-$oxidation}} burns  all fatty acids
in order to produce energy and to recover Acetyl-CoA.
\item  ATP consumption  expresses  the energy (ATP) the cell consumes for living.
\item  {\bf {\em Degradation of metabolites}} (Acetyl Co-A, S/MU-FA, PUFA) is
used with a broad meaning including cell growth induced dilution,
leaks or transfers to non-represented pathways, and effective
molecular degradation. Negligible on the timescale of the metabolic
processes, these processes can not be neglected on the genetic
timescale.
\end{itemize}

 \medskip
\noindent{{\bf Functioning modes.}} There are two functioning
antagonist modes of the fatty acid metabolism in liver: lipogenesis
that produce reserves, fatty acid oxidation that burns reserves and
produces energy. The choice of the functioning mode depends on
nutrition conditions: a lack of food (i.e. a sustained low level of
glucose) stimulates lipolysis and oxidation; normal feed (normal
glucose level)  induces lipogenesis.

 \medskip
\noindent{{\bf Fatty acids and genetic control.}} A good part of the
known regulation mechanisms implies transcription factors such as
nuclear receptors  PPAR$\alpha$ (peroxisome proliferator activated
receptor $\alpha$) and LXR$\alpha$ (liver X receptor $\alpha$).
 The latter is known to activate the transcription of SREBP-1 (Sterol response element
binding protein 1) known to trans-activate different genes involved
in fatty acids synthesis and desaturation.

Concerning the metabolites, it has been established that fatty acids
can up-regulate or down-regulate the expression of different genes
controlling their metabolism.  The regulatory effect is mainly due
to PUFA: the interaction of S/MU-FA  with genes is supposed weak.
More precisely, it has been proposed that PUFA  regulates the
activity of SREBP-1 and of several members of the steroid-thyroid superfamily of
nuclear receptors such as PPAR$\alpha$ and LXR$\alpha$ (for reviews see \cite{cle03,peg03,dup04,jump04}).


 \medskip
\noindent{{\bf Genetic variables.}}  Since the  genetic interactions
between metabolites and fluxes is not direct, we consider the
following abstractions for the genetic regulation variables whose
corresponding symbols are given in Table \ref{table:symbols}:
\begin{itemize}
\item the active form of the nuclear receptor PPAR,
\item  the active form
of the nuclear receptor LXR, representing  in a very  simplified way the regulation path LXR$\alpha$-SREBP-1.
\item a representative abstract enzyme  for each  set of enzymes that are
involved in  S/MU-FA synthesis and oxidation, PUFA oxidation and
ketone bodies exit respectively. Abstract enzymes production
 is controlled by LXR (modelling the LXR$\alpha$-SREBP-1 pathway) and PPAR.
\end{itemize}

{\small

\begin{table}[!ht]
\begin{minipage}{8.8 cm}
\begin{tabular}{|p{5.3cm}|l|l|}
\hline
Variable (Concentration) & Symbol  & $\D{\, {\mbox{\tiny product}}}{t}$ \\
\hline
Acetyl Co-A & $\A$ & $\Phi_\A$ \\
Saturated and monounsaturated fatty acids
(S/MU-FA)  & $\Fun$  & $\Phi_{\Fun}$  \\
Poly-unsaturated fatty acids (PUFA) & $\Fdeux$ & $\Phi_{\Fdeux}$  \\
Energy ATP & $\T$ & $\Phi_{\T}$\\
Active form of PPAR & $\PP$ & $\Psi_1$\\
Active form of the regulation path LXR-SREBP & $\L$ & $\Psi_2$ \\
Enzymes of S/MU-FA synthesis & $\Eun$ & $\Psi_3$ \\
Enzymes of S/MU-FA oxidation & $\Edeux$ & $\Psi_4$ \\
Enzymes of PUFA oxidation & $\Etrois$  & $\Psi_5$ \\
Enzymes of Ketone body exit & $\Equatre$ & $\Psi_6$\\
\hline
\end{tabular}

\begin{center}
\begin{tabular}{|l|l|}
\hline
Parameter & Symbol   \\
\hline
Glucose & $\G$ \\
\hline
\end{tabular}
\end{center}

\end{minipage}
\begin{minipage}{6.9cm}
 \begin{tabular}{|l|l|}
\hline
Primitive flux & Symbol    \\
\hline
Glycolysis &  $\Gly$  \\
Krebs cycle &  $\Krebs$ \\
Ketone bodies exit &  $\Kout$ \\
Lipogenesis &  $\Syn$ \\
$\beta-$oxidation of  S/MU-FA &  $\Oxiun$  \\
$\beta-$oxidation of PUFA&  $\Oxideux$  \\
 S/MU-FA fatty acids  intake/outake &  $\Finun$ \\
PUFA fatty acids intake/outake &  $\Findeux$  \\
ATP consumption &  $\Deg\T$ \\
Degradation of a metabolite $\V$ &  $\Deg\V$  \\
($\V$ = $\A$, $\Fun$, $\Fdeux$) &  \\
\hline
\end{tabular} \end{minipage}
\caption{Symbols for the variables, their production (expressed as
time derivatives), parameter  and primitive fluxes of the
genetically regulated fatty acid metabolism.}\label{table:symbols}
\end{table}
}

\subsection{Regulations for the mixed model of fatty acid metabolism}

\begin{figure}[!ht]
\begin{center}
\epsfig{file=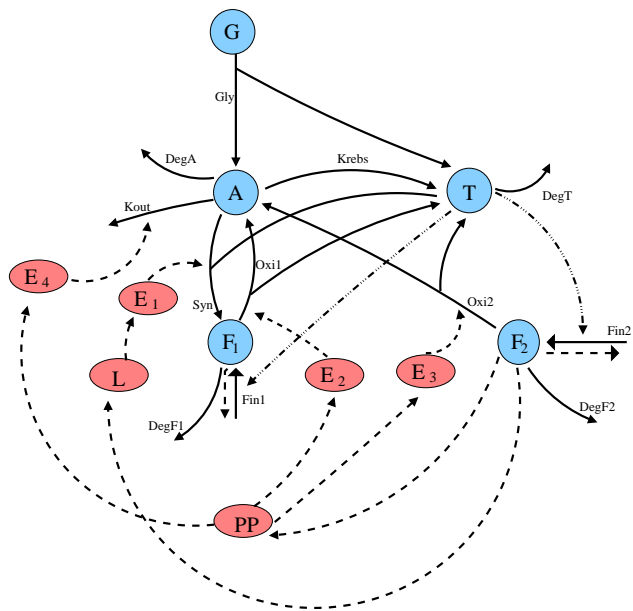,height=8cm} \caption{A model for
genetic regulations of fatty acid metabolism. Dashed arrows stand
for genetic timescale actions from the origin on to target. Plain
arrows stand for metabolic fluxes. Dot arrows stand for energetic
regulations implying $\T$. In this model, notice that a metabolite
$\Fdeux$ (that is, polyunsaturated fatty acids PUFA) regulate the
genetic regulators $\L$ (LXR$\alpha$-SREBP-1 pathway)) and $\PP$ (PPAR-$\alpha$).
}\label{fluxesGenReg}
\end{center}
\end{figure}

We have first defined a sketch of the model. We now intend to add
the regulation relations between fluxes and variables. These are
described by the sign of the variations of a flux when we increase
the activity of the regulator. In mathematical terms, regulation is
summarized by the set of signs of the partial derivatives of the
primitive fluxes with respect to the variables.
 We consider two classes of regulation, namely metabolic  and genetic
 regulations.

\medskip
\noindent{{\bf Metabolic and energetic (ATP) regulations.}}
Metabolic biochemistry has intrinsic regulation: substrates
stimulate and products inhibit reactions. The latter effect is a
consequence of the more general Le Chatelier principle that says
briefly that effects turn against their causes. In control theory
this means that feedback is negative, which is in favor of
uniqueness and stability of equilibrium.
Other negative feedback regulations are responses to the energetic
balance (ATP/ADP or ATP/AMP ratios) via direct biochemical
regulation, or more complex signalling pathways \cite{carling}. An
increase of ATP favours catabolic processes (lipolysis and
oxidation) while a decrease favours synthetic pathways.




We can classify metabolic  regulations as following:

\begin{itemize}
\item {\bf {\em Substrate effect}}
An increase of substrate increases the associated flux. This implies
the following relations in the model:

\medskip

\centerline{{\small $\DP{\Syn}{A} > 0$, $\DP{\Gly}{G} > 0$,
$\DP{\Oxiun}{F_{1}} > 0, \DP{\Oxideux}{F_{2}} > 0$, $\DP{\Kout}{A} >
0$, $\DP{\Krebs}{A} > 0$, $\DP{\Deg\T}{\T} > 0$, $\DP{\Syn}{T} >
0$.}}


Also, degradation reactions are modeled as $\Deg\V(\V)=\delta_\V
\V,\, \delta_\V>0$ where $\V$ denotes any variable  $\A$, $\Fun$,
$\Fdeux$. Hence $\DP{\Deg\V}{\V}>0$.

\item {\bf {\em Passive or active transport effects}}
Intake/outtake fluxes $\Finun$ and $\Findeux$ are conventionally
directed to the inside. Hence, they decrease when the internal
concentrations of fatty acids increase: $\DP{\Finun}{\Fun} < 0$ and
$\DP{\Findeux}{\Fdeux} < 0$.

\item {\bf {\em Product negative feed-back}}
Fluxes producing ATP are negatively controlled by ATP.
Thus,
$\DP{\Gly}{\T} < 0$, $\DP{\Oxiun}{\T} < 0$, $\DP{\Oxiun}{\T} < 0$,
$\DP{\Krebs}{\T} < 0$.

\item {\bf{\em Energy effect on fat intake}}
Fat intake is needed to produce energy by oxidation. A drop in
energy (ATP) stimulates  fat intake. This means that
$\DP{\Finun}{\T}<0$ and $\DP{\Findeux}{\T}<0$.
\end{itemize}





\medskip
\noindent{{\bf Genetic regulations.}} The role of PUFA in genetic
control has been discussed in recent publications
\cite{peg03,dup04,jump04}. Although the precise mechanisms have not
been proven yet, some well established facts can be used for
modeling:

\begin{itemize}
\item {\bf {\em Fatty Acid synthesis down-regulation}}.  PUFA inhibit lipogenesis
via the LXR$\alpha$-SREBP-1 regulation path. Concerning the
mechanisms of these interactions there are some hypothesis.  PUFA
could regulate the nuclear abundance of transcription factors such
as SREBP-1 via the turnover of its mRNA and also via its proteolytic
processing which is specific of SREBP's family \cite{jump04}. It has
also been suggested that PUFA can bind to and modify the activity of
nuclear receptors PPAR and LXR: active PPAR and LXR are heterodimers
with  RXR (retinoid X receptor): PUFA could prevent nuclear receptor
LXR from forming a heterodimer with RXR, and therefore blocks its
activity as a transcription
factor \cite{jump04}..  
\item {\bf {\em Oxidation up-regulation}}. PUFA stimulate their
oxidation as well as the oxidation of $\Fun$ since they activate
PPAR. The detailed mechanism is not known: it either cooperative
stimulation of transcriptional effect of  PPAR or indirect (active
PPAR is a heterodimer with RXR; preventing LXR/RXR formation fatty
acids shifts the equilibrium toward PPAR/RXR formation).
\item  {\bf {\em Ketone exit up-regulation}}.
The mitochondrial HMG-CoA synthase, a key enzyme of the ketone body
formation is known to be transactivated by PPAR$\alpha$; in vivo
PPAR$\alpha$ activation leads to an increase of ketone bodies exit
\cite{lee04}.
\end{itemize}

We translate this biological information into several relations
between the variables. Notice that since these regulations imply
genetic interactions, they occur only on long  (genetic) timescale
$\tau_\G$.

\begin{itemize}
\item PUFA ($\Fdeux$) activates PPAR ($\PP$) and inhibits active-LXR and SREBP-1 ($\L$): $\DP{\Psi_1}{\Fdeux} > 0$, $\DP{\Psi_2}{\Fdeux} < 0$.
\item LXR and SREBP-1 ($\L$) triggers $\Eun$ production (where $E_1$ models S/MU-FA synthesis enzymes): $\DP{\Psi_3}{\L} > 0$.
\item PPAR ($\PP$) triggers the production of $\Edeux$ (S/MU-FA oxydation enzymes), $\Etrois$ (PUFA oxydation enzymes) and $\Equatre$ (ketone exit enzymes):  $\DP{\Psi_4}{\PP} > 0$,  $\DP{\Psi_5}{\PP} > 0$, $\DP{\Psi_6}{\PP} > 0$.
\item Degradation effects occurs on each genetic variable.
\item Abstract enzymes $E_i$ stimulate the corresponding fluxes. $\DP{\Syn}{\Eun} > 0$, $\DP{\Oxiun}{\Edeux} > 0$, $\DP{\Oxideux}{\Etrois} > 0$, $\DP{\Kout}{\Equatre} > 0$.
\end{itemize}

\subsection{Differential model for the regulated fatty acid metabolism}

\begin{table}[bht]
{\footnotesize{$\left\{ \begin{array}{lcl}
\D{\A}{t}      & =&       \Gly(\G,\T) + n_1 \Oxiun( \Fun , \T, \Edeux )  + n_2 \Oxideux( \Fdeux , \T, \Etrois ) \\
& & - \Krebs(\A,\T)  - \Kout( \A , \Equatre ) - n_1 \Syn( \A, \T, \Eun )  - \delta_\A \A   \\
\D{\Fun}{t}    &= &  \Syn( \A, \T, \Eun ) -\Oxiun( \Fun , \T, \Edeux ) +\Finun(\Fun,\T)-\delta_{\Fun} \Fun   \\
\D{\Fdeux}{t}   & = &  -\Oxideux( \Fdeux , \T, \Etrois ) +\Findeux(\Fdeux,\T)-\delta_{\Fdeux} \Fdeux    \\
\D{\T}{t}   & =  &  \alpha_\G \Gly(\G,\T) + \alpha_\K \Krebs(\A,\T)
+ \alpha_{\Oun} \Oxiun( \Fun , \T, \Edeux )  \\
&& + \alpha_{\Odeux} \Oxideux( \Fdeux , \T, \Etrois ) - \alpha_S \Syn( \A, \T, \Eun )- \Deg\T(\T) \\
\D{\PP}{t}  &  = &  \widetilde \Psi_1 ( \Fdeux) \,-\,\delta_{\PP}\PP   \\
\D{\L}{t}  & =  & \widetilde\Psi_2 ( \Fdeux) \,-\,\delta_{\L}\L  \\
\D{\Eun}{t}  & = &   \widetilde\Psi_3 ( \L ) \,-\,\delta_{\Eun}\Eun   \\
\D{\Edeux}{t}  & =&   \widetilde\Psi_4 ( \PP )  \,-\,\delta_{\Edeux}\Edeux  \\
\D{\Etrois}{t}  & = &   \widetilde\Psi_5 ( \PP)  \,-\,\delta_{\Etrois}\Etrois \\
\D{\Equatre}{t} &  = &   \widetilde\Psi_6 ( \PP) \,-\,\delta_{\Equatre}\Equatre \\
 \end{array}
\right.$}}

{\scriptsize{$
\begin{array}{|c|ccccccccccccccc|}
\hline
\DP{\, \mbox{flux}}{ \, \mbox{var.}} &  \Gly   &  \Krebs  &  \Kout  &  \Syn    &  \Oxiun  &  \Oxideux &  \Finun &  \Findeux   &  \Deg\T & \widetilde\Psi_1 & \widetilde\Psi_2 & \widetilde\Psi_3 & \widetilde\Psi_4 & \widetilde\Psi_5 & \widetilde\Psi_6 \\
\hline
\A  &  0       &  +  &  +  &  +     &  0  &  0 &  0 &  0    &  0 & 0 & 0 & 0 & 0 & 0 & 0 \\
 \Fun  &  0     &  0  &  0  &  0    &  +  &  0 &  - &  0    &  0 & 0 & 0 & 0 & 0 & 0 & 0 \\
\Fdeux  &  0       &  0  &  0  &  0   &  0  &  + &  0 &  -      &  0 & + & - & 0 & 0 & 0 & 0 \\
\T   &  -        &  -  &  0  &  +     & -  &  - &  - &  -    &  + & 0 & 0 & 0 & 0 & 0 & 0 \\
\PP   &  0      &  0  &  0  &  0    &  0  &  0 &  0 &  0     &  0 & 0 & 0 & 0 & + & + & + \\
  \L    &  0       &  0  &  0  &  0    &  0  &  0 &  0 &  0     &  0 & 0 & 0 & + & 0 & 0 & 0 \\
 \Eun  &  0       &  0  &  0  & +      &  0  &  0 &  0 &  0    &  0 & 0 & 0 & 0 & 0 & 0 & 0 \\
 \Edeux   &  0       &  0  &  0  &  0     & +  &  0 &  0 &  0    &  0 & 0 & 0 & 0 & 0 & 0 & 0 \\
  \Etrois   &  0       &  0  &  0  &  0     &  0  & +  &  0 &  0    &  0 & 0 & 0 & 0 & 0 & 0 & 0 \\
  \Equatre   &  0      &  0  &  +  &  0    &  0  & 0  &  0 &  0    &  0 & 0 & 0 & 0 & 0 & 0 & 0 \\
 \G    &  +        &  0  &  0  &  0    &  0  &  0 &  0 &  0    &  0 & 0 & 0 & 0 & 0 & 0 & 0 \\
 \hline
\end{array}
$ }} \caption{Differential model for the regulated fatty acid
metabolism: equations and constraints. The flux of each metabolic
variable is obtained as a mass balance of primitive fluxes. The
dependance of primitive metabolic fluxes and of genetic production
terms on variables is given in the table above.
}\label{eq:RegDiffModel}
 \end{table}

The graphical representation of the model is shown in
Fig.\ref{fluxesGenReg}. Table \ref{eq:RegDiffModel} summarizes a
differential model including the above described relations among
metabolic and genetic variables.   Let us summarize the notations:

\begin{itemize}
\item $\vect{X}=(\A,\Fun,\Fdeux,\T)$ is the set of metabolic variables;
\item $\vect{Y}=(\PP,\L,\Eun,\Edeux,\Etrois,\Equatre)$ is the set of genetic variables;
\item $\vect{p}=\G$ is the parameter of the model; we suppose that
it takes values inside a compact interval, $G \in [0, G_{max}]$.
\item $\Phi:\R^{10}_+\to\R^4$ and $\Psi:\R^{10}_+\to\R^6$ are defined such that:
$$\left\{\begin{array}{rcl} \frac{d\vect{X}}{dt} & = & \Phi(\vect{X},\vect{Y},\vect{p}) \\
 \frac{d\vect{Y}}{dt} & = & \Psi(\vect{X},\vect{Y},\vect{p}) \\ \end{array}\right. .$$
\end{itemize}

The differential model was built as follows:
\begin{itemize}
\item The production $\Phi_\A$,  $\Phi_{\Fun}$, $\Phi_{\Fdeux}$, $\Phi_\T$
of each metabolic variable is obtained as the sum of primitive
fluxes that produce or consume the metabolite.
\item Primitive fluxes are treated as single reactions with simple
stoechiometry. Thus, the fluxes $\Gly$, $\Krebs$, $\Oxiun$,
$\Oxideux$, $\Syn$ are considered to have the stoechiometries $\G
\rightarrow \A + \alpha_G T$, $\A \rightarrow \alpha_K T$, $\Fun
\rightarrow n_1 \A + \alpha_{O1} T$, $\Fdeux \rightarrow n_1 \A +
\alpha_{O2} T$, $n_1 \A + \alpha_S T \rightarrow \Fun$ respectively.

\item Degradation reactions of metabolites
are supposed to be linear: $\Deg\V(\V)=\delta_\V \V$ where $\V$
denotes any variable $\A$, $\Fun$, $\Fdeux$.
\item The functions $\Psi_i$ expressing variations of the genetic variables
($\PP$, $\L$, $\Eun$, $\Edeux$, $\Etrois$, $\Equatre$) were not
detailed because mechanisms are still unknown.
Instead, each function $\Psi_i$ has been decomposed into a
non-negative production term $\widetilde\Psi_i$ and a linear
degradation term. Qualitative information on the production term is
translated into a set of signs of its partial derivatives as in
Table \ref{eq:RegDiffModel}; for the time being these are the only a
priori constraints to the model.
\end{itemize}



\subsection{Reduced models for the metabolic variables}
Let us construct the two reduced models and state equations for the
metabolic variables: first, the genetic partial equilibrium model
and its associated state equations; second,  the \gq model and its
state equations.

 In order to obtain the genetic partial equilibrium model
 we must eliminate the genetic variables from their equilibrium
equations, i.e. we must solve the subsystem (genetic partial
equilibrium):

{\small
\begin{equation}\label{subsysgen}
\left\{\begin{array}{lllll}
\widetilde \Psi_1 ( \Fdeux) \,-\,\delta_{\PP}\PP & = 0 &, \qquad & \widetilde\Psi_4 ( \PP )  \,-\,\delta_{\Edeux}\Edeux &= 0\\
 \widetilde\Psi_2 ( \Fdeux) \,-\,\delta_{\L}\L  &= 0&, \qquad&    \widetilde\Psi_5 ( \PP)  \,-\,\delta_{\Etrois}\Etrois &= 0 \\
    \widetilde\Psi_3 ( \L ) \,-\,\delta_{\Eun}\Eun &= 0 &, \qquad&    \widetilde\Psi_6 ( \PP) \,-\,\delta_{\Equatre}\Equatre &= 0. \\
\end{array}\right.
\end{equation}
}

 Notice that in the subsystem \eqref{subsysgen}, the global variable $\Fdeux$ is a
 parameter. If a solution of this subsystem exists, then the
 genetic variables $\PP,\L,\Eun,\Edeux,\Etrois,\Equatre$ are
 expressible  as functions of $\Fdeux$. It can be easily shown that
 this is possible in a unique way, meaning that the Unique Genetic Partial Equilibrium
 Condition \ref{uniquepeq} is fulfilled:

\begin{proposition}\label{p.existenceequilgen}
For any nonnegative value of $\Fdeux$, the partial equilibrium
equations of genetic variables (Eq.\eqref{subsysgen}) admit a unique
solution, i.e. the Condition \ref{uniquepeq} is fulfilled.
\end{proposition}

\noindent{{\bf Proof.}} Solving the partial equilibrium equations
gives readily the values of genetic variables: {\small
$\PP_{\red}(\Fdeux) = \frac1{\delta_{\PP}}\widetilde
\psi_1(\Fdeux)$, $\L_{\red}(\Fdeux) = \frac1{\delta_{\L}}\widetilde
\psi_2(\Fdeux)$, ${\Eun}_{\red}(\Fdeux) =
\frac1{\delta_{\Eun}}\widetilde \psi_3(\L_{\red}(\Fdeux) )$,
${\Edeux}_{\red}(\Fdeux)   = \frac1{\delta_{\Edeux}}\widetilde
\psi_4(\PP_{\red}(\Fdeux) )$, ${\Etrois}_{\red}(\Fdeux)  =
\frac1{\delta_{\Etrois}}\widetilde \psi_5(\PP_{\red}(\Fdeux) )$ and
${\Equatre}_{\red}(\Fdeux)  = \frac1{\delta_{\Equatre}}\widetilde
\psi_6(\PP_{\red}(\Fdeux))$. } \cqfd

\medskip

\begin{proposition}\label{SigneVarImplicit}
At genetic partial equilibrium, the derivatives of the values of the
genetic variables with respect to $\Fdeux$    satisfy {\small
$$\D{\PP_{\red}}{\Fdeux} >0, \, \D{\L_{\red}}{\Fdeux} <0, \,
\D{\Eun_{\red}}{\Fdeux} < 0, \, \D{\Edeux_{\red}}{\Fdeux} >0, \,
\D{\Etrois_{\red}}{\Fdeux} >0, \, \D{\Equatre_{\red}}{\Fdeux} >0.$$
}
\end{proposition}

\noindent{{\bf Proof.}} We use the chain rule formula. For example:
{\small $\D{\PP_{\red}}{\Fdeux}= \frac1{\delta_{\PP}} \DP{\widetilde
\Psi_1}{\Fdeux} >0.$ } \cqfd

\medskip

The next step of the reduction is to express the primitive fluxes at
partial equilibrium as functions of metabolic variables only. For
instance, $\Oxideux_{\red}(\Fdeux,\T)= \Oxideux ( \Fdeux , \T,
{\Etrois}_{\red}(\Fdeux))$.

\begin{proposition}\label{Prop:equilibrium}
At genetic partial equilibrium, the fluxes $\Syn$, $\Oxiun$,
$\Oxideux$ and $\Kout$ become functions of $\A,\Fun,\Fdeux,\T$ only,
denoted by $\Syn_{\red}$, $\Oxiun_{\red}$, $\Oxideux_{\red}$,
$\Kout_{\red}$. The dependence of these functions on $\Fdeux$
satisfy: {\small $$ \DP{\Syn_{\red}}{\Fdeux}  < 0, \qquad
\DP{\Oxiun_{\red}}{\Fdeux}  > 0,  \qquad
 \DP{\Oxideux_{\red}}{\Fdeux}  > 0,
\qquad  \DP{\Kout_{\red}}{\Fdeux}  > 0.$$ }
\end{proposition}

\noindent{{\bf Proof.}}
 By definition $\Syn_{\red}(\A,\T) = \Syn(\A,\T,\Eun_{\red}(\Fdeux))$. Then   $
\DP{\Syn_{\red}}{\Fdeux}= \DP{\Syn}{\Eun} \D{\Eun_{\red}}{\Fdeux}$.
By Lemma \ref{SigneVarImplicit} and Table \ref{fluxesGenReg} it
follows that $\DP{\Syn_{\red}}{\Fdeux}<0$.

The sign of the other derivatives is computed in a similar way.
\cqfd

\medskip


\begin{table}[t]
{\footnotesize{ \hspace{-2.5cm}$ \left\{
\begin{array}{lcl}
\D{\A}{t} & = & \Gly(\G,\T) + n_1 \Oxiun_{\red,gnr}(\Fun,\Fdeux,\T)
+
n_2 \Oxideux_{\red,gnr}(\Fdeux,\T)   \\
  &&    - \Krebs(\A,\T) -   \Kout_{\red,gnr}(\A,\Fdeux)
  - n_1 \Syn_{\red,gnr}(\A,\Fdeux,\T) - \delta_\A \A   \\
\D{\Fun}{t} & = &
\Syn_{\red,gnr}(\A,\Fdeux,\T)-\Oxiun_{\red,gnr}(\Fun,\Fdeux,\T)+
\Finun(\Fun,\T) -  \delta_{\Fun} \Fun   \\
\D{\Fdeux}{t} & = & -\Oxideux_{\red,gnr}(\Fdeux,\T)+
\Findeux(\Fdeux,\T) -\delta_{\Fdeux} \Fdeux     \\
\D{\T}{t} & = & \alpha_\G \Gly(\G,\T) + \alpha_\K
\Krebs(\A,\T)+ \alpha_{\Oun} \Oxiun_{\red,gnr}(\Fun,\Fdeux,\T) \\
& & +   \alpha_{\Odeux} \Oxideux_{\red,gnr}(\Fdeux,\T)  -
\alpha_{\S}\Syn_{\red,gnr}(\A,\Fdeux,\T)  - \Deg\T(\T)  \\
\end{array}
\right.$}}  \hspace{-0.5cm}{\scriptsize{
$\begin{array}{|c|ccccccccc|} \hline \DP{\, \mbox{flux}}{ \,
\mbox{variable}} &  \Gly   & \Krebs  &  \Kout  & \Syn  &
 \Oxiun  &  \Oxideux &  \Finun &  \Findeux  &    \Deg\T \\
\hline
\A  &  0     &  +  &  +  &  +    &  0  &  0 &  0 &  0  &  0 \\
 \Fun  &  0     &  0  &  0  &  0    &  +  &  0 &   -  &  0  & 0 \\
&&&&&&&&&\\
\Fdeux \begin{array}{c} {\mbox{\small gnr}}  \\
 {\mbox{\small peq}} \end{array}  &  0     &  0  &
 \begin{array}{c}0 \\ + \end{array}  &  \begin{array}{c} 0 \\-  \end{array}
 &  \begin{array}{c} 0 \\ +  \end{array} &  + &  0 &  -     &  0  \\
&&&&&&&&&\\
\T   &  -     &  -  &  0  &  +   & -  &  - &  - &  -   &  + \\
 \G    &  +       &  0  &  0    &   0  &  0  &  0 &  0 &  0   & 0 \\
 \hline
\end{array}
$}}

\smallskip

\begin{minipage}{8cm}
\begin{center}
\epsfig{file=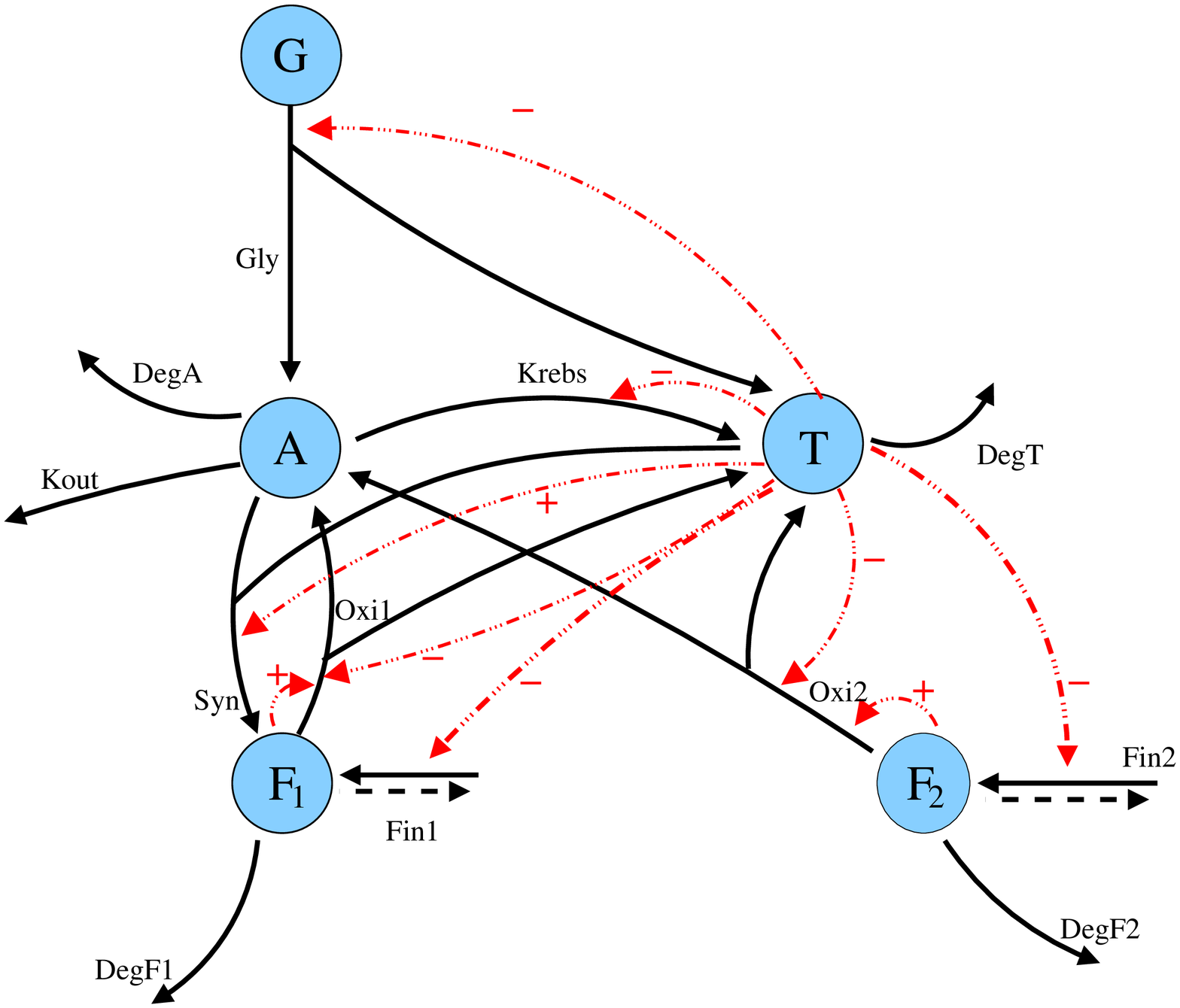,width=6cm}

a) \gq model.
\end{center}
\end{minipage}
\begin{minipage}{8cm}
\begin{center}
\epsfig{file=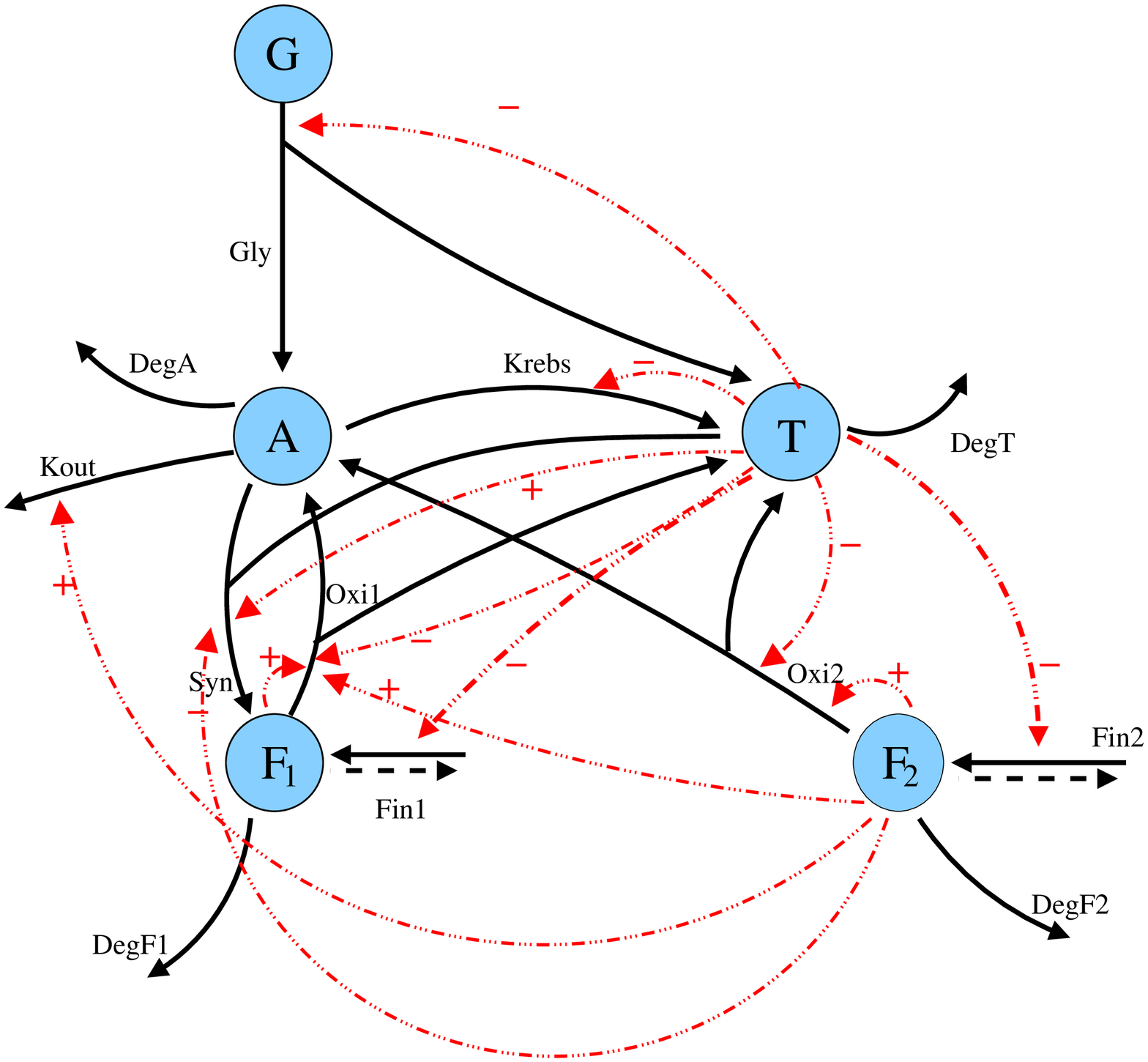,width=6cm}

b) genetic partial equilibrium model.
\end{center}
\end{minipage}

\caption{Reduced models for the metabolic variables of the regulated
fatty acid metabolism. The reduced primitive fluxes and their
regulations are also represented graphically. } \label{IDSMLipid}
\end{table}

\medskip
In \gq states genetic variables are constant, equal to their initial
values:

\begin{equation}
\Eun(t)=\Eun(0), \Edeux(t)=\Edeux(0), \Etrois(t)=\Etrois(0),
\Equatre(t)=\Equatre(0) \label{gnrequations}
\end{equation}

 Primitive fluxes are functions of metabolic variables only
and their derivatives follow directly from the Table
\ref{fluxesGenReg}:

\begin{proposition}\label{Prop:qstat}
In \gq states, the fluxes $\Syn$, $\Oxiun$, $\Oxideux$ and $\Kout$
become functions of $\A,\Fun,\Fdeux,\T$ only, denoted by
$\Syn_{gnr}$, $\Oxiun_{gnr}$, $\Oxideux_{gnr}$, $\Kout_{gnr}$. The
dependence of these functions on $\Fdeux$ is as following:

\centerline{{ $ \DP{\Syn_{gnr}}{\Fdeux}  = 0, \qquad
\DP{\Oxiun_{gnr}}{\Fdeux} = 0,  \qquad
 \DP{\Oxideux_{gnr}}{\Fdeux}  >    0,
\qquad  \DP{\Kout_{gnr}}{\Fdeux}  = 0$ }}
\end{proposition}

Table \ref{IDSMLipid} summarizes the reduced models in the two
situations together with the table of constraints (signs of the
partial derivatives of the reduced primitive fluxes with respect to
the metabolic variables).

\section{Equilibrium and quasi-stationary states: existence and uniqueness}

The process of elimination of variables simplifies the study of
equilibrium states of the model. We have eliminated the genetic
variables in order to obtain the state equations for metabolic
variables. These state equations have to be solved. The solutions of
state equations are equilibria or quasi-stationary states. In this
section we focus on the number of solutions. As discussed in the
introduction if equilibria are unique, modifications of the
variables induced by (slow) changes of the parameter of the model
(food) are smooth equilibrium shifts. The same is true for quick
changes if quasi-stationary states are unique. In order to find
solutions of the state equations we proceed by further elimination
of variables. The technical details are given in Section 4.


\subsection{Existence of an equilibrium and of an quasi-stationary state }

The concentration of metabolites results from the balance of
production fluxes and degradation or consumption fluxes.

Until now, we have assumed  the following conditions on the
elementary fluxes:

\begin{condition}[Flux global constraints] \label{cflux}
\begin{itemize}
\item The fluxes are differentiable functions of the concentrations and satisfy differential constraints (signs of partial derivatives),
summarized in Table \ref{eq:RegDiffModel}.
\item Degradation terms
are  linear: $\Deg\V(\V)=\delta_\V \V$ where $\V$ denotes any
variable $\P$, $\A$, $\Fun$, $\Fdeux$.\end{itemize}
\end{condition}

These hypotheses are very mild. In order to go on with the analysis,
we now add some more assumptions which are natural and not
restrictive:

\begin{condition}[Boundary and
asymptotic conditions] \label{bflux}
\begin{itemize}
\item
In the absence of substrates all fluxes vanish.
\item
All fluxes except degradation  saturate at high concentrations of
metabolites.
\item
ATP consumption is an increasing function of ATP with no saturation
effect, that is $\lim_{T \to \infty} \Deg\T$ $=+\infty$. This is
consistent with the fact that cells can not store ATP.
\item
There exists a recovery effect on each metabolic variable. By
recovery effect we mean that if a variable is zero, then at least
one elementary flux that produces the variable is activated. In
particular, if the cell contains no PUFA, then PUFA enter the cell.
\end{itemize}
\end{condition}

By a mathematical argument, we can prove that under these
assumptions, our model has at least a quasi-stationary state and at
least an equilibrium. Details are given in Section 6. Let us suppose
that the glucose concentration can change between 0 and a maximal
value $G_{max}$.

\begin{theorem}[Existence of equilibrium] \label{existenceFixedPoint}
Let us suppose that the  conditions \ref{cflux} and \ref{bflux} are
satisfied for all glucose concentrations within an interval $0 \leq
\G \leq G_{max}$. Then, the \gq model (Eq.\eqref{gnrequations} and
Table~\ref{IDSMLipid}) and the genetic partial equilibrium model
(Eq.\eqref{subsysgen} and Table~\ref{IDSMLipid}) for fatty acid
metabolism admit at least an equilibrium state for every $0 \leq \G
\leq G_{max}$.
\end{theorem}

According to  Proposition \ref{equilibria}, we derive a result about
the full model.

\begin{prediction}
Under the conditions of Theorem \ref{existenceFixedPoint} the
regulated fatty acids metabolism model described in Table
\ref{eq:RegDiffModel} has at least a quasi-stationary state and at
least an equilibrium.
\end{prediction}

\subsection{Uniqueness of equilibrium}

Various methods using the {\em interaction graph} provide sufficient
criteria for the uniqueness of equilibrium of a differential model
\cite{thom81,snou98,gouz98,soul03}. In this section, we show that
those methods do not apply to our case and we propose a different
method.

\medskip

\noindent{{\bf Interaction graphs for metabolic variables.}}
Oriented interaction graphs can be defined for systems of
differential equations \cite{soul03}.
 The interaction graph gathers information relative to the (direct or
indirect) action of a variable on another one and is thus important
in the theory of response \cite{radulescu05}. The interaction graph
can be computed at genetic partial equilibrium or at fixed genetic
variables (in \gq states) as follows. There is an arc from $X_i$ to
$X_j$ whenever $\DP{\Phi_j^{red}}{X_i} \neq 0$ meaning that $X_i$
has an influence on the flux of $X_j$. The sign of the regulation
arc is the sign of the derivative $\DP{\Phi_j^{red}}{X_i}$. At fixed
genetic variables the interaction graph gathers purely metabolic
influences between metabolites; it corresponds to response on
timescales that are too short to allow for genetic readjustments. At
genetic partial equilibrium the interaction graph gathers both
metabolic and genetically mediated influences between metabolites.

In order to build the interaction graphs, we need to compute the
signs of the derivatives of the fluxes with respect to the
metabolites. This is done in the  Table \ref{TableIntGraph} by using
Table \ref{IDSMLipid}. The sign of the fluxes of metabolites is
computed like in the following example (corresponding to genetic
partial equilibrium):
$$ \begin{array}{lclcl}
 \DP{\Phi_{\Fun}^{peq}}{\Fun} & =  & \DP{}{\Fun}[\Syn_{\red}(\A,\Fdeux,\T)-
 \Oxiun_{\red}(\Fun,\Fdeux,\T)+
\Finun(\Fun,\T) -  \delta_{\Fun} \Fun] &=& \\
&= &- \, \underbrace{
\DP{}{\Fun}\Oxiun_{\red}(\Fun,\Fdeux,\T)}_{>0}+ \,
\underbrace{\DP{}{\Fun}\Finun(\Fun,\T)}_{<0} -  \,
\underbrace{\delta_{\Fun}}_{>0} < 0.  & &
\end{array}$$

\begin{table}[!ht]
\begin{minipage}{6cm}
 $$\begin{array}{|c|cccc|}
\hline \DP{\, \mbox{flux}}{ \, \mbox{variable}} &
  \Phi_\A  &  \Phi_{\Fun}  & \Phi_{\Fdeux}   &  \Phi_\T  \\
\hline
\A  &   - &   +  & 0   &  (I)  \\
\Fun  &  + &  -  & 0   &  +  \\
&&&&\\
\Fdeux \begin{array}{c} {\mbox{\small gnr}}  \\  {\mbox{\small peq}}
\end{array}  &
   \begin{array}{c} + \\ (II) \end{array} & \begin{array}{c}0 \\ - \end{array} & -   &  + \\
&&&&\\
\T   &    (III) &  (IV)  & (V)   &  -   \\
\G    & + &  0  & 0  &  +   \\
 \hline
\end{array}
$$
\end{minipage}
\begin{minipage}{4.5cm}
\begin{center}
\epsfig{file=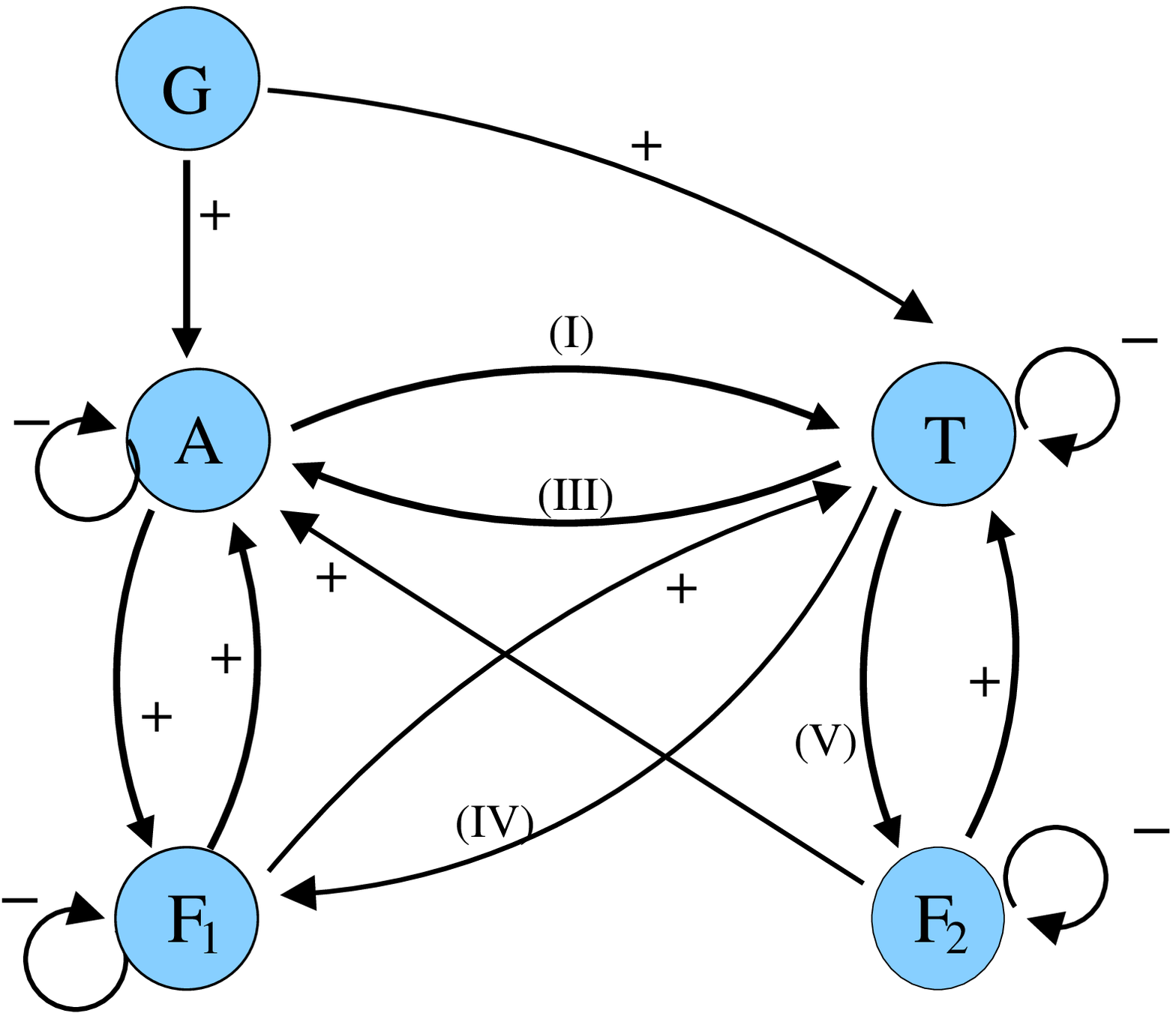,width=4cm}

a) non-genetically regulated model (fixed genetic variables,
metabolic influences)
\end{center}
\end{minipage}
\begin{minipage}{4.5cm}
\begin{center}
\epsfig{file=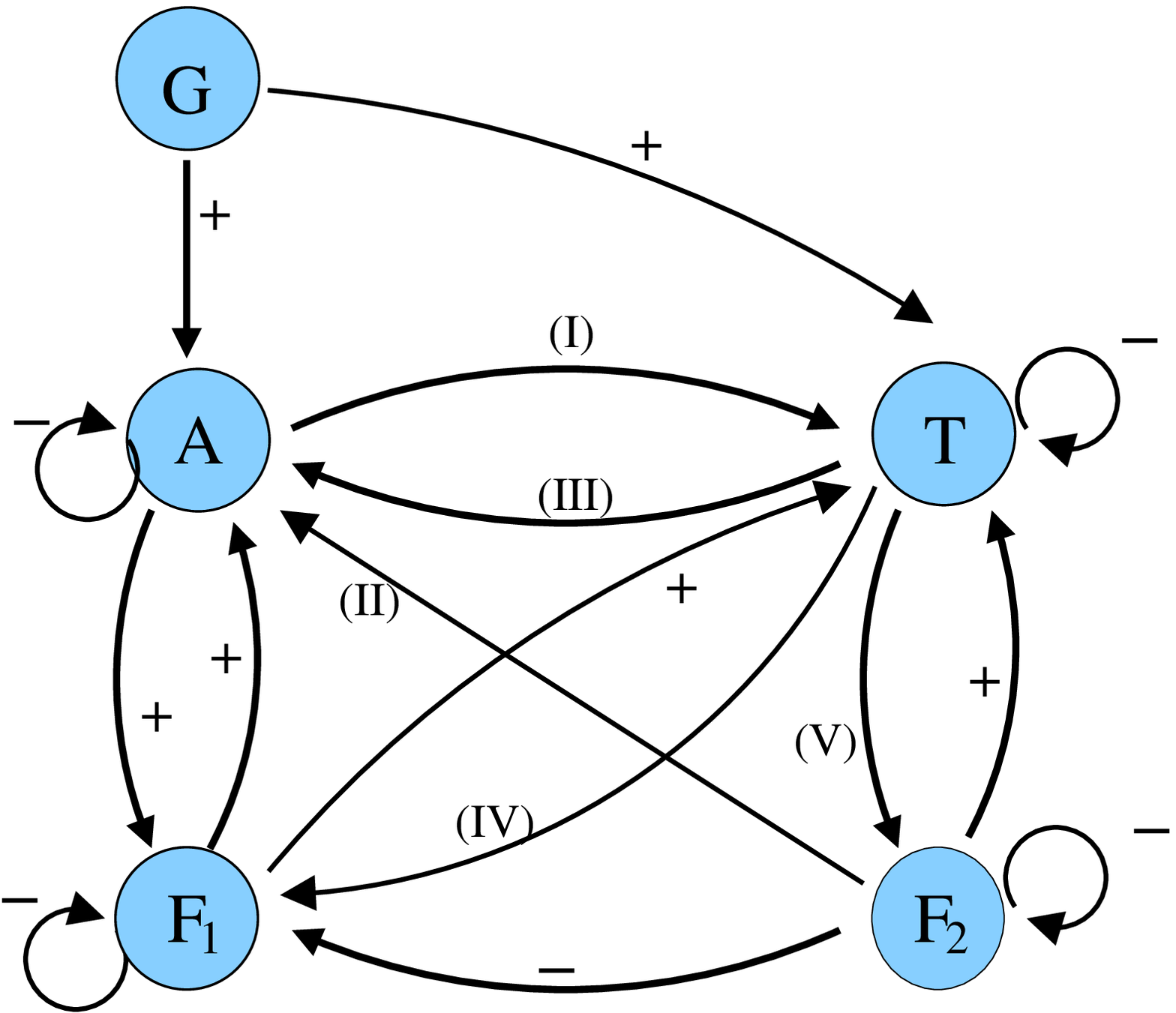,width=4cm}

b) genetic partial equilibrium model (genetic and metabolic
influences)
\end{center}
\end{minipage}
\caption{Signs of the partial derivatives of the fluxes and
interaction graphs for the models derived from the regulated fatty
acid metabolism model at fixed genetic variables and genetic partial
equilibrium. The roman numerals indicate partial derivatives and
regulation arcs whose signs depend on the point where they are
computed.}\label{TableIntGraph}
\end{table}

Simple topological conditions for the uniqueness of equilibrium do
not apply to any of these interaction graphs. Thomas rule asserts
that the absence of positive loops in the interaction graph for all
values of the node variables and external parameters is a sufficient
condition for the uniqueness of equilibrium
\cite{thom81,snou98,gouz98,soul03}. Unfortunately, it appears that
Thomas rule can not be applied
 in our case as stated in the following proposition.

\begin{proposition}
The interaction graphs of the \gq and of the genetic partial
equilibrium models for the
 regulated fatty acid metabolism have both at least a positive
loop for any value of the variables and of the external parameter
$\G$.
\end{proposition}

 \noindent{{\bf Proof.}} There exists a positive loop between $\A$ and $\Fun$
 (see Table~\ref{TableIntGraph}). \cqfd

 \medskip

In the following, we develop another method to prove the uniqueness
of equilibrium. This method proceeds by successive  eliminations of
variables until we are left with only one equation. Then we write
down a sufficient condition for having a unique solution to that
last equation. The success of the method depends on the appropriate
gathering of variables that can be simultaneously eliminated in an
unique way. This is described in full generality in Section 6.

\medskip

\noindent{\bf Uniqueness of equilibrium and of the quasi-stationary
state} A maximal set of variables that can be  eliminated in a
unique way from the state equations is $\{  \A, \F_1,\F_2\}$. In
order to eliminate them, we consider the following state equations,
at fixed $T$ and $G$ (those are both considered as parameters here):
\begin{equation}
\label{sysTnonreg}
 \Phi_{\A}^{red}(\G, \A,\Fun,\Fdeux,\T) = 0 ,\,\,
\Phi_{\Fun}^{red}(\A,\Fun,\Fdeux,\T) = 0 ,\,\,
\Phi_{\Fdeux}^{red}(\Fdeux,\T) = 0.
\end{equation}

Let us recall that $\Phi^{red}$ is either $\Phi_{gnr}$ or
$\Phi_{\red}$ depending on the situation: at fixed genetic variables
or at genetic partial equilibrium.

\begin{proposition}\label{TheoTnonreg}
Suppose that the Conditions \ref{cflux} and \ref{bflux} are
fulfilled for $0 \leq G \leq G_{max}$ and $T \geq 0$. Then the
system \eqref{sysTnonreg} admits a unique solution for any given
pair of values $(\G,\T)\in [0, G_{max}]\times \R_+$, both at \gqty
and at genetic partial equilibrium.

The functions  $\A^{(1)}(\G,\T)$, ${\Fun}^{(1)}(\G,\T)$,
${\Fdeux}^{(1)}(\G,\T)$ expressing this solution are differentiable
in $(\G,\T)$.
\end{proposition}



To ensure uniqueness of equilibrium, we need to eliminate $\T$ from
the state equation
$$\Phi_T^{(1)}(\T,\G)=\Phi_T^{red}(\G,\A^{(1)}(\G,\T),{\Fun}^{(1)}(\G,\T),{\Fdeux}^{(1)}(\G,\T),\T)=0.$$
A sufficient condition for the unique elimination of $\T$  is
$\D{\Phi_T^{(1)}}{T} < 0$. In Section 6 we show that
$\D{\Phi_T^{(1)}}{T}$ is a linear combination with positive
coefficients of  derivatives with respect to $\T$; this allows us to
restate the uniqueness conditions as follows:



\begin{condition}[Strong lipolytic condition] \label{ucondition}
 The following inequality is satisfied for any
$(\G,\T) \in [0,G_{max}] \times \R_+$:

$\DP{{\A^{(1)}}}{\T}( \alpha_\K \DP{\Krebs}{\A} -
\frac{\alpha_\K}{n_1}\DP{\Syn}{\A} )+ \DP{{\Fun}^{(1)}}{\T}
\alpha_{\Oun}  \DP{\Oxiun}{\Fun}+
\DP{{\Fdeux}^{(1)}}{\T}(\alpha_{\Oun} \DP{\Oxiun}{\Fdeux} -
\alpha_{S} \DP{\Syn}{\Fdeux} + \alpha_{\Odeux}
\DP{\Oxideux}{\Fdeux}) + \DP{\Phi_T}{T}<0$.
\end{condition}

\begin{theorem}[Uniqueness of equilibrium] \label{suffcond}
Suppose that the  Conditions \ref{cflux}, \ref{bflux},
\ref{ucondition} are fulfilled at \gqty and at genetic partial
equilibrium, for every $\G \in [0,G_{max}]$.

Then, the state equations \eqref{sysTnonreg} for metabolic variables
admit a unique solution for every $\G \in [0,G_{max}]$, both at
\gqty and at genetic partial equilibrium. The concentration of
metabolites and ATP at equilibrium are differentiable functions of
$\G$:
$\A^{(2)}(\G)$,$\Fun^{(2)}(\G)$,$\Fdeux^{(2)}(\G)$,$\T^{(2)}(\G)$.

In other words, the equilibrium and the quasi-stationary states of
the full model for genetic regulations of fatty acid metabolism  are
unique.
\end{theorem}

\begin{prediction}[Shift mechanism]
When the Conditions 1,2,3,4 are satisfied, the mechanism allowing
the change of functioning modes (lipogenesis and oxidation) is a
shift, both for rapid response (quasi-stationarity) and for slow
static response (equilibrium).
\end{prediction}

\subsection{Computable version and biological significance of the strong lipolytic condition}

In practice, the strong lipolytic condition \ref{ucondition}  is not
readily exploitable because it contains the derivatives of implicit
functions ${\A}^{(1)}$, ${\Fun}^{(1)}$, ${\Fdeux}^{(1)}$. In order
to give a computable version of it, we express the above derivatives
by using control coefficients and metabolic elasticities.

\medskip

\noindent{{\bf Control coefficients, Metabolic elasticities.}} In
metabolic control  language, a control coefficient quantifies the
dependency of a flux on an enzyme activity. The general idea of
control coefficients is that they permit to compare the strength of
fluxes variations one with respect to the other. Following
\cite{cornish}, we call {\em control coefficient of a flux} the
derivative of the logarithm of the flux with respect to the
logarithm of the enzyme concentration. In our problem, although they
are not enzymes, $\Fdeux$ and $\T$ play regulatory roles on the
fluxes through genetic and metabolic factors. Notice that these
effects are absent in classical metabolic control analysis (this
focuses  on the effect of enzymes).
 Thus we choose to call {\em non-logarithmic control
coefficients} all the following quantities:

\begin{itemize}
\item {genetic (non-logarithmic) control coefficients}:
{\small $$ \rodeux = \DP{\Oxideux_{\red,gnr}}{\Fdeux}, \quad \roun =
\DP{\Oxiun_{\red,gnr}}{\Fdeux}, \quad \rosyn  =
-\DP{\Syn_{\red,gnr}}{\Fdeux}, \quad \rk  =
\DP{\Kout_{\red,gnr}}{\Fdeux}.
$$}

\item {ATP (non-logarithmic) control coefficients}:
{\small
$$ \tinun  = - \DP{\Finun}{\T}, \quad
\tindeux  =  - \DP{\Findeux}{\T}, \quad \tsoyn  =
\DP{\Syn_{\red,gnr}}{\T}, \quad \toun  =  -
\DP{\Oxiun_{\red,gnr}}{\T}.$$

$$ \todeux  =  - \DP{\Oxideux_{\red,gnr}}{\T}, \quad
\tkr  = - \DP{\Krebs}{\T}, \quad \tgly  = - \DP{\Gly}{\T}.$$ }


 \end{itemize}

These quantities were defined such that they are all positive (see
Proposition \ref{signelast}). This sign choice simplifies the
identification of balances in the interaction graph. For instance
the interactions in the interaction graphs with undetermined signs
can be expressed as sums and differences of positive control
coefficients:

{\small $$\begin{array}{cl}
(II)  & \DP{\Phi_\A}{\Fdeux} = n_1 \roun + n_1 \rosyn +
n_2 \rodeux - \rk,\\
(III)  & \DP{\Phi_\A}{\T} = \tkr - \tgly - n_1 \tsoyn - n_1 \toun - n_2 \todeux, \\
\end{array} \qquad \begin{array}{cl}
(IV) &  \DP{\Phi_{\Fun}}{\T} = \tsoyn + \toun - \tinun, \\
(V) &   \DP{\Phi_{\Fdeux}}{\T} =  \todeux - \tindeux.\\
\end{array}$$}

The non-logarithmic control coefficients can be defined for all
values of the metabolic variables. In particular, at \gqty the
relations $(\roun)_{gnr} = 0$, $(\rosyn)_{gnr} = 0$, $(\rk)_{gnr} =
0$ express the absence of genetic regulation. The coefficient
$\rodeux$ is the sum of a positive metabolic and a positive genetic
contribution. The genetic contribution vanishes at fixed genetic
variables,
 hence $0 <(\roun)_{gnr} < (\roun)_{peq}$.

Following \cite{cornish}, we call {\em elasticity} the derivative of
the logarithm of the rate with respect to the logarithm of the
substrate concentration. In our setting we call {\em non-logarithmic
elasticities} the following  coefficients: they quantify how rates
and fluxes of a  metabolite depend on this metabolite. Recall that
the rates $\Phi_V$  are sums
 of primitive fluxes in Table \ref{IDSMLipid}.

 {\small $$ \chiauntot  = -\DP{\Phi_{\Fun}}{\Fun},\quad
\chiadeuxtot  =   -\DP{\Phi_{\Fdeux}}{\Fdeux}, \quad  \chimtot  =
-\DP{\Phi_{\A}}{\A}, \quad \chimsyn =  \DP{{\Syn_{\red}}}{\A}, \quad
\chimkr =  \DP{{\Krebs}}{\A},  \quad \chiaunoxi =
\DP{{\Oxiun_{\red}}}{\Fun}.$$ }

Then the last interaction with undetermined sign in the interaction
graph can be expressed as $(I):  \DP{\Phi_\T}{\A} = \alpha_\K
\chimkr  -  \alpha_{\S}\chimsyn$.

\medskip

It will be useful in the following to introduce the following {\em
elasticity ratios} which evaluate the contribution of one primitive
flux to the total elasticity of a metabolite.

{\small
$$\rhoaunoxi  =\frac{\chi_{\Fun}^{\Oxiun}}{\chi_{\Fun}^{tot}}, \qquad
 \rhomsyn = \frac{n_1 \chi_{\A}^{\Syn}}{\chi_{\A}^{tot}}.$$
}

Also it will be useful to consider the genetic control coefficient
$\rodeux$ as an elasticity and to define its associated elasticity
ratio: {\small
$$ \rhoadeuxoxi  = \frac{\rodeux}{\chi_{\Fdeux}^{tot}}.$$ }

The values of the above defined quantities must satisfy the
following:

\begin{proposition}\label{signelast}
The genetic control coefficients, the ATP control coefficients and
the elasticities are non negative. The elasticity ratios are well
defined, non negative and less than 1.
\end{proposition}

\noindent{{\bf Proof.}} From Table \ref{IDSMLipid}, we can easily
prove that genetic control coefficients and ATP control coefficients
are all positive. At \gqty, some of the coefficients vanish
($\roun=\rosyn=\rk=0$). From this analysis and Table
\ref{TableIntGraph}, we get that all elasticities and the implicit
ATP control coefficient are positive and the elasticity ratio are
well defined, non negative and smaller than 1. \cqfd

\medskip

Let us now restate the results of the previous sub-section in terms
of control coefficients and elasticities. First, let us suppose that
the stoechiometry condition is fulfilled:

\begin{condition}[Stoechiometric condition]\label{scon}
The stoechiometric coefficients satisfy the inequalities:
$\alpha_{S} < \alpha_{\Oun} < n_1 \alpha_G$, $n_2 \alpha_{\Oun} <
n_1 \alpha_{\Odeux}$.
\end{condition}

The stoechiometric condition can be checked from biochemical data.
Indeed $n_1,n_2$ are the numbers of Acetyl-coA and
$\alpha_{\Oun},\alpha_{\Odeux}$ are the numbers of ATP molecules
produced (in the average by different fatty acids of the same type)
by oxidation of a molecule of S/MU-FA, or PUFA respectively.
$\alpha_{S}$ is the average number of ATP necessary for the
synthesis of a molecule of S/MU-FA. $\beta$-oxidation produces 5
molecules of ATP for each released molecule of Acetyl-coA, thus
$\alpha_{\Oun} = 5(n_1 - 1)$, $\alpha_{\Odeux} = 5(n_2 - 1)$. PUFA
have in the average longer chains than de novo synthesized fatty
acids, meaning that $n_2 > n_1$. We deduce $n_2 \alpha_{\Oun} <
n_1 \alpha_{\Odeux}$. Synthesis consumes less ATP than
oxidation (for example, for  palmitic acid
$\alpha_S=23$, $\alpha_{\Oun}=35$);  by generalization, we get $\alpha_S < \alpha_{\Oun}$. Finally, $\alpha_\G=7$,
$\alpha_K=12$ (these represent the number of ATP molecules produced
by glycolysis and Krebs cycle per each molecule of Acetyl-CoA), from
which $\alpha_{\Oun} < n_1 \alpha_\G$.

\medskip
Let us define the following combinations of control coefficients:
\begin{eqnarray}
A  & = & X  \rhoaunoxi , \quad  \quad X  = n_1 (\alpha_{\Oun} - \alpha_{S} \rhomsyn + n_1 \alpha_{K} \rho_A^{Krebs} ), \notag \\
 B  & = &  B_1
{R_{\Fdeux}^{\Syn}}/{\chiadeuxtot} + B_2
{R_{\Fdeux}^{\Kout}}/{\chiadeuxtot} + B_3 R_{\Fdeux}^{\Oun}/\chiadeuxtot  +
  B_4 \rhoadeuxoxi, \notag \\
 B_1 & = & X - n_1(\alpha_{\Oun} - \alpha_{S})(1 - \rhomsyn\rhoaunoxi), \quad \quad  B_2 = \alpha_{\Oun}(1 - \rhomsyn \rhoaunoxi ) - {X}/{n_1}, \notag \\
B_3 & = &  X (1 - \rhoaunoxi),  \quad \quad  B_4  =  n_1 \alpha_{\Odeux}(1 - \rhomsyn \rhoaunoxi) +
  {n_2}/{n_1} X  - n_2 \alpha_{\Oun} (1 - (\rhomsyn)^2
  \rhoaunoxi),\notag \\
C  & = & [X/n_1  + (n_1  \alpha_{G} -  \alpha_{\Oun})( 1 - \rhomsyn
\rhoaunoxi) R_T^{Gly} + [\alpha_{\Oun} + n_1 \alpha_{K}  + X R_T^{Oxi1}\notag \\
  & &   [n_2/n_1 X  + n_2 \alpha_{S} \rhomsyn  (1-\rhoaunoxi) +
(n_1 \alpha_{\Odeux} -n_2 \alpha_{\Oun}) (1 - \rhomsyn \rhoaunoxi) ]
R_T^{Oxi2} + \notag \\
 & &   -
(\alpha_{S} +X/n_1) \rhoaunoxi + \rhomsyn ] R_T^{Krebs}  + [n_1(\alpha_{S}-\alpha_{\Oun})+X](1-\rhoaunoxi)R_T^{Syn},\notag \\
D &  =& [X/n_1  - \alpha_{\Oun} (1 - \rhomsyn)] R_T^{Krebs} + n_2
\alpha_{S} \rhomsyn (1-\rhoaunoxi) R_T^{Oxi2} + n_1
(\alpha_{\Oun}-\alpha_{S})(1 - \rhomsyn) \rhoaunoxi R_T^{S}. \notag\\
& & \label{ABCD}
\end{eqnarray}

The strong lipolytic condition is equivalent to the following, more
explicit condition:





\begin{proposition}\label{ABC}
The strong lipolytic response condition reads:
\begin{equation}\label{cond3}
A ( \tinun - \toun ) +  B ( \tindeux - \todeux ) + C > D
\end{equation}

Furthermore, if the stoechiometric condition \ref{scon} is
fulfilled, then the combinations of control coefficients
$X,A,B_1,B_4,C,D$ defined by Eq.\ref{ABCD} are positive.

\end{proposition}

In fact, fluxes of $F_2$ are much smaller than fluxes of $F_1$.
Further simplification of the condition is reasonable:

\begin{proposition}\label{ABCbis}
If $|B ( \tindeux - \todeux )| <<< A | \tinun - \toun |$, then the
strong lipolytic response condition reads:
\begin{equation}\label{cond4}
 A( \tinun - \toun )  + C > D \,\text{where} \, A,C,D>0.
\end{equation}
\end{proposition}




\medskip

\noindent {\bf Comments.}

\begin{itemize}
\item
In Eqs.\eqref{cond3},\eqref{cond4} all control coefficients are
functions of $\G$ and $\T$ because the strong lipolytic condition
\ref{ucondition} has to be checked for all $(\G,\T) \in
[0,G_{max}]\times\R_+$.
\item The conditions in Eqs.\eqref{cond3},\eqref{cond4} may seem more complicated
than the Condition \ref{ucondition}. Nevertheless, they are readily
computable and they provide the biological significance of the
strong lipolytic condition. Eq.(\ref{cond4}) is fulfilled if
$\tinun - \toun $ is large enough, which means that the energy
variation has a sufficiently strong effect on the arrival of fatty
acids inside the cell. This justifies the name strong lipolytic
response condition.
\end{itemize}

\begin{prediction}
Under the hypothesis of  \ref{ABCbis}, the strong lipolytic
condition means that the energy
variation has a sufficiently strong effect on the arrival of fatty
acids inside the cell.
\end{prediction}

\section{Validation, prediction and illustration of the model}

Our model of regulated fatty acid metabolism can be considered at
different levels.

At a {\em qualitative level}, the model consists of a set of
differential equations, together with the differential constraints
in Table \ref{eq:RegDiffModel}. The functions giving the fluxes are
not specified, neither the numerical constants involved in these
functions. In order to validate the model or make some predictions,
we provide sufficient qualitative conditions under which the model
has a certain behavior.

If the behavior has been proven experimentally, the conditions
should be added  to the qualitative model as  extra constraints to
provide a valid model.  If the behavior is a hypothesis not yet
proven, the satisfiability of the sufficient condition provides
predictive models. For instance the strong lipolytic condition
guarantees the uniqueness of equilibrium. We do not know
experimentally whether the equilibrium is unique or not. This could
be tested by response experiments, by looking at the absence or at
the presence of (as in the case of the operon lactose of E.coli)
hysteresis in the response curves.

In this section, our aim is to determine sufficient conditions
either to render  from the known behaviors of the system or to  have
predictions on this behavior. Ideally, the conditions should accept
biological interpretation. The type of behaviors that we intend to
discuss within this approach are about signs of variations of
metabolite concentrations in fasting/refeeding protocols (rather
standard in biological studies of metabolism).

{\em Quantitative versions} of our model can be obtained by
replacing the undetermined functions in Table \ref{eq:RegDiffModel}
by specific functions containing numerical constants. The advantage
is that we can probe the dynamical behavior, much easier than in
qualitative models. Specifying realistic functions and constants is
an enormous task for such complex biological systems. In vitro
measurements of the kinetical constants are rarely available and are
not always reliable. Furthermore, low complexity abstractions are
only very approximate models. Thus, only robust features of dynamics
of the model (that are stable against changes of the parameters or
of the forms of functions) are meaningful. We shall use quantitative
versions of the model as illustrations of robust dynamical
behaviors.

\subsection{Fatty acids concentration increase at fasting}

Let us suppose that fluxes in the qualitative model satisfy the
Conditions 2,3,4. Then, a unique equilibrium state and a unique
quasistationary state exist for any value of the glucose
concentration $G\in[0,G_{max}]$.

The next result is about the dependance of PUFA concentration on
glucose. In order to state our result, let us notice that at
equilibrium or at quasi-stationarity the  control coefficients
$\tindeux, \todeux$ depend only on $\G$.

\begin{proposition}\label{prediBio}
Suppose that the Conditions 1-5 are satisfied. Let $\Fdeux^{(2)} (\G)$ be the value of $\T$
at equilibrium or quasi-stationarity (function of
$\G$). Then the sign of $\D{{\Fdeux}^{(2)}}{\G}$ is equal to the sign of $\todeux-
\tindeux$.
\end{proposition}

This proposition can be stated in biological terms as follows.

\begin{prediction}[response of PUFAs during fasting]
 Suppose that the Conditions 1-5 are satisfied.  The following predictions
are valid for rapid (at quasi-stationarity) as well as for slow (at
equilibrium) response:
\begin{itemize}
\item  If $\left( \tindeux - \todeux \right)_{eq,qs} > 0$ for any $0
\leq \G \leq G_{max}$, then PUFAs increase during fasting and
decrease during feeding.
\item If $\left( \tindeux - \todeux \right)_{eq,qs} < 0$
for any $0 \leq \G \leq G_{max}$, then
 PUFAs  decrease during fasting and increase during feeding.
\end{itemize}
\end{prediction}

In this case, the value of  $\tindeux - \todeux$ implies two distinct behaviors. This means
that  the qualitative constraints associated to the differential are not precise enough
to decide which behavior occurs. However, in this case, some biological information
is avalaible about the behavior of PUFAs during fasting. We will use this information
to refine the model with an additional qualitative constraints.

More precisely, Lee et al. \cite{lee04} studied for wild-type and PPAR-/- mutant
murine liver, the fatty acids profiles in triglycerides (TG), which
are the predominant ($>50\%$) hepatic fatty acids and also in
phospholipids (PL) which go into cellular membranes.
 Let us recall that TG and PL are  storage forms
of fatty acids and that PL contribute much less than TG to the total
fatty acid mass. These authors \cite{lee04} show that for wild type
hepatocytes after 72h of fasting  fatty acids profiles do not change
significantly in PL, but there is a strong increase of TG and of
their fatty acids constituents, in particular PUFA. Based on these
experimental findings, we make the hypothesis of a mass increase
during fasting, of regulating PUFA in the hepatic cell,  and look
for sufficient conditions ensuring this behavior in our qualitative
model. This is consistent with the regulation role of PUFA: in order
to trigger oxidation a persistent increase of PUFA concentration is
needed inside the cell. A qualitative reasoning using an extended
genetically regulated model of lipogenesis \cite{radulescu05} gives
further support to this hypothesis.

Consequently, we have to add $\left( \tindeux - \todeux \right)_{eq,qs} > 0$ as
a qualitative  constraint  to the
model to fit with behaviors observed in \cite{lee04}. By this way, we make a refinement of the model.

\begin{condition}[Experimental constraint added to the model]
A necessary condition for the observed behavior of PUFAs at fasting
(increase) is $\left( \tindeux - \todeux \right)_{eq,qs} > 0$,
meaning that at equilibrium (or quasi-stationarity) the intake
control overcomes the oxidation control for PUFA.
\end{condition}

\noindent {\bf Remark} The strong lipolytic response condition,
which is sufficient for the uniqueness of equilibrium asks that the
intake control overcomes the oxidation control for S/MU-FA fatty
acids (the major contribution to the total mass of fatty acids).
Fasting experiments imply that the strong lipolytic is satisfied for
all fatty acids, including PUFA. Nevertheless, this is not an
experimental proof for the uniqueness of equilibrium, because one
needs the strong lipolytic condition to be satisfied, not only at
equilibrium (or quasi-stationarity), but in all $T$ constrained
states as well. Unfortunately, these states are not accessible
experimentally.

\medskip

The next result is a dynamical one. This may seem paradoxical,
because in this paper we study equilibria and equilibria shifts
which is a statical problem. Nevertheless, the definition of
quasi-stationarity is based on dynamical timescales. Since
quasi-stationarity is reached before equilibrium, comparing these
two states provides some information on the dynamics of the mixed
differential model.


In order to formulate the next result let us denote by $B_{eq}$,
$B_{qs}$ the values at equilibrium and at quasi-stationarity of the
combination of control coefficients $B$ by Eq.\eqref{ABCD}.

\begin{proposition}\label{lemmeCompT}
If the Conditions 1-6 are satisfied and if furthermore
$B_{eq}>B_{qs}$ then $\left| \D{{\Fdeux}^{(2)}}{\G} \right|_{qs} >
\left| \D{{\Fdeux}^{(2)}}{\G} \right|_{eq}$.
\end{proposition}

This result  means that when $\G$ decreases (fasting), the value of
$\Fdeux$ at quasi-stationarity is greater than the value at
equilibrium.  Thus, we can state the following prediction of the
model.

\begin{prediction}[Overshoot of fatty acid concentration] \label{corrCompT}
Under the hypothesis of Prop. \ref{lemmeCompT}, the curves
representing PUFA concentration during fasting must show an
overshoot: the increase in concentration is greater immediately at
quasi-stationarity than later at equilibrium.
\end{prediction}

\noindent {\bf Comments.}

\begin{itemize}
\item
The condition $B_{eq}>B_{qs}$ is equivalent to $B_1 \left(
R_{\Fdeux}^{\Oxiun} \right)_{eq} + B_2 \left( R_{\Fdeux}^{\Kout}
\right)_{eq} + B_3 \left( R_{\Fdeux}^{\Syn} \right)_{eq} +
  B_4 \DP{\Oxideux}{E3}\DP{E3}{\Fdeux} >0 $, with $B_1>0$, $B_4
  \DP{\Oxideux}{E3}\DP{E3}{\Fdeux} >0$. This means that even if
  $B_2,B_3$ are negative the oxidation control term is strong
  enough to win. At fasting, this is a plausible supposition.
 \item
 The existence of an overshoot of faty acids concentration during
 fasting depends on the dynamical accesibility of the
 quasi-stationary state. This state is accesible for a discontinous
step-like glucose input, but may not be accesible if the glucose
input drops slowly (see also the sections 4.4, 6.1).
\end{itemize}






\subsection{Genetic regulation reinforces energy homeostasis}

With the same methods, we prove in Section 6  how the values at
equilibrium respond to the variations of the entering node $G$: not
surprisingly, we predict that ATP decreases at fasting.

\begin{proposition}\label{prediBio1}
If the Conditions 1-5 are satisfied, then  $\D{{\T}^{(2)}}{\G} >0$ at at
equilibrium or quasi-statio\-na\-ri\-ty, where  ${\T}^{(2)}({\G})$ is the
value of $\T$ at equilibrium or quasi-stationarity (function of
$\G$). \end{proposition}

In biological terms, comparing the value of $\D{{\T}^{(2)}}{\G}$ at
quasi-stationarity (representing a model with no genetic regulation)
and at equilibrium allows  understanding the role of genetic regulations.

\begin{prediction}[response of ATP during fasting]
If the Conditions 1-5 are satisfied, then ATP decreases during
fasting and increases during feeding. This prediction is valid for
rapid (at quasi-statio\-na\-ri\-ty) as well as for slow (at equilibrium)
response.\end{prediction}

\begin{proposition}\label{lemmeCompT2} Under the hypothesis of Prop. \ref{lemmeCompT},
then $\left( \D{{\T}^{(2)}}{\G} \right)_{qs} > \left(
\D{{\T}^{(2)}}{\G} \right)_{eq} > 0$.
\end{proposition}

Let us emphasize that the derivative $\D{{\T}^{(2)}}{\G}$ quantifies
the energy buffering effect: the lower is this derivative, hence the
lower is the variation of $\T$ for a fixed variation of $\G$, the
stronger is the energy buffering effect. Thus, we can state the
following:

\begin{prediction}[Role of genetic regulations in energy homeostasis] \label{corrCompT2}
Under the hypothesis of Prop. \ref{lemmeCompT}, genetic regulation
reinforces the energy buffering effect.
\end{prediction}


\subsection{PPAR knock-out reduces energy buffering and increases PUFA at fasting}

Our qualitative model can predict the effects of various genetic
perturbations. Let us consider here the effect of PPAR knock-out.
Without PPAR, there is no longer a genetic control on oxidation,
therefore we expect to have less energy buffering on fasting. Less
obvious is what happens to the concentration of PUFA. We can predict
the behavior of PUFA, under the same hypothesis as above. Let
${\Fdeux}^{(2)}({\G})$ be the value of PUFA concentration  as a
function of $\G$. Also, let $B_{WT,eq}$, $B_{PPAR-/-,eq}$ be the
values at equilibrium in wild type and mutants
 of the coefficient $B$ defined at Proposition 3.4.

Then we have:

\begin{proposition}\label{pparko}
If conditions 1-6 and $B_{WT,eq}>B_{PPAR-/-,eq}$  are satisfied,
then
\begin{enumerate}
\item
$\left( \D{{\T}^{(2)}}{\G} \right)_{eq,PPAR-/-} > \left(
\D{{\T}^{(2)}}{\G} \right)_{eq,WT}$,
\item
$\left| \D{{\Fdeux}^{(2)}}{\G} \right|_{eq,PPAR-/-} > \left|
\D{{\Fdeux}^{(2)}}{\G} \right|_{eq,WT}$.
\end{enumerate}
\end{proposition}

\begin{prediction}[PPAR mutants] \label{corrpparko}
Under the hypotheses of Prop. \ref{pparko}  we have:
\begin{enumerate}
\item  PPAR knock-out reduces energy buffering.
\item  PUFA concentration increase under fasting is stronger in PPAR
knocked-out cells compared to the same increase in wild type cells.
\end{enumerate}
\end{prediction}

\noindent {\bf Comment.} The condition $B_{WT,eq}>B_{PPAR-/-,eq}$ is
equivalent to $B_1 \left(  R_{\Fdeux}^{\Oxiun} \right)_{WT,eq} + B_2
\left( R_{\Fdeux}^{\Kout} \right)_{WT,eq} +
  B_4 \DP{\Oxideux}{E3} \left( \DP{E3}{\Fdeux} \right)_{WT,eq} >0 $, with $B_1>0$, $B_4
  \DP{\Oxideux}{E3} \left( \DP{E3}{\Fdeux} \right)_{WT,eq} >0$. This means that even if
  $B_2$ is negative the oxidation genetic control term is large enough to compensate.


Experiments on transgenic mice showed that after a 72h-fast, fatty
acids concentration increases at a higher extent in PPAR knocked-out
cells with respect to wild type cells \cite{barnouin04}. This is
coherent with the observations by Lee et al.\cite{lee04} that for
the same length of fasting time the hepatic accumulation of
triacylglycerol is $2.8$ fold higher in PPAR knocked-out than in
wild-type mice.  Hence,  the global behavior of fatty acids is
consistent with our predictions.

\medskip

\noindent{\bf Remark.} Data from \cite{lee04} show a rather
selective behavior among different hepatic TG PUFA in PPAR
knocked-out mice: concentrations (in mg/g of liver) of
$\alpha$-Linolenic acid are amplified on fasting to 4-fold higher
levels than in WT mice, but Arachidonic, Docosahexaenoic and
Eicosapentaenoic acids are depleted to non-detectable levels in PPAR
mutants. Consequently,  the  behavior of some important regulating
PUFA in mutants can not be explained by our model, at least not
within the assumptions that we have made. In fact, our model is too
crude to explain the contradictory behavior of part of the PUFA in
mutants. A more complex model including more variables for PUFA and
genes, could answer the questions raised by the experiment. Such a model
should separate essential fatty acids from long-chain PUFA synthetized
from the essential fatty acids, and include the genes involved in
the synthesis of long-chain PUFA.






\begin{figure}[ht]
\begin{center}
\centerline{\includegraphics[width=9cm]{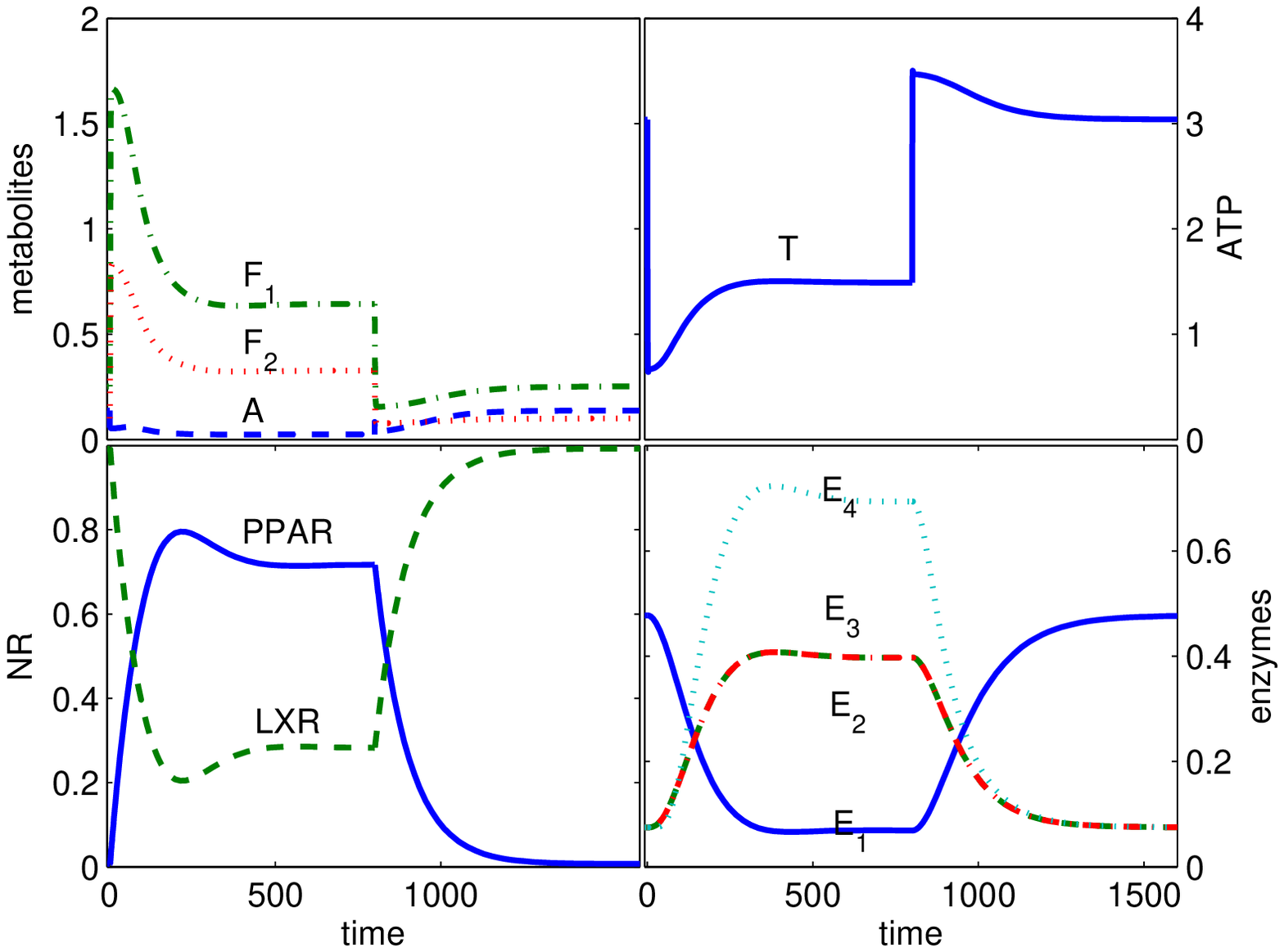}
\includegraphics[width=9cm]{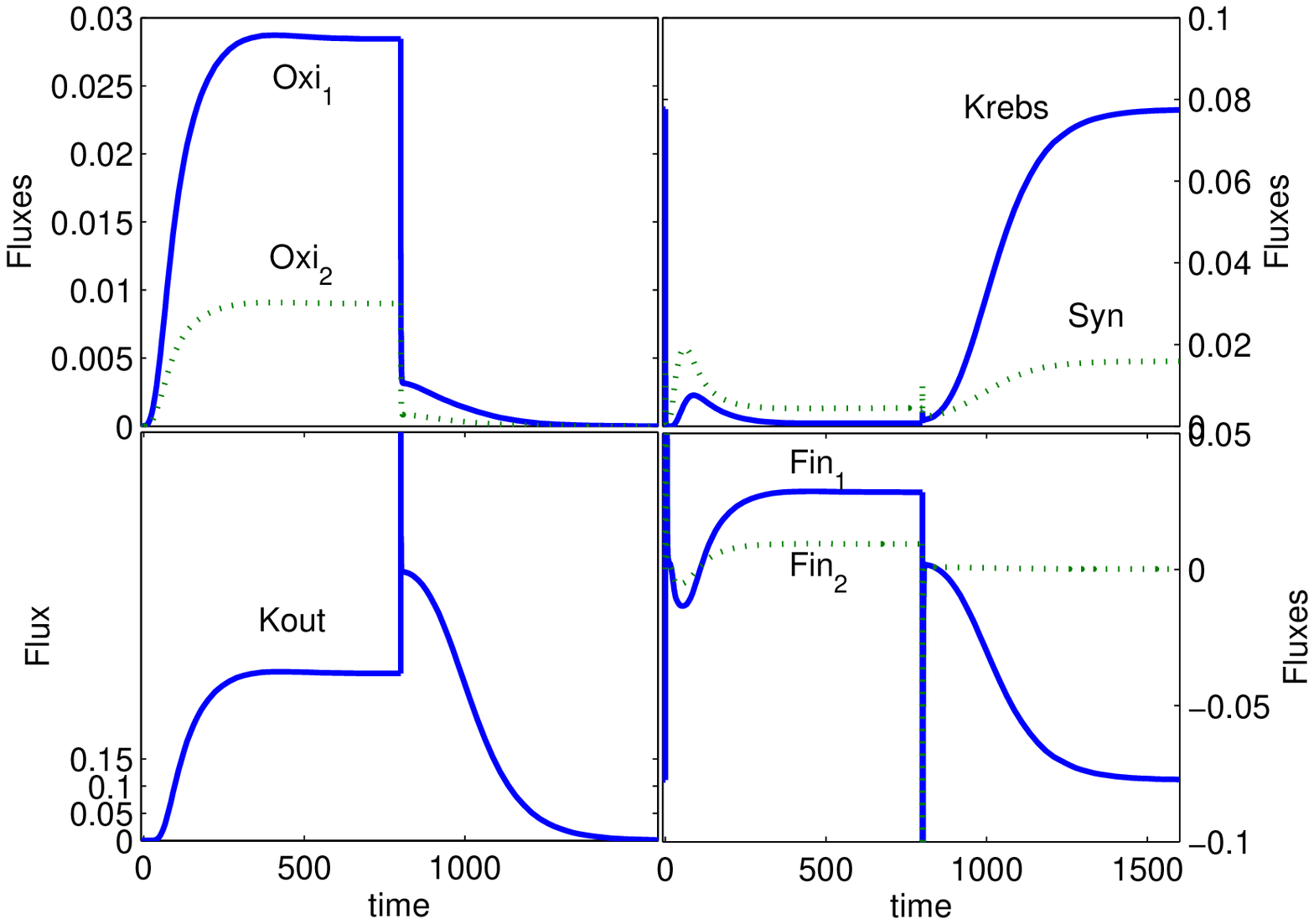}}
\centerline{a)  \hskip8truecm  b)}

\caption{Starving/refeeding protocol. a) Concentrations of
metabolites and main regulators. Starving  begins at $t=0$ and
refeeding at $t=750$ ($\G = 0$ for $0<t<750$ and $\G = 10$ for
$t<0,\,t>750$). Fatty acids ($\Fun$ S/MU-FA, $\Fdeux$ PUFA) increase
at fasting, with overshoots, decrease at refeeding, with
undershoots. Acetyl-coA $\A$ and $\T$ (ATP, energy) have the
opposite behavior. By PPAR and LXR we mean the active forms of
nuclear receptors. These stimulate oxidation and ketonic exits at
fasting (LXR falls, PPAR raises) and synthesis at refeeding (LXR
raises, PPAR falls). b) Fluxes. At fasting ($\G = 0$) there is
practically no synthesis, and the Krebs cycle functions at a very
low level. The fatty acid intake fluxes Fin are positive, oxidation
and ketonic exit are strong. Refeeding inverses the
Oxi,Kout/Syn,Krebs flux ratios, and changes the sign of the fatty
acids intake fluxes (which become negative).}

\label{Figconcentrations}
\end{center}
\end{figure}

\begin{figure}[ht]
\begin{center}
\centerline{\includegraphics[width=9cm]{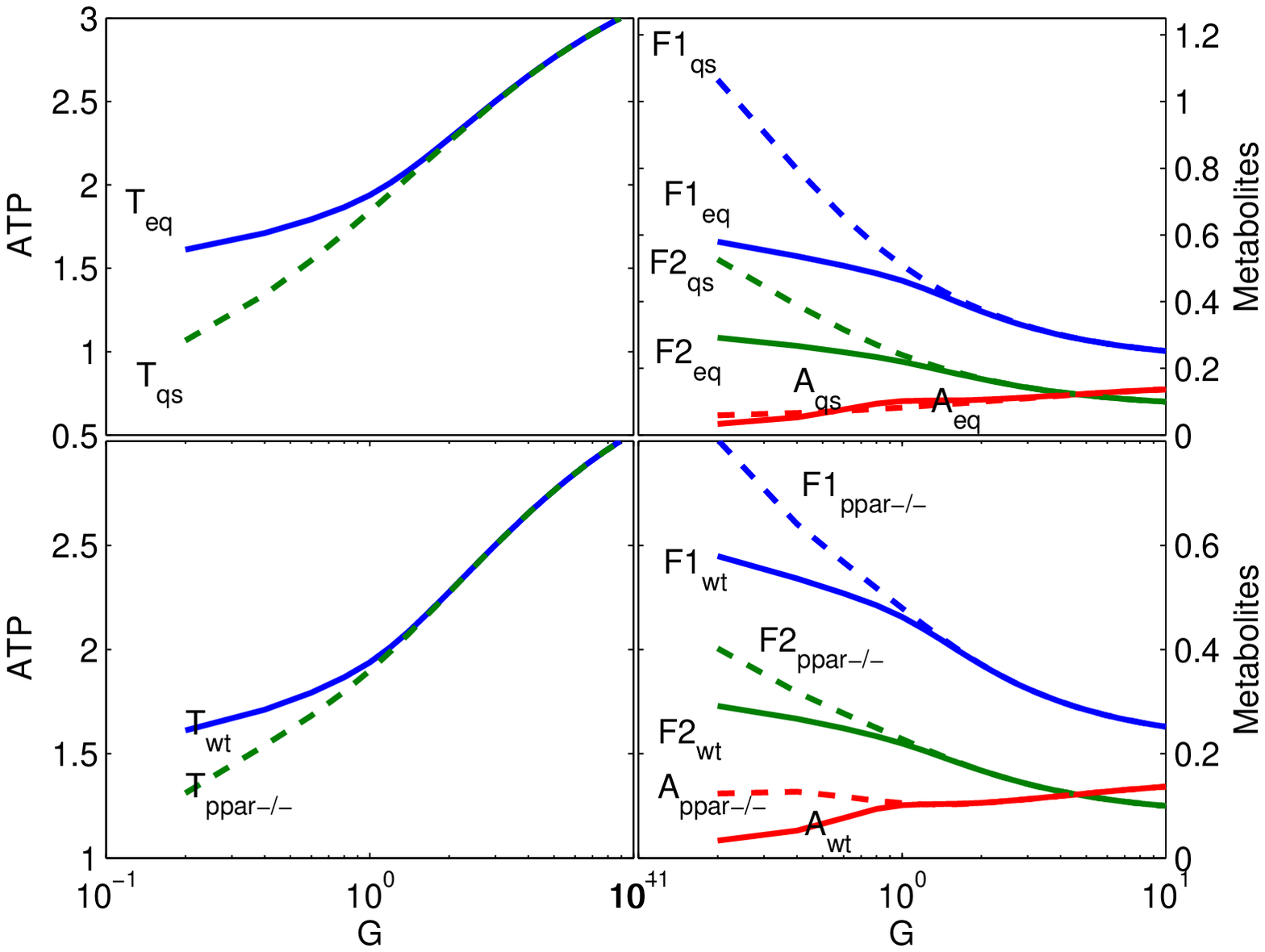}
\includegraphics[width=9cm]{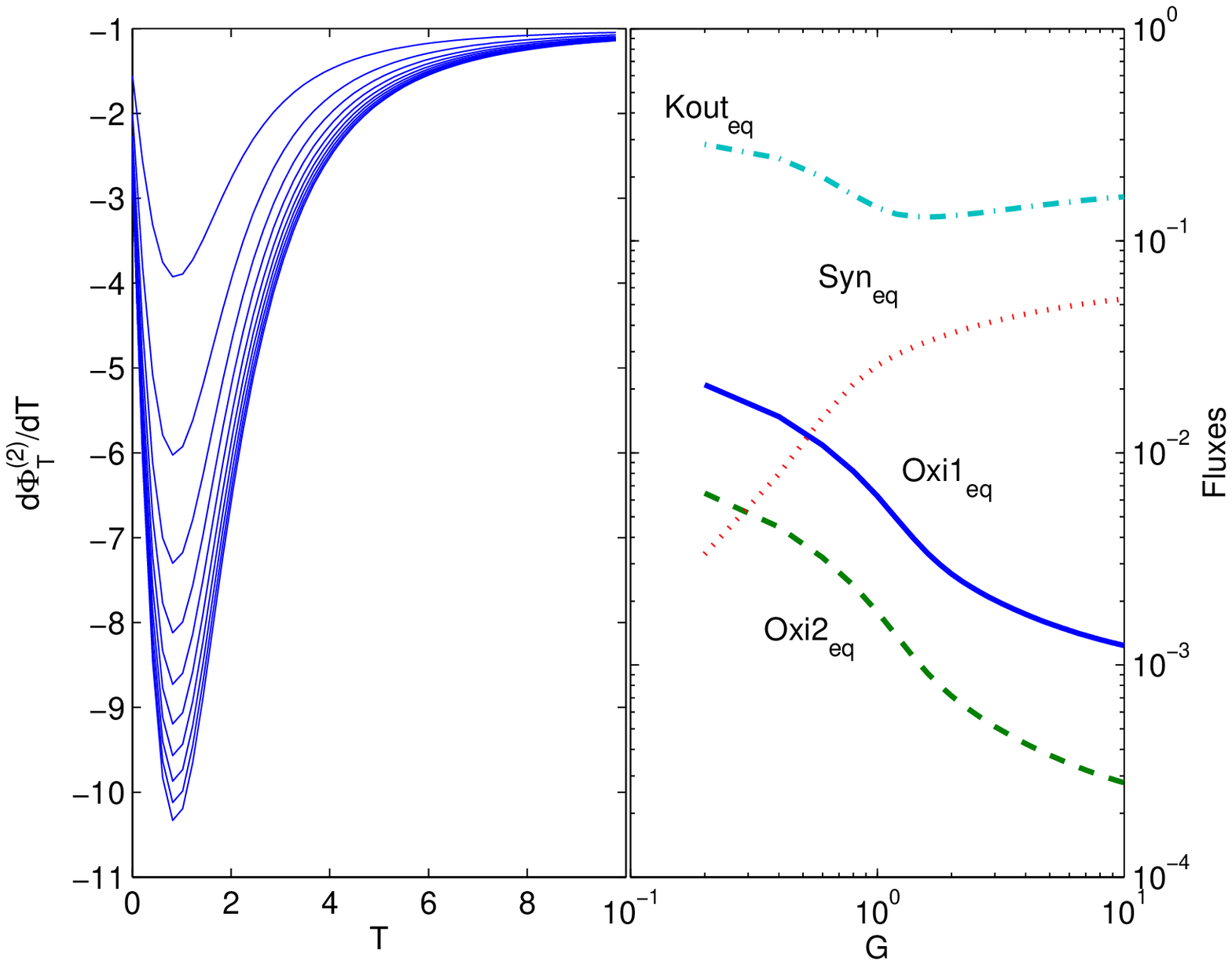}}

\centerline{a)  \hskip6truecm  b) \hskip4truecm  c)}

\caption{a) Response curves for varying food (glucose $\G$).
Quasi-stationarity (qs) corresponds to passing from a conventionally
chosen normal, fixed feeding level $\G = 10$ to a smaller, variable
value of $\G$, with enzyme levels unchanged from the normal feeding
equilibrium. Genetic effect on energy balance : the energy $\T$ is
lower at quasistationarity (no genetic regulation). Fatty acids
increase with fasting (decreasing $G$); this behavior is more
pronounced at quasistationarity. b) The strong lipolytic condition
meaning that $\DP{ \Phi_T^{(1)}  }{T}<0$ has been plotted as a
function of $T$, and for various values of $G$ ($0 \leq G \leq 10$).
c) Mode commutation; when food increases, the model commutes from a
oxidation dominated mode to a synthesis dominated
mode.}\label{Figresponse}
\end{center}
\end{figure}

\begin{figure}[ht]
\begin{center}
\includegraphics[width=10cm]{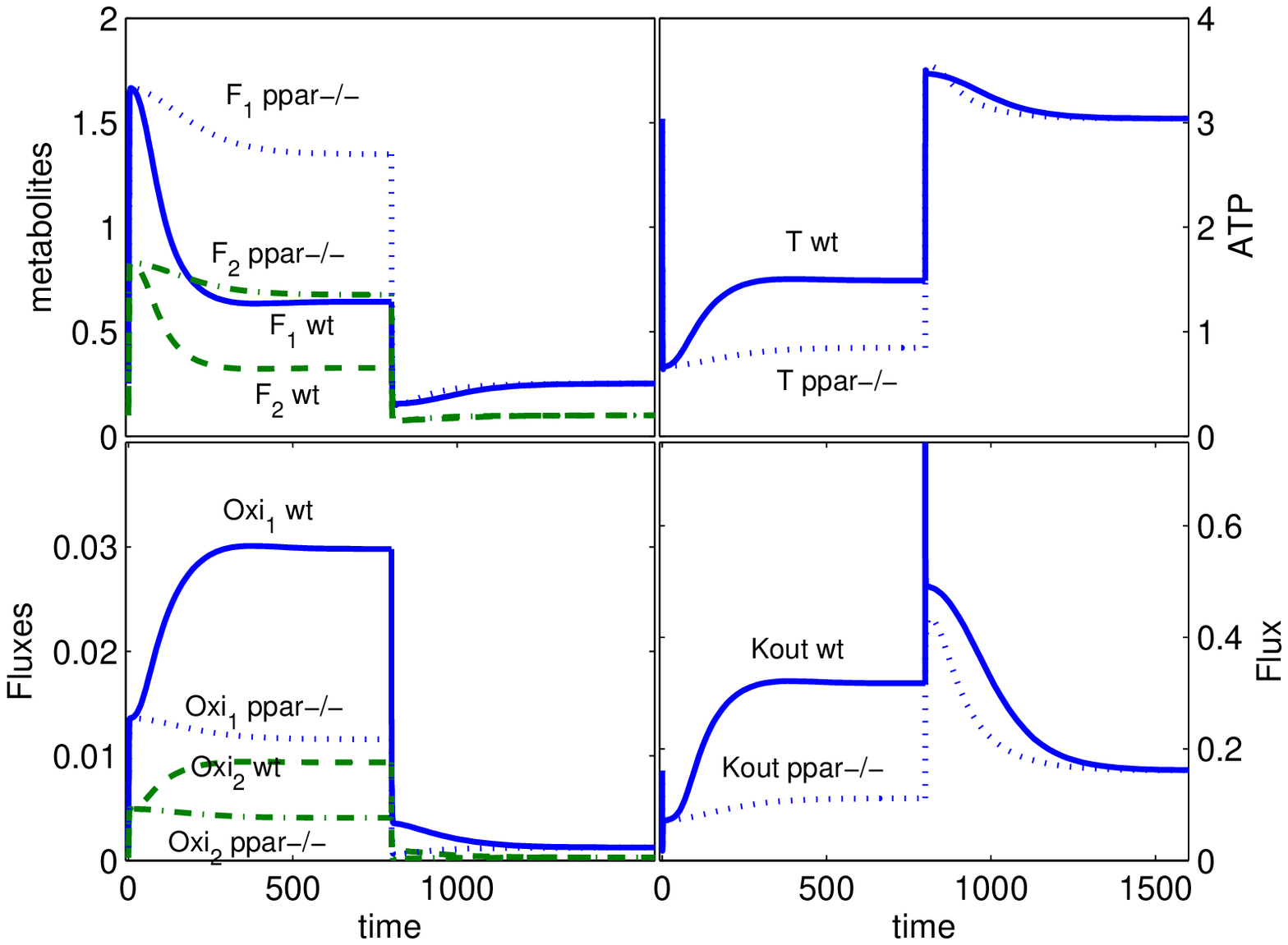}

\caption{Starving/refeeding protocol: dynamics of metabolites,
energy and fluxes for a PPAR-/- mutant compared to wild-type.
Fasting starts at $t=0$, refeeding starts at
$t=750$.}\label{figpparmutant}
\end{center}
\end{figure}

\subsection{Illustration of dynamics}

In order to illustrate the predictions of the previous section and
the dynamical behavior of the model, we move here to another level
of abstraction. The forms of the functions describing how primitive
fluxes depend on concentrations are chosen, including numerical
constants. The choice is rather generic, by no means precise.

\begin{itemize}
\item {\bf {\em Michaelis-Menten regulation functions.}}
The characteristic features of those functions are the following:
their value at $0$ is $0$; their derivative is positive and strictly
decreasing to $0$ at infinity. A typical analytical expression is:
$f(x)\,=\,\frac{kx}{K+x}$, where $k$ is the asymptotic value at
infinity and $k/K$ is the slope at the origin. This type of function
models the dependency of an primitive flux on its substrate.

\item {\bf {\em Repression functions.}} They are positive, strictly
decreasing functions, tending to $0$ at infinity and have an
inflexion point. A typical analytical expression is $f(x)\,=\,\frac
k{1+Kx^a}$, where $k$ is the maximal value (attained at $0$); $K$
controls the position of the inflexion point $x_\theta$: $x_\theta =
\left( \frac{a-1}{K(a+1)}\right)^{1/a}$. $a>1$ (also called Hill
coefficient) is the logarithmic slope at the inflexion point.

\item {\bf {\em Activation functions.}}
They are positive, strictly increasing bounded functions and have an
inflection point.
 A typical expression is $f(x)\,=\, b + \frac{k x^a}{K+x^a}$
with $K>1$. At $0$, the value is $b$; the function tends to $k+b$ at
infinity. $K$ controls the position of the inflexion point:
$x_\theta = \left( \frac{K(a-1)}{a+1}\right)^{1/a}$. $a>1$ is the
logarithmic slope at the inflexion point.

\item {\bf {\em Energy dependence functions.}} Some reactions start
when energy is missing and rapidly decrease to zero when energy
increases. Other, like synthesis, are energy stimulated. We suppose
that energetic regulation is  sigmoidal, rather than hyperbolic. We
use the following repressing function $f(x)\,=\,\frac l{L+x^2}$ and
the following activating function $g(x)\,=\,\frac {lx^2}{L+x^2}$.

\item {\bf {\em Degradation functions.}} All products are supposed to degrade
at a rate proportional to their concentration.

\item {\bf {\em Consumption of ATP.}} Concerning ATP, we suppose
that the consumption is linear.

\item  {\bf {\em Intake/outake of fatty acids.}} The case of intake/outake of fatty acids is special. Both reactions are reversible. We suppose that outake is proportional to the internal concentration of  fatty acids and intake occurs only when energy is missing inside the cell.
\end{itemize}

\noindent{{\bf Equations for metabolic fluxes}}
$$ \begin{array}{lll}
\Gly(\G,\T)  =  \fr{l_\Gly}{L_\Gly+\T^2}\fr{k_\Gly \G}{K_\Gly+\G}  &
\Krebs(\A,\T)  =  \fr{l_\Krebs}{L_\Krebs+\T^2}\fr{k_\Krebs
\A}{K_\Krebs+\A}
&  \Syn(\A,\T,\Eun)  =  \fr{k_{\Syn} \Eun \A}{K_\Syn+ \A} \frac{l_\Syn \T^2}{L_\Syn+\T^2} \\
\Oxideux(\Fdeux,\T,\Etrois)  =  \fr{l_{\Oxideux}}{L_{\Oxideux}+\T^2}\fr{k_{\Oxideux} \Etrois \Fdeux}{K_{\Oxideux}+\Fdeux} &\Oxiun(\Fun,\T,\Edeux)  =  \fr{l_{\Oxiun}}{L_{\Oxiun}+\T^2}\fr{k_{\Oxiun} \Edeux \Fun}{K_{\Oxiun}+\Fun} & \Deg\T= \delta_{\T} \T\\
{\Finun}(\Fun,\T)  =  -k_{{\Finun}}
\Fun+\fr{l_{\Finun}}{L_{\Finun}+\T^2} &{\Findeux}(\Fdeux,T)  =
-k_{\Findeux} \Fdeux+\fr{l_{\Findeux}}{L_{\Findeux}+\T^2}  &
\Kout(\A,\Equatre)  =  \fr{k_{\Kout} \Equatre \A}{K_\Kout+ \A}
 \end{array}$$


 \noindent{{\bf  Equations for genetic variables}}
All controls are positive activations except the control of $\Fdeux$
on $L$:
$$\begin{array}{lll} \widetilde\psi_1(\Fdeux)  =
\fr{k_{\PP}\Fdeux^{a_{\PP}}}{K_{\PP}+\Fdeux^{a_{\PP}}} + b_{\PP} &
\widetilde\psi_2(\Fdeux)  =  \fr{k_{\L}}{1+K_L\Fdeux^{a_\L}}
&\widetilde\psi_3(\L) =
\fr{k_{\Eun}L^{a_{\Eun}}}{K_{\Eun}+L^{a_{\Eun}}} + b_{\Eun} \\
\widetilde\psi_4(\PP) =
\fr{k_{\Edeux}\PP^{a_{\Edeux}}}{K_{\Edeux}+\PP^{a_{\Edeux}}} +
b_{\Edeux} &\widetilde\psi_5(\PP)  =
\fr{k_{\Etrois}\PP^{a_{\Etrois}}}{K_{\Etrois}+\PP^{a_{\Etrois}}} +
b_{\Etrois}
&\widetilde\psi_6(\PP)  = \fr{k_{\Equatre}\PP^{a_{\Equatre}}}{K_{\Equatre}+\PP^{a_{\Equatre}}} + b_{\Equatre}\\
\end{array}$$


The final system in given in Table \ref{systemexplicit}.

\begin{table}[ht]
{ $$\left\{\begin{array}{rcl}
 \fr{d\A}{dt} & = &   \fr{k_\Gly
\G}{K_\Gly+\G} + n_1 \fr{l_{\Oxiun}}{L_{\Oxiun} +
\T^2}\fr{k_{\Oxiun} \Edeux \Fun}{K_{\Oxiun} + \Fun} + n_2
\fr{l_{\Oxideux}}{L_{\Oxideux} + \T^2}\fr{k_{\Oxideux}
\Etrois \Fdeux}{K_{\Oxideux}+\Fdeux} - \fr{l_\Krebs}{L_\Krebs+\T^2}\fr{k_\Krebs \A}{K_\Krebs+\A} \\
&& - \fr{k_{\Kout} \Equatre \A}{K_\Kout+ \A} -
n_1 \fr{k_{\Syn}  \Eun \A}{K_\Syn+ \A} \frac{l_\Syn \T^2}{L_\Syn+\T^2} - \delta_\A \A    \\
\fr{d\Fun}{dt} & = &  \fr{k_{\Syn} \Eun \A}{K_\Syn+ \A} \frac{l_\Syn
\T^2}{L_\Syn+\T^2}
 -k_{\Finun} \Fun+\fr{l_{\Finun}}{L_{\Finun}+\T^2} -
 \fr{l_{\Oxiun}}{L_{\Oxiun}+\T^2}\fr{k_{\Oxiun} \Edeux \Fun}{K_{\Oxiun}+\Fun} - \delta_{\Fun}\Fun \\
\fr{d\Fdeux}{dt} & = & -k_{\Findeux}
\Fdeux+\fr{l_{\Findeux}}{L_{\Findeux}+\T^2} -
\fr{l_{\Oxideux}}{L_{\Oxideux}+\T^2}\fr{k_{\Oxideux} \Etrois
\Fdeux}{K_{\Oxideux}+\Fdeux}
- \delta_{\Fdeux}\Fdeux  \\
\fr{d\T}{dt} & = &  \alpha_\G \fr{l_\Gly}{L_\Gly+\T^2}\fr{k_\Gly \G}{K_\Gly+\G}+ \alpha_\K \fr{l_\Krebs}{L_\Krebs+\T^2}\fr{k_\Krebs \A}{K_\Krebs+\A} + \alpha_{\Oun}\fr{l_{\Oxiun}}{L_{\Oxiun}+\T^2}\fr{k_{\Oxiun} \Edeux \Fun}{K_{\Oxiun}+\Fun} \\
&& + \alpha_{\Odeux}
\fr{l_{\Oxideux}}{L_{\Oxideux}+\T^2}\fr{k_{\Oxideux}
\Etrois \Fdeux}{K_{\Oxideux}+\Fdeux} - \alpha_{S} \fr{k_{\Syn}  \Eun \A}{K_\Syn+ \A} \frac{l_\Syn \T^2}{L_\Syn+\T^2}- \delta_\T\T  \\
\fr{d\PP}{dt} & = &  \fr{k_{\PP}\Fdeux^{a_{\PP}}}{K_{\PP}+\Fdeux^{a_{\PP}}}\,-\,\delta_{\PP}\PP + b_{\PP} \\
\fr{d\L}{dt} & = &   \fr{k_{\L}}{1+ K_L\Fdeux^{a_\L} }   \,-\,\delta_{\L}\L\\
\fr{d\Eun}{dt} & = &  \fr{k_{\Eun}L^{a_{\Eun}}}{K_{\Eun}+L^{a_{\Eun}}}\,-\,\delta_{\Eun}\Eun  + b_{\Eun}\\
\fr{d\Edeux}{dt} & = & \fr{k_{\Edeux}\PP^{a_{\Edeux}}}{K_{\Edeux}+\PP^{a_{\Edeux}}} \,-\,\delta_{\Edeux}\Edeux + b_{\Edeux} \\
\fr{d\Etrois}{dt} & = &  \fr{k_{\Etrois}\PP^{a_{\Etrois}}}{K_{\Etrois}+\PP^{a_{\Etrois}}}\,-\,\delta_{\Etrois}\Etrois + b_{\Etrois}\\
\fr{d\Equatre}{dt} & = & \fr{k_{\Equatre}\PP^{a_{\Equatre}}}{K_{\Equatre}+\PP^{a_{\Equatre}}} \,-\,\delta_{\Equatre}\Equatre + b_{\Equatre} \\
\end{array}\right.$$}
\caption{A generic example for the model of the regulated
metabolism.}\label{systemexplicit}
\end{table}

\medskip
\noindent {\bf Simulation of fasting/refeeding protocols} In
Fig.\ref{Figconcentrations} we have simulated a fasting/refeeding
protocol. One can notice the increase of fatty acids on fasting with
an overshoot (as predicted in Prop. \ref{lemmeCompT} and its
Cor.\ref{corrCompT}).

 The dynamics has two timescales: a quick
increase up to the maximum, then a slow decrease. The concentration
of Acetyl-coA is decreasing at fasting. The energy (ATP) has an
abrupt fall, then it recovers slowly as a result of oxidation. The
behavior of nuclear receptors correspond to what is experimentally
observed \cite{jump04}: LXR diminishes and PPAR is amplified at
fasting. This leads to variations of the enzymes: oxidation
$\Edeux,\Etrois$ and ketone exit enzymes $\Equatre$ are amplified,
the synthesis enzyme $\Eun$ is diminished. Again, this fits with
experimental observations \cite{jump04}. Furthermore, fluxes have
textbook behavior \cite{salway}. Oxidation and ketone exist occurs
during fasting, while synthesis occurs during normal feeding. During
fasting fatty acids enter the cell $\Finun, \Findeux > 0$. At normal
feeding, $\Finun$ changes sign (de novo synthesized fatty acids
exit) and $\Findeux$ vanishes. The ketone overshoot at refeeding can
be explained by enzyme inertia. The high fasting level of enzyme
$\Equatre$ can not drop immediately. As large amounts of Acetyl-coA
are again available, this boosts the ketone production.

The reader should be warned of a possibility not studied in this
paper. Two timescales dynamics, involving the observability of the
quasistationary states and the rapid overshoots and undershoots in
the dynamics of metabolites and fluxes, can be avoided. For
instance, glucose homeostasis could be responsible of slow instead
of steep decrease of glucose at fasting. A slow input signal will
drive the system quasi-statically, avoiding quasistationarity and
rapid transients. As we do not study glucose dynamics and
homeostasis, our model can not tell how glucose behave in time. This
information should be provided by experiment.

In Fig.\ref{Figresponse} we have simulated response curves of
various metabolites, fluxes and energy when food $\G$ is changing.
It can be noticed that fatty acids concentrations decrease with food
and that Acetyl-coA concentration increases with food. Energy $\T$
is increasing with food. There is a buffering effect, that preserves
energy against variations of food: energy $\T$ is not zero when food
$\G$ is zero. As discussed in Prop. \ref{lemmeCompT2}, a strong
buffering effect means a weak slope of the dependence of $\T$  on
$\G$. Genetic regulation decreases this slope, therefore increases
buffering as can be seen by comparing the curves in
Fig.\ref{Figresponse}a) at quasi-stationarity and at equilibrium.
The antagonistic relation between synthesis and oxidation is
illustrated in Fig.\ref{Figresponse}c): when food $\G$ decreases,
the synthesis dominated regime changes to an oxidation dominated
regime.

\medskip
\noindent {\bf Satisfiability of the uniqueness and strong lipolytic
conditions} Our sufficient uniqueness conditions, allow, with small
computational effort, to check the uniqueness of equilibrium for a
given set of parameters. We have checked  the strong lipolytic
response condition for various sets of parameters. We have noticed
that this condition is robust (see Fig.\ref{Figresponse}b)).

\medskip
\noindent {\bf Simulation of the effect of PPAR mutations} We have
modeled a PPAR mutant by considering that the enzymes
$\Edeux,\Etrois,\Equatre$ controlled by PPAR have constant,
unadjustable values. We have considered that the values of
$\Edeux,\Etrois,\Equatre$ are those for the normal feeding
equilibrium state in wild type hepatocytes. In
Fig.\ref{figpparmutant} the dynamics during a fasting/refeeding
protocol are compared in the mutant and wild type case. The main
feature of
 mutants is the difficulty to recover energy at fasting (see
 Fig.\ref{figpparmutant}~b)). This is the consequence of
  inefficient oxidation (notice the low oxidation fluxes in
 Fig.\ref{figpparmutant}~c)). Similarly, in mutants the ketone production
 is decreased ( see
 Fig.\ref{figpparmutant}~d)) and the fatty acids increase at fasting is
 more pronounced  just like we have predicted qualitatively in
 Prop. \ref{pparko} and its Cor.\ref{corrpparko}(see
 Fig.\ref{figpparmutant}~a)).


\section{Discussion}

Let us first summarize some of the characteristics of our model.
\begin{itemize}
\item It is an {\em integrative model}, because it takes into account all the main
processes of carbohydrate and fatty acid metabolism in liver
(glycolysis, lipogenesis, Krebs cycle,
 fatty acids mobilization, oxidation, ketogenesis) together with
their various regulation (metabolic, genetic).
\item Our model is {\em not
distributed}: dynamical variables cope for average values in a
tissue and no space information is taken into account.
\item The model that we propose is a {\em low complexity abstraction}. It has
just enough complexity to represent the basic features of metabolism
in the main nutritional states. Whenever possible, complex metabolic
chains of reactions were modeled as a single global reaction
preserving the overall balance of products and reactants. Our main
concern was to keep the model as qualitative as possible.
\item {\em Our predictions are not dependent on specific numerical values} of
kinetic constants, on specific reaction mechanisms, or on specific
forms of the functions relating fluxes to concentrations. We rather
replace this information by sufficient qualitative conditions that
are chosen as biologically significant as possible.
\item A qualitative approach has been
used to discuss {\em response properties}. The introduction of the
two types of states (quasi-stationary for rapid response and
equilibrium for slow response) allowed us to distinguish between
quick metabolic  and slow genetic response.
\end{itemize}

Our model copes with the main experimental findings on the behavior
of regulated fatty acid metabolism in hepatocytes. Under fasting,
the model shifts from a synthesis dominated regime to an
oxidation/lipolysis dominated regime. This shift stabilizes energy,
replacing food supply by reserve consumption. At short times, the
shift is performed by metabolic  control of synthesis, lipolysis and
oxidation. At longer times, the regulatory effect of an increase of
intracellular PUFA on the nuclear receptors PPAR and LXR reinforces
this control. Refeeding  shifts the system in the opposite
direction.

Our model is sufficiently general to apply to various higher
organisms. Nevertheless, there are some biases in this model.
More precisely, we have considered only hepatocytes, which is
justified by the fact that in various species like chicken, rodents
and humen, fatty acid synthesis and oxidation occur in the same
organ (liver).
This is not so for other species such as pigs, for which synthesis takes place mainly in adipocytes. 

We have also proposed a methodology to build small complexity
abstractions that integrate various qualitative aspects of regulated
metabolism. These abstractions are by no means rigid. On the
contrary, they are evolutive and can integrate new experimental
results. As an example of possible evolution of the model presented
here we should mention the role of various PUFA, already discussed.

\section{Method}

From here on, we develop the mathematical setting and arguments
which lead to the results announced before.

\subsection{Switches, shifts and obervability of equilibria}\label{secTheoryExp}

\begin{figure}[ht]
\begin{minipage}{5cm}

{\small A shift: the $0$-level curve is a sigmoid}

\end{minipage}
$\,$
\begin{minipage}{5cm}
\begin{center}
\includegraphics[width=4cm]{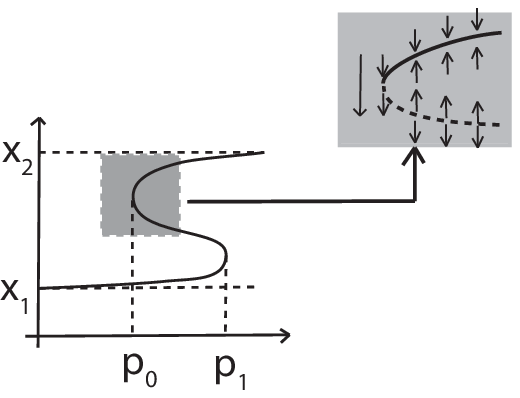}
\end{center}

{\small A  switch: the $0$-level curve is not a graph. Around
${\vect{p}}_0$ and ${\vect{p}}_1$ a saddle-node bifurcation occurs.}
\end{minipage}
$\,$
\begin{minipage}{5cm}

{\small Hysteresis effect: the experimental curves.}
\end{minipage}
\caption{Switches and shifts}\label{f.hysteresis}
\end{figure}

\noindent {\bf Equilibrium switches and shifts.} The two types of
equilibrium changes shift and switch can be mathematically described
as follows. Let $\vect{Z}=(\vect{X},\vect{Y})$ and
$F(\vect{Z},\vect{p})$,
 $\vect{Z} \in U \subset {\mathbb R}^n$,  $\vect{p} \in I \subset {\mathbb R}$,   be a
differentiable vector field defined on an open set $U$ of ${\mathbb
R}^n$ depending smoothly on a  parameter evolving in an open
interval $I$. We suppose that for any $\vect{p}$ in a subinterval
$J\subset I$, the vector field $F_{\vect{p}}: \vect{Z} \to
F(\vect{Z},\vect{p})$ admits at least a singular (equilibrium)
point, that is: there exist $\vect{Z}\in U$ such that
$F_{\vect{p}}(\vect{Z})=0$. We call the $0$-level (equilibrium)
curve:
\begin{equation}\label{pointsequilibre}
{\mathcal{L}}\,=\,\{\,({\vect{p}},\vect{Z})\in J\times
U\,,\,F_{\vect{p}}(\vect{Z})=0\,\}.
\end{equation}
$\mathcal{L}$ is a differentiable curve in $J\times U$. According to
its shape, we get a switch or a shift. More precisely,
\begin{enumerate}
\item A {\em shift} is characterized by the following features:
\begin{itemize}
\item $\mathcal{L}$ is a graph $\vect{Z}=L({\vect{p}})$. Equivalently it means that for each value of ${\vect{p}}\in J$ there exist a {\em unique} equilibrium point,
\item $\mathcal{L}$ is a sigmoid. That is there exist a threshold value ${\vect{p}}_0$, an interval $]{\vect{p}}_0-\delta, {\vect{p}}_0+\delta[$, two equilibrium values $\vect{Z}_1$ and $\vect{Z}_2$ and $\epsilon>0$ such that for ${\vect{p}}<{\vect{p}}_0-\delta$, $\norm{L({\vect{p}})-\vect{Z}_1}<\epsilon$ and for ${\vect{p}}>{\vect{p}}_0+\delta$, $\norm{L({\vect{p}})-\vect{Z}_2}<\epsilon$.
\end{itemize}
See an illustration in Fig. \ref{f.hysteresis}.

The  sigmoid shape of the level curve has two consequences: a {\em
"jump" effect} and  {\em reversibility}. If we start with a certain
value of the parameter ${\vect{p}}_I$, say
${\vect{p}}_I<{\vect{p}}_0-\delta$, and increase smoothly the
parameter up to a final value ${\vect{p}}_F>{\vect{p}}_0+\delta$, we
first observe an asymptotic state close to $\vect{Z}_1$ and a sudden
"jump" to an asymptotic state close to $\vect{Z}_2$. We insist on
that the apparent "jump" is due to  the steepness of $\mathcal{L}$
at ${\vect{p}}_0$ which induces a small $\delta$: in reality there
is no discontinuity. Now if we start with the value ${\vect{p}}_F$
and decrease smoothly the parameter, we note exactly the reverse: a
jump at ${\vect{p}}_0$ from an asymptotic state close to
$\vect{Z}_2$ to an asymptotic state close to $\vect{Z}_1$, a
property so-called reversibility.

Notice that both properties are often used to identify shift like
phenomena.

\item A {\em switch} has the following properties:
\begin{itemize}
\item $\mathcal{L}$ is not a graph. In particular, there
exist two parameter values ${\vect{p}}_0$ and ${\vect{p}}_1$ for
each of which the vector field has two singularities (equilibria)
and for any ${\vect{p}}\in]{\vect{p}}_0,{\vect{p}}_1[$, the vector
field $F_{\vect{p}}$ has three singularities (one unstable and two
stable). For ${\vect{p}}<{\vect{p}}_0$ or ${\vect{p}}>{\vect{p}}_1$
there is only one singularity (stable).
\item ${\vect{p}}_0$ and ${\vect{p}}_1$ are bifurcation points of saddle node type, that
is  an attracting and a repelling singularity collide and disappear.
\item There exist $\delta>0$, two equilibrium values $\vect{Z}_1$ and $\vect{Z}_2$
and $\epsilon>0$ such that for ${\vect{p}}<{\vect{p}}_0-\delta$,
$\norm{L({\vect{p}})-\vect{Z}_1}<\epsilon$ and for
${\vect{p}}>{\vect{p}}_1+\delta$,
$\norm{L({\vect{p}})-\vect{Z}_2}<\epsilon$.
\end{itemize}
To understand the hysteresis effect implied by the shape of the
curve ${\mathcal L}$ it is worth considering the experimental curve,
that is, the curve of observed equilibria when moving the parameter.
If we start with a certain value of the parameter ${\vect{p}}_I$,
say ${\vect{p}}_I<{\vect{p}}_0-\delta$, and increase smoothly the
parameter up to a final value ${\vect{p}}_F>{\vect{p}}_0+\delta$, we
first observe an asymptotic state close to $\vect{Z}_1$ and a sudden
"jump" to an asymptotic state close to $\vect{Z}_2$. To the contrary
to the previous shift situation, this jump is a real discontinuity.
Notice that the jump occurs at the parameter value $p_1$. Now if we
start with the value ${\vect{p}}_F$ and decrease smoothly the
parameter, we note a jump at ${\vect{p}}_0$ from an asymptotic state
close to $\vect{Z}_2$ to an asymptotic state close to $\vect{Z}_1$.
We do not have reversibility, because the jump occurs for different
critical values of ${\vect{p}}$. The curves in both cases are
depicted in Fig. \ref{f.hysteresis}.
\end{enumerate}

\medskip

\noindent {\bf Experimental curves and equilibria.} Consider now the
differential system:

\begin{equation}\label{systemF}
\D{\vect{Z}}{t}= \vect{F}(\vect{Z}, p), \quad \vect{Z}\in \R_+^n, p
\in I \subset \R
\end{equation}

The points on the 0-level curve of the vector field $\vect{F}$ are
equilibria of the dynamical system (Eq.\eqref{systemF}). In the
following we consider the simple situation when for any
 $p \in I$ there is a unique attractive equilibrium $\vect{Z}_0(p)$, of open
 attractive basin $B(p)$. Let us recall that the attraction basin
 consists of all $\cal{Z}$ such that the trajectory $\vect{u}(t,\vect{Z},p)$
 of the system \eqref{systemF}, starting at $\vect{Z}$, approaches $\vect{Z}_0(p)$
 for large times
 $| u(t,\vect{Z},p) - \vect{Z}_0(p) | \to 0 \quad (t \to
 \infty)$.

A typical experimentation consists in starting in an equilibrium
$\vect{Z}_0(p_1)$ and suddenly changing the value of the parameter
$p$ from $p_1$ to $p_2$. Suppose that the following condition is
fulfilled $\vect{Z}_0(p_1) \in B(p_2)$. This condition is
automatically fulfilled if for instance $\vect{Z}_0(p_2)$ is
globally attractive, i.e. $B(p_2) = \R_+^n$. Then, the observed
state $\vect{u}(t,\vect{Z}_0(p_1),p_2)$ will approach the state
$\vect{Z}_0(p_2)$ on the 0-level curve after a long enough time.
Suppose now that the 0-level curve $\vect{Z}=L(p)$ is such that the
component $L_i(p)$ satisfies $\D{L_i}{p} >0$ for any $p \in I$.
Then, it exists $T$ such that $u_i(t,\vect{Z}_0(p_1),p_2) >
(Z_0)_i(p_1)$ for $t > T$, meaning that we observe an increase of
the component $i$ of the state of the system between the beginning
and the end of the experimentation.

Hence, {\em the shape of the 0-level curve informs on variations of
products during an experiment.} This justifies that we can predict
experimental behaviors of the system from the study of its
equilibria.

\medskip

\noindent {\bf Observability of quasi-stationarity.} Let us suppose
now that the system \eqref{systemF} has two time scales, one slow
and one fast. This can be taken into account by supposing that there
is a small parameter $\epsilon$ representing the fast time scales
and that the system \eqref{systemF} reads $\D{\vect{X}}{t}=
\vect{\Phi}(\vect{X},\vect{Y}, p),
 \D{\vect{Y}}{t}= \epsilon \vect{\Psi}(\vect{X},\vect{Y}, p)$,
where  $\vect{X}\in \R^{n_f}, \vect{Y}\in \R^{n_s}$, $n_f+n_s=n$.
Considering the time scale $\tau = \epsilon t$ we arrive to the more
classical form:

\begin{equation}\label{systemFsf}
 \epsilon  \D{\vect{X}}{\tau}= \vect{\Phi}(\vect{Z}, p),
 \D{\vect{Y}}{\tau}= \vect{\Psi}(\vect{Z}, p)
\end{equation}

A result due to Tikhonov \cite{tikhonov} and reformulated by
Fenichel \cite{fenichel79} implies that the system \eqref{systemFsf}
has a remarkable behavior. Suppose that the following two conditions
are satisfied:

\begin{itemize}
\item
For fixed $\vect{Y}=\vect{Y}_0$, the system  $\D{\vect{X}}{\tau}=
\vect{\Psi}(\vect{X},\vect{Y}_0, p)$ has an attractor
$X_0(\vect{Y}_0,p)$ that satisfies the relation
$\vect{\Psi}(\vect{X}_0(\vect{Y}_0,p),\vect{Y}_0, p) =0$. The
equation $\vect{X}=\vect{X}_0(\vect{Y}_0,p)$ defined the slow
manifold $\Sigma_0(p)$.
\item
The Jacobian matrix $D_\vect{X}\vect{\Psi}$ admits $n_f$ eigenvalues
with  strictly negative real parts at
$(\vect{X}_0(\vect{Y}_0,p),\vect{Y}_0)$.
\end{itemize}
Then the system \eqref{systemFsf} admits an attractive invariant
manifold $\Sigma(\epsilon,p)$ close to the slow manifold
$\Sigma_0(p)$. Thus, trajectories sufficiently close to the slow
manifold converge to it. The smaller $\epsilon$ is, the quicker is
the convergence. Notice that the slow manifold defines the
quasi-stationary states. The Tikhonov-Fenichel result means that the
system rapidly tends to quasi-stationarity.

\subsection{Existence of equilibrium (Theorem \ref{existenceFixedPoint})}
 The proof of Theorem \ref{existenceFixedPoint}  uses the following more general theorem:

\begin{theorem}\label{fixedpoint}
Let $\vect{\Phi}(\vect{X}) = \vect{G(\vect{X})} -
\vect{\Lambda}(\vect{X}) $ be a smooth vector field on ${\mathbb
R}^n_+$ (${\mathbb R}^n_+$ represents all the vectors of ${\mathbb
R}^n$ having non-negative coordinates) such that :
\begin{enumerate}
\item $\vect{G}$ is  bounded,
\item For all $\vect{X}=(X_1,\dots, X_n)$ such that $X_i=0$ and $X_j\neq 0$ for all $j\neq i$, $\vect{G}$  satisfies $G_i(\vect{X})>0$,
 \item $\vect{\Lambda}=(\Lambda_1(X_1),\ldots,\Lambda_n(X_n)) : {\mathbb R}^n_+ \to {\mathbb R}^n_+$,
 and $\Lambda_i$ are differentiable and satisfy  $\Lambda_i(0)=0$ and  $\lim_{\|\vect{X}\| \to +\infty}
 \Lambda_i (\vect{X})=+\infty$, for all $1\leq i \leq n$.
\end{enumerate}
Then the equation $\vect{\Phi}(\vect{X})=0$ has at least one
solution in ${\mathbb R}^n_+$.
\end{theorem}
 The proof of the Theorem is based on the following standard mathematical lemma.

 \begin{lemma}\label{l.indice}
 Let $\DD$ be a smooth ball in $\R^n$ and let $\S$ be the boundary of $\DD$. Let $\Phi$ be a
 differentiable vector field defined on a neighborhood of $\DD$. If $\vect{\Phi}$ points inward $\DD$
 at any point of $\S$ then $\vect{\Phi}$ admits a zero in the interior of $\DD$.
 \end{lemma}
 \noindent{{\bf Sketch of the proof of Lemma \ref{l.indice}.}}
By the Poincar\'e-Hopf formula a sufficient condition for having a
zero in the interior of $\DD$ is to have a non-zero index for the
vector field on $\S$. Since $\vect{\Phi}$ points inward $\DD$ on
$\S$, we can construct a smooth change of variables which conjugates
$\vect{\Phi}$ on a neighborhood of $\DD$ to a vector field
$\vect{\Phi}'$ defined on a neighborhood of the unit $n-$ball ${\bf
B}^n$, such that on a neighborhood of the unit $n-$sphere ${\bf
\S}^n$, $\vect{\Phi'}$ coincides with the radial vector field
$\vect{X}\mapsto-\vect{X}$. For this vector field $\vect{\Phi'}$, we
can compute its index, which is $1$ or $-1$ according to the parity
of $n$. The Lemma is proved since the index is a differential
invariant. \cqfd

\medskip

\noindent{{\bf Proof of Theorem \ref{fixedpoint}.}} From Lemma
\ref{l.indice}, it is enough to find a smooth ball in the positive
orthant on the boundary of which the vector field $\vect{\Phi}$
points inwards.

For $R>0$, let us consider the intersection domain of the closed
$n$-ball of radius $R$ with the positive orthant:
$\Delta\,=\,\{\vect{X}\in\R^n_+\,,\,\|\vect{X}\|\leq R\,\}$. This
domain is a topological ball; let us denote $\Sigma$ its boundary.
If $\vect{X}\in \Sigma$ and none of its components is $0$, then for
$R$ large enough, $\vect{\Phi}(\vect{X})$ points inward $\Delta$,
because $G$ is bounded and $\Lambda_i(\vect{X})$ tend to infinity
with $\vect{X}$, hence $\Phi_i(\vect{X})<0$, for all $1 \leq i \leq
n$. On the other hand, if only one of the components of $\vect{X}$
is $0$, then by hypothesis (2), $\vect{\Phi}(\vect{X})$ points
inward $\Delta$. Since the set of points where the property of
pointing inwards is open, we can find a smooth ball $\DD$ contained
in $\Delta$ and sufficiently close to it, such that on the boundary
of $\DD$, the $\vect{\Phi}$ points inward $\DD$. \cqfd

\medskip

Actually, the Theorem implies naturally a stronger result which will
be useful in the proof of uniqueness of equilibrium.

\begin{corollary}\label{Corfixedpoint}
Under the hypotheses of Theorem \ref{fixedpoint}, let
$\vect{X}=(\vect{X_1}, \vect{X_2})$ be any partition of the
variables. We write $\Phi_1(\vect{X_1}, \vect{X_2})$ for the
projection of $\Phi(\vect{X_1}, \vect{X_2})$ in the vector space
spanned by the coordinates of $\vect{X_1}$. Given $\vect{X_2}$, the
system of equations
  $\Phi_1(\vect{X_1}, \vect{X_2})=0$, where $\vect{X_2}$ is considered
as a constant parameter vector, admits a solution in $\vect{X_1}$
with non negative entries.
\end{corollary}

\noindent{{\bf Proof.}} Theorem \ref{fixedpoint} applies to the
vector field
 $\Phi_1(\vect{X_1}, \vect{X_2})$, where $\vect{X_2}$ is considered
as a constant parameter vector, since $\Phi_1$ satisfies the
hypotheses of the Theorem \ref{fixedpoint} as soon as $\Phi$
satisfies them. \cqfd

\medskip

\noindent{{\bf Proof of Theorem \ref{existenceFixedPoint}.}} The
proof runs by applying Theorem \ref{fixedpoint} to the following
vectors:
$$\vect{\Lambda} =
\begin{pmatrix}
\Deg\Fun  \\
\Deg\Fdeux                \\
\Deg\T \\
\Deg\A
\end{pmatrix} \qquad \vect{G}=\begin{pmatrix}
\Syn_{\red} - \Oxiun_{\red} + {\Finun}  \\
- \Oxideux + {\Findeux}      \\
\alpha_\G \Gly - \alpha_S Syn + \alpha_\K \Krebs +
\alpha_{\Oun} \Oxiun+ \alpha_{\Odeux} \Oxideux   \\
\Gly + n_1 \Oxiun + n_2 \Oxideux  -\Krebs -\Kout -n_1 \Syn
\end{pmatrix}$$

Notice that equilibrium states of the implicit models are the zeroes
of the vector field $\vect{\Phi}(\vect{X}) = \vect{G(\vect{X})} -
\vect{\Lambda}(\vect{X})$.

Let us verify hypotheses of Theorem \ref{fixedpoint}. First,
$\vect{G}$ is differentiable because all fluxes are differentiable.
Then, $\vect{G}$ is bounded because it is composed of primitive
fluxes which saturate at high concentrations of metabolites
(Condition \ref{bflux}). Finally, to verify the condition
$G_i(\ldots,X_i=0,\ldots)>0$, it is enough to notice that each
coordinate can be decomposed into the difference of the fluxes which
produce the variable and the fluxes which consume the variable. The
second ones are zero when the variable is zero by zero substrate
effect (Condition \ref{bflux}). The sum of the producing fluxes is
strictly positive by recovery effect (Condition \ref{bflux}).

The assumptions on $\vect{\Lambda}$ are  satisfied by linearity of
degradation terms and by unboundedness of ATP consumption term
(Condition \ref{bflux}).\cqfd




\subsection{Box reduction of systems of non-linear equations}\label{Maths}

Let $\vect{\Phi}: \R^n \times \Delta  \to \R^n$, where $\Delta$ is a
compact subset of  $\R^q$, be a differentiable vector field.
$\vect{\Phi}$ defines the following system of linear equations
parametrized by $p$:

\begin{equation} \label{system}
 {\cal S}: \vect{\Phi}(\vect{X},\vect{p})=0
\end{equation}





\noindent{{\bf Box of a system of equations.}} We call box of the
system \eqref{system} a subset $\vect{X}^{(i)}$ of the set of
variables $\vect{X}$, such that $\vect{X} =
(\vect{X}^{(i)},\vect{X}^{(e)})$ is a partition of the set of
variables. The variables $\vect{X}^{(i)}$, $\vect{X}^{(e)}$ are
called internal and external  variables, respectively.
To each partition of the variables, let us consider the
corresponding partition of the vector field components
$\vect{\Phi}=(\vect{\Phi}^{(i)},\vect{\Phi}^{(e)})$.

We call {\em box equilibration} the elimination of internal
variables from the equations defined by the internal part of the
vector field:
$\vect{\Phi}^{(i)}(\vect{X}^{(i)},\vect{X}^{(e)},p)=0$.

\medskip

\noindent{{\bf Sequence of box equilibration.}} After a box
equilibration the internal variables can be expressed as functions
of the external variables. A {\em sequence of box equilibrations} is
the finite iteration of the following operations:

\begin{enumerate}
\item
Define $\vect{X}_1=\vect{X}$ and
$\vect{\Phi_1}(\vect{X}_1,\vect{p})=\vect{\Phi}(\vect{X},\vect{p})$,
${\cal D}_1(\vect{p})=\R^n$.
\item
At $k$-th iteration, divide the variables and the vector field
components into internal and external parts $\vect{X}_k =
(\vect{X}_k^{(i)}, \vect{X}_k^{(e)})$, $\vect{\Phi}_k =
(\vect{\Phi}_k^{(i)}, \vect{\Phi}_k^{(e)})$.
\item If the external part is not empty then:
\begin{itemize}
\item
solve
 $\vect{\Phi}_k^{(i)}(\vect{X}_k^{(i)},\vect{X}_k^{(e)},\vect{p})=0$
with the constraint  $(\vect{X}_k^{(i)},\vect{X}_k^{(e)}) \in {\cal
D}_k(\vect{p})$ and express the internal variables as functions of
the external variables $\vect{X}_k^{(i)}=\vect{{\cal
M}_k}(\vect{X_k^{(e)}},\vect{p})$. Notice that the solution might
not be unique, that is $\vect{{\cal M}_k}$ is not necessarily
univocal. We restrict our discussion to the case when the number of
solutions is finite and bounded, such as for polynomial systems.
 Also, notice
that one has a solution for $\vect{X_k^{(e)}}$ in a maximal domain
${\cal D}_{k+1}(\vect{p})$. If ${\cal D}_{k+1}(\vect{p})$ is empty
then stop: there is no solution.
\item
define  $\vect{X_{k+1}}=\vect{X_k^{(e)}}$, and $\vect{ \Phi_{k+1} }
= \vect{ \Phi_{k}^{(e)} } (\vect{{\cal M}_k}( \vect{ X_n^{(e)} },
\vect{p}), \vect{ X_n^{(e)} }, \vect{p} )$.
\end{itemize}
\item
If the external part is empty  then solve
$\vect{\Phi}_k^{(i)}(\vect{X}_k^{(i)},\vect{p})=0$ and  stop.
Conventionally, in this case ${\cal D}_{k+1}(\vect{p})$ is
considered non-empty iff the equation has a solution.
\item
go to step 2.
\end{enumerate}


A sequence of box equilibrations is {\em complete} if all components
are equilibrated i.e.
$$\vect{X} =\vect{X}_1^{(i)} \oplus \vect{X}_2^{(i)} \oplus \ldots
\oplus \vect{X}_{N_b}^{(i)}.$$

After a complete sequence of box equilibrations one should be able
to express metabolite levels as functions of the external
parameters: $\vect{X} = \vect{\cal M}(\vect{p})$, where $\vect{\cal
M}$ results from a composition of the functions $\{\vect{{\cal
M}_k}\}_{k=1,N_b}$.

An example of (incomplete) sequence of box equilibrations is the
reduction (at equilibrium) of genetic variables in the mixed
metabolic/genetic differential system. There is only one box whose
internal variables are the genetic variables $\vect{Y}$. These are
eliminated from the equations
$\vect{\psi}(\vect{X},\vect{Y},\vect{p})=0$. The functions ${\cal
M}_1(\vect{X},\vect{p})$ are $\vect{Y}^{peq}(\vect{X},\vect{p})$,
and the reduced fluxes are
$\vect{\Phi}_2(\vect{X},\vect{p})=\vect{\Phi}(\vect{X},\vect{Y}^{peq}(\vect{X},\vect{p}),\vect{p})$.

\medskip

\noindent{{\bf Existence and uniqueness of solutions.}}

Box equilibrations perform nothing else than the substitution method
for non-linear systems of equations. The existence and properties of
solutions relatively to box equilibrations are straightforward.

\begin{proposition}
\begin{itemize}
\item  A solution of the system \eqref{system} exists for a value of the parameter
$p$ if there is a complete sequence of box equilibrations with
non-empty domains ${\cal D}_{k+1}(\vect{p})$.
\item The function $\vect{\cal M}$ is univocal (to one $\vect{p}$ corresponds
a single value of $\vect{\cal M}$) if all the domains ${\cal
D}_{k+1}(\vect{p})$ are non-empty and each one of the function
$\vect{{\cal M}_k}$ is univocal on its
 maximal domain ${\cal D}_{k+1}(\vect{p})$ for a complete sequence of box
equilibrations.
\end{itemize}
\label{boxcondition}
\end{proposition}

This  property is useful to prove the existence and uniqueness of
solutions of systems of non-linear equations. It is enough to choose
a complete sequence of box equilibrations and to show that at each
step the functions $\vect{{\cal M}_k}$  are univocal on non-empty
domains ${\cal D}_{k+1}(\vect{p})$.

It is difficult to give a "only if" version of the property. Indeed,
even if we find a box such that the equations
$\vect{\Phi}_k^{(i)}(\vect{X}_k^{(i)},\vect{X}_k^{(e)},\vect{p})=0$
have multiple solutions in $\vect{X}_k^{(i)}$ it is not excluded
that some of these solutions are incompatible with the rest of the
equations: after all the box equilibrations we may still have an
unique solution.

\medskip

\noindent{{\bf Sketch of the proof of Proposition \ref{TheoTnonreg}
and Theorem \ref{suffcond}.}} According to
Proposition~\ref{boxcondition} a sufficient condition for uniqueness
of equilibrium is to find a complete sequence of box equilibrations
for the state equations of metabolic variables
(Eq.\eqref{IDSMLipid}). To simplify notations, in this section we
write the reduced fluxes without the subscripts $peq,gnr$. Notice
that in order to get the state equations a first box elimination of
the genetic variables (shown to be univocal) has already been
performed at equilibrium. For \gq states, the reduction of the
genetic variables has been performed by replacing them by constants.

Let us consider a box decomposition corresponding to the following
partition of variables:
\begin{itemize}
\item Box 1: $\{\A, \Fun,\Fdeux\}$.
\item Box 2: $\{\T\}$.
\end{itemize}
The main steps of the proof are the following:
\begin{itemize}
\item Step 1: the equilibration equations for Box 1 with  internal variables
$\A$, $\Fun$, $\Fdeux$ is:
 \begin{eqnarray}
\Phi_\A(\G,\A,\Fun,\Fdeux,\T)& =& 0
\label{b2eq1} \\
\Phi_{\Fun}(\A,\Fun,\Fdeux,\T) &=&0 \label{b2eq2} \\
\Phi_{\Fdeux}(\Fdeux,\T) &=& 0 \label{b2eq3}
\end{eqnarray}
In Lemma \ref{propbox2}, we prove that for all $(\G, \T) \in
[0,G_{max}] \times \R_+$, this system has a unique solution
$(\A^{(1)}(\G,\T),{\Fun}^{(1)}(\G,\T),{\Fdeux}^{(1)}(\G,\T)) \in
\R_+^3$, which ends the proof of Proposition \ref{TheoTnonreg}.
\item Step 2: the equilibration equation for Box 2 with internal variable $\T$ is:
\begin{equation}
\Phi_\T(\G,\A^{(1)}(\G,\T),{\Fun}^{(1)}(\G,\T),
{\Fdeux}^{(1)}(\G,\T),\T) = 0 \label{b2}
\end{equation}
In Lemma \ref{lemmebox3}, we shall provide a sufficient condition
for this equation to have a unique positive solution
$\T=\T^{(2)}(G)$. The unique solution of the state equations is thus
$\A=\A^{(2)}(G)=\A^{(1)}(\G,\T^{(2)}(G))$,
$\Fun={\Fun}^{(2)}(\G)={\Fun}^{(1)}(\G,\T^{(2)}(G))$,
$\Fdeux={\Fdeux}^{(2)}(\G)={\Fdeux}^{(1)}(\G,\T^{(2)}(G))$. This
solution depends on being at equilibrium or at quasi-stationarity,
because the expressions of the reduced fluxes $\Phi_\A$,
$\Phi_{\Fun}$, $\Phi_{\Fdeux}$ depend on being at equilibrium or at
quasi-stationarity.
\item Step 3: finally, we show that
the strong lipolytic condition is the sufficient condition of Lemma
\ref{lemmebox3}.
\end{itemize}

All the formal manipulations of this sequence where performed using
Wolfram Research Mathematica version 5.2 software.







To treat the case of Box 1, we use the following result which is a
direct consequence of Gale-Nikaido-Inada theorem \cite{par83}. This
theorem can be seen as a generalization to higher dimensions of the
monotonicity of functions on ${\mathbb R}$. Let us recall that a
principal minor of a matrix $M=(m_{i,j})_{i,j \in \{1, \dots, n\}}$
is defined as $\det M_{I}$, where $I \subset \{1, \dots, n\}$ and
$M_I=(m_{i,j})_{i,j \in I}$.

\begin{lemma}[Gale-Nikaido]\label{GaleNikaido}
If $(x,y,z) \rightarrow (f_x,f_y,f_z)$ is a differentiable mapping
from ${\mathbb R}^3_+$ to  ${\mathbb R}^3$, of Jacobian $J$, such
that all the principal minors of $-J$ are positive, then this
mapping is globally univalent. In particular the system
$f_x(x,y,z)=0$, $f_y(x,y,z)=0$, $f_z(x,y,z)=0$ has a unique solution
if a solution exists.
\end{lemma}

\begin{lemma}\label{propbox2}
For all $(\G,\T) \in [0,G_{max}] \times \R_+$, the system of
Eqs.~\eqref{b2eq1},\eqref{b2eq2} and \eqref{b2eq3} has a unique
solution in $(\A,\Fun,\Fdeux) \in \R_+^3$, both at genetic partial
equilibrium and at \gqty. This expresses $\A,\Fun,\Fdeux$ as
univocal functions of $\G$ and $\T$, denoted by $\A^{(1)}(\G,\T)$,
${\Fun}^{(1)}(\G,\T)$ and ${\Fdeux}^{(1)}(\G,\T)$.
 \end{lemma}

\noindent{{\bf Proof.}}
We apply Corollary \ref{Corfixedpoint} to prove that the system of
equations (\ref{b2eq1}), (\ref{b2eq2})  and (\ref{b2eq3}) has a
solution for every fixed $(\G,\T)$.

Let us consider the mapping $(\A,\Fun,\Fdeux) \rightarrow
(\Phi_{\A},\Phi_{\Fun},\Phi_{\Fdeux})$ on ${\mathbb R}_{+}^3$. To
prove the uniqueness of the solution to the system, we apply Lemma
\ref{GaleNikaido} to this mapping by ensuring that all the principal
minors of $-J^{(1)}$ are positive, where $J^{(1)}$ is the Jacobian
of this mapping:
\[
J^{(1)}=
\begin{pmatrix}
-\chimtot & & \DP{\Phi_\A}{\Fun} &&& \DP{\Phi_\A}{\Fdeux} \\
 \DP{\Phi_{\Fun}}{\A} && -\chiauntot &&& \DP{\Phi_{\Fun}}{\Fdeux} \\
0 && 0 &&& -\chiadeuxtot
\end{pmatrix}
=  \begin{pmatrix} -\chimtot & & n_1 \chiaunoxi &&& n_2 \rodeux
+ n_1 \roun + n_1 \rosyn - \rk \\
\chimsyn  && -\chiauntot &&& - \roun - \rosyn \\
0 && 0 &&& -\chiadeuxtot
\end{pmatrix}.
\]

The principals minors of  $-J^{(1)}$ are all positive: $\chimtot
>0$,
$\chimtot \chiauntot - n_1 \chimsyn \chiaunoxi = \chimtot \chiauntot
(1 - \rhomsyn \rhoaunoxi ) > 0$, $\chiadeuxtot \chimtot \chiauntot
(1 - \rhomsyn \rhoaunoxi )
> 0$, as a consequence of Prop.  \ref{signelast}. This is valid both
at genetic partial equilibrium and at \gqty.


Hence,  Lemma \ref{GaleNikaido} applies, so that there exist
functions
 $\A^{(1)}(\G,\T)$,
 ${\Fun}^{(1)}(\G,\T)$ and
 ${\Fdeux}^{(1)}(\G,\T)$ that are the unique
solutions of the system  of Eqs.~(\ref{b2eq1}), (\ref{b2eq2})  and
(\ref{b2eq3}) for each $(\G,\T)$. These functions are differentiable
on ${\mathbb R}_{+}^2$
 by the implicit function theorem.
 \cqfd

\medskip

\noindent{{\bf Proof of Proposition \ref{TheoTnonreg}.}} Proposition
\ref{TheoTnonreg} derives directly from Lemma \ref{propbox2}. \cqfd

\begin{lemma}\label{lemmebox3}
Eq.(\ref{b2})  has a unique solution in $\T$ as soon as the
following inequality holds for every $(\G,\T) \in [0,G_{max}] \times
\R_+$:
\begin{equation}\label{cond2bis}
\DP{{\Phi_{\T}}}{\T} + (\alpha_\K \chimkr - \alpha_S \chimsyn)
\DP{{\A^{(1)}}}{\T} +
 \alpha_{\Oun}  \chiaunoxi \DP{{\Fun}^{(1)}}{\T} +  (\alpha_{\Oun}
\roun + \alpha_{\Odeux}  \rodeux + \alpha_S \rosyn
)\DP{{\Fdeux}^{(1)}}{\T}  < 0.
\end{equation}


\end{lemma}

\noindent{{\bf Proof.}} Let ${\Phi_\T}^{(1)}(\G,\T) =
\Phi_\T(\G,\A^{(1)}(\G,\T),{\Fun}^{(1)}(\G,\T),
{\Fdeux}^{(1)}(\G,\T),\T)$. The biological hypotheses imply that
Theorem \ref{existenceFixedPoint} applies
that is, for every $\G$, there exists $(a,f_1,f_2,t) \in {\mathbb
R}_{+}^{4}$ such that
$$\Phi_\A(a,f_1,f_2,t)= \Phi_{\Fun}(a,f_1,f_2,t)=\Phi_{\Fdeux}(f_2,t) = \Phi_\T(\G,a,f_1,f_2,t)=0.$$

By uniqueness in the previous Lemmas, we get $a=\A^{(1)}(\G,t)$,
$f_1={\Fun}^{(1)}(\G,t)$ and $f_2={\Fdeux}^{(1)}(\G,t)$. Hence
${\Phi_\T}^{(1)}(\G,t)=0$ and the function ${\Phi_\T}^{(1)}(\G,T)$
has a root in $T$ for every $\G$.

Moreover, the function ${\Phi_\T}^{(1)}$ is differentiable on
${\mathbb R}_{+}^2$. From the definition of the function
${\Phi_\T}^{(1)}$ it follows: {\small
 \begin{equation}
\DP{{\Phi_{\T}}^{(1)}}{\T}  =  \DP{{\Phi_{\T}}}{\T} + (\alpha_\K
\chimkr - \alpha_S \chimsyn)  \DP{{\A^{(1)}}}{\T} + 
\alpha_{\Oun}  \chiaunoxi \DP{{\Fun}^{(1)}}{\T} + (\alpha_{\Oun}
\roun + \alpha_{\Odeux}  \rodeux + \alpha_S \rosyn
)\DP{{\Fdeux}^{(1)}}{\T}
\label{eqphiT2}
\end{equation}}

Hence if the inequality \ref{cond2bis} is satisfied, then
$\DP{{\Phi_{\T}}^{(1)}}{\T}$ is negative. In other words,
${\Phi_{\T}}^{(1)}$ is monotonic so that it has a unique zero. \cqfd

\medskip

\noindent{\sf{\bf Proof of Theorem \ref{suffcond} }} Theorem
\ref{suffcond}  derives directly from Lemma \ref{lemmebox3}. \cqfd

\begin{lemma}\label{CalculDeriveeT}
The derivatives $\DP{{\A}^{(1)}}{\T}$, $\DP{{\Fun}^{(1)}}{\T}$,
$\DP{{\Fdeux}^{(1)}}{\T}$ can be expressed by means of fluxes and of
control coefficients in the following way: {\small
\begin{eqnarray}\label{Tderivatives} - \frac{\det(J^{(1)})}{\chiauntot}
\DP{\A^{(1)}}{\T} & = & \chiadeuxtot \{ \tgly - \tkr + n_2[
\rhoadeuxoxi \tindeux +
(1-\rhoadeuxoxi) \todeux ] +  n_1 [(1-\rhoaunoxi)(\toun + \tsoyn) + \notag \\
&& \rhoaunoxi \tinun] \}   + (\tindeux - \todeux) [-\rk  + n_1
(\roun + \rosyn )(1-\rhoaunoxi) ] \notag
 \\
- \frac{\det(J^{(1)})}{\chimtot} \DP{{\Fun}^{(1)}}{\T} & = &
\chiadeuxtot \{ \rhomsyn (\tgly- \tkr  + n_2 \todeux) + n_1 [\tinun
-(1-\rhomsyn)(\toun + \tsoyn) ] \} \notag
\\ & &
+(\todeux -\tindeux) [ n_1 (\roun +\rosyn ) - n_2 \rodeux + \rhomsyn
\rk ] \notag  \\
\DP{{\Fdeux}^{(1)}}{\T}  & = &   (\chiadeuxtot)^{-1}  (\todeux
-\tindeux) \end{eqnarray} } where $ - \det(J^{(1)})  = \chiadeuxtot
\chimtot\chiauntot (1 - \rhomsyn \rhoaunoxi )$.

\end{lemma}

\noindent{{\bf Proof.}} The lemma follows straightforwardly from
{\small
\begin{eqnarray*} \DP{}{\T}
\begin{pmatrix}
{\A}^{(1)} \\
{\Fun}^{(1)} \\
{\Fdeux}^{(1)}
\end{pmatrix}
&=& - {(J^{(1)})}^{-1} \DP{}{\T}
\begin{pmatrix}
\Phi_{\A} \\
\Phi_{\Fun} \\
\Phi_{\Fdeux}
\end{pmatrix}.
\end{eqnarray*} }
\cqfd

\begin{lemma}\label{slrcondition}

\begin{equation}
\DP{{\Phi_{\T}}^{(1)}}{\T}  = [  A ( \toun - \tinun  ) +  B (
\todeux -\tindeux   ) + C - D ]/[n_1(1-\rhomsyn \rhoaunoxi)]
\end{equation}
where $A,B,C,D$ are combinations of control parameters defined in
Eq.~\eqref{ABCD}.

Furthermore, if the stoechiometric condition \ref{scon} is
fulfilled, then $X>0,A>0,B_1>0,B_4>0,C>0,D>0$.
\end{lemma}

\noindent{{\bf Proof.}} The proof is a lengthy but straightforward
formal manipulation of Eqs.\eqref{eqphiT2},\eqref{Tderivatives}. We
have
 gathered control coefficients into as large as possible positive
combinations.

As an illustration of how the stoechiometric condition was used let
us consider the sign of $D$. From $\alpha_{\Oun}>\alpha_{S}$, $X/n_1
- \alpha_{\Oun} (1 - \rhomsyn) = (\alpha_{\Oun} - \alpha_{S})
\rhomsyn + n_1 \alpha_{K} \rho_A^{Krebs}  > 0$,  $\rhomsyn \leq 1$,
it follows that $D>0$. \cqfd

\medskip
\noindent{\sf{\bf Proof of Proposition \ref{ABC}.}} It follows
directly from Lemma \ref{slrcondition}.  \cqfd

\begin{lemma}\label{variationT}
Let $T={\T}^{(2)}(\G)$ be the solution of the equation \eqref{b2}.
If the strong lipolytic response condition and the stoechimetric
conditions are satisfied then $\D{\T^{(2)}}{\G}
> 0$.
\end{lemma}

\noindent{{\bf Proof.}}  The partial derivatives with respect to
$\G$ of the metabolites after equilibration of the box 1 can be
obtained from: {\small \[ \DP{}{\G}
\begin{pmatrix}
{\A}^{(1)} \\
{\Fun}^{(1)} \\
{\Fdeux}^{(1)}
\end{pmatrix}
= - {(J^{(1)})}^{-1} \DP{}{\G}
\begin{pmatrix}
\Phi_{\A} \\
\Phi_{\Fun} \\
\Phi_{\Fdeux}
\end{pmatrix}
= - {(J^{(2)})}^{-1}
\begin{pmatrix}
R_G^{Gly}  \\
0 \\
0
\end{pmatrix}
= \frac{R_G^{Gly}}{\chimtot\chiauntot (1 - \rhomsyn \rhoaunoxi )}
\begin{pmatrix}
\chiauntot  \\
\chimsyn \\
0
\end{pmatrix}.
\]
}

From this and from the definition of $\Phi_{\T}^{(1)}$ it follows

\begin{equation}
\begin{array}{c}\DP{{\Phi_{\T}}^{(1)}}{\G}= \frac{R_G^{Gly}}{n_1 \chimtot (1 - \rhomsyn \rhoaunoxi )}
\{n_1 \alpha_\K \chimkr + \chimtot [ n_1 \alpha_\G (1 - \rhoaunoxi
\rhomsyn) + \alpha_{\Oun} \rhoaunoxi \rhomsyn - \alpha_S \rhomsyn]
\}
\end{array} \label{eqdiffTG}\end{equation}

If $\alpha_S < \alpha_{\Oun} < n_1 \alpha_\G$ (which is true by the
stoechiometry condition 5), then $0 \leq \alpha_{\Oun}(1 - \rhomsyn
 ) =   \alpha_{\Oun} (1 - \rhoaunoxi \rhomsyn) + \alpha_{\Oun}
\rhoaunoxi \rhomsyn - \alpha_{\Oun} \rhomsyn < n_1 \alpha_\G (1 -
\rhoaunoxi \rhomsyn) + \alpha_{\Oun} \rhoaunoxi \rhomsyn - \alpha_S
\rhomsyn$, and consequently $\DP{{\Phi_{\T}}^{(1)}}{\G}>0$.

The strong lipolytic response condition is equivalent to having
$\DP{{\Phi_{\T}}^{(1)}}{\T}<0$. If this is satisfied, then
 $\D{{\T}^{(2)}}{\G}=  -\left( \DP{{\Phi_{\T}}^{(1)}}{\G}\right)/\left(\DP{{\Phi_{\T}}^{(1)}}{\T}\right) >0$. \cqfd

\medskip
\noindent{\sf{\bf Proof of Proposition \ref{prediBio1}.}}  It
follows directly from Lemmas \ref{variationT}. \cqfd

\begin{lemma}\label{variationF2}
Let $\Fdeux^{(2)} (\G) = \Fdeux^{(1)}  ( \G , {\T}^{(2)}(\G)) $. If
the strong lipolytic condition is satisfied,  then the sign of
$\D{{\Fdeux}^{(2)}}{\G}$ is equal to the sign of $\todeux-
\tindeux$.
\end{lemma}


\noindent{{\bf Proof.}} The chain rule gives
$$\D{{{\Fdeux}}^{(2)}}{\G} = \DP{{{\Fdeux}}^{(1)}}{\G} +
\DP{{{\Fdeux}}^{(1)}}{\T} \D{{\T}^{(2)}}{\G}.
 $$

From the proof of Lemma \ref{variationT},
$\DP{{{\Fdeux}}^{(1)}}{\G}=0$. It follows from Lemma
\ref{CalculDeriveeT} that the sign of $\DP{{{\Fdeux}}^{(1)}}{\T}$ is
the same as the sign of $ \todeux-\tindeux$. Moreover,  if  the
strong lipolytic condition and the stoechimetric condition are
satisfied, then ${\D{{\T}^{(2)}}{\G}}
>0$. \cqfd

\medskip
\noindent{\sf{\bf Proof of Proposition \ref{prediBio}.}}  It follows
directly from Lemmas \ref{variationT},\ref{variationF2}. \cqfd

\medskip

\noindent{{\bf Proof of Propositions  \ref{lemmeCompT} and
\ref{lemmeCompT2}.}} The differences between equilibrium and
quasi-stationarity occur at two levels:
\begin{enumerate}
\item At  quasi-stationarity $F_2$ does not regulate the genetic
variables:
\begin{equation}\label{nocontrolqs}
\left( \rosyn \right)_{qs}= \left( \roun  \right)_{qs} = \left( \rk
 \right)_{qs} = 0.
\end{equation}

\item At quasi-stationarity the control of $F_2$ on its oxidation is
only a metabolic substrate effect. Genetic control is added at
equilibrium. We have $\rodeux= \DP{\Oxideux}{\Fdeux} +
\DP{{\Etrois}}{\Fdeux} \DP{\Oxideux}{\Etrois}$ with
$\DP{\Oxideux}{\Etrois} >0$, and
$\chiadeuxtot=\rodeux-\DP{\Findeux}{\Fdeux}$,
$\rhoadeuxoxi=(1-\frac{\DP{\Findeux}{\Fdeux}}{\rodeux})^{-1}$, with
$\DP{\Findeux}{\Fdeux}<0$. Furthermore, $\left(
\DP{{\Etrois}}{\Fdeux} \right)_{eq}
>0$ and $\left( \DP{{\Etrois}}{\Fdeux} \right)_{qs}=0$.
Hence:
\begin{equation}\label{controloxi2f2}
\left( \rodeux \right)_{eq} > \left( \rodeux \right)_{qs},\quad
 \left(
\chiadeuxtot \right)_{eq} > \left( \chiadeuxtot \right)_{qs},\quad
\left( \rhoadeuxoxi \right)_{eq} > \left( \rhoadeuxoxi \right)_{qs}.
\end{equation}
\end{enumerate}

We deduce from Eq.(\ref{eqdiffTG}) that $\DP{{\Phi_{\T}}^{(1)}}{\G}$
is the same at equilibrium and at quasi-stationarity. From Lemma
\ref{slrcondition}, it follows :

\begin{equation} \label{dpphiT}
\begin{array}{c}\DP{{\Phi_{\T}}^{(1)}}{\T} = {\cal R} - \frac{B }{n_1(1-\rhomsyn \rhoaunoxi)}(\tindeux -
\todeux)  ,\end{array}
\end{equation}
where ${\cal R}$ is a term not changing from quasi-stationarity to
equilibrium and the expression of $B$ is detailed in Prop.
\ref{ABC}.

%

 From $\D{{\T}^{(2)}}{\G}=
-\left( \DP{{\Phi_{\T}}^{(1)}}{\G} \right)/\left(
\DP{{\Phi_{\T}}^{(2)}}{\T}\right)$, from $B_{qs}< B_{eq}$, and from
Eqs.(\ref{nocontrolqs},\ref{controloxi2f2},\ref{dpphiT}),
 it follows that $\left( \D{{\T}^{(2)}}{\G} \right)_{qs} >
\left( \D{{\T}^{(2)}}{\G} \right)_{eq}$. From $\D{\Fdeux^{(2)}}{\G}
= - \D{{\T}^{(2)}}{\G} (\tindeux - \todeux) / \chiadeuxtot $ and
Eq.(\ref{controloxi2f2}) it follows that $\left|
\D{{\Fdeux}^{(2)}}{\G} \right|_{qs} > \left| \D{{\Fdeux}^{(2)}}{\G}
\right|_{eq}$. \cqfd

\medskip
\noindent{{\bf Proof of Proposition \ref{pparko}}} We follows
closely the proof of Prop. \ref{lemmeCompT}. The differences between
$PPAR-/-$ and $WT$ cells occur at two levels: {\small
\begin{eqnarray}\label{nocontrolko}  \left( \roun \right)_{PPAR-/-}
= \left( \rk
 \right)_{PPAR-/-} = 0  & \\  \notag
\left( \rodeux \right)_{WT,eq}  > \left( \rodeux
\right)_{PPAR-/-,eq},\,
 \left(
\chiadeuxtot \right)_{WT,eq} > \left( \chiadeuxtot
\right)_{PPAR-/-,eq}, & \\  \label{controloxi2f2ko} \left(
\rhoadeuxoxi \right)_{WT,eq} > \left( \rhoadeuxoxi
\right)_{PPAR-/-,eq}
\end{eqnarray}}

If $B_{WT,eq}
> B_{PPAR-/-,eq}$, it follows (along the same lines as the proof of Prop.
\ref{lemmeCompT}) that \\
\noindent
 {\small $\left( \D{{\T}^{(2)}}{\G} \right)_{PPAR-/-,eq} >
\left( \D{{\T}^{(2)}}{\G} \right)_{WT,eq}$}. From {\small
$\D{\Fdeux^{(2)}}{\G} = - \D{{\T}^{(2)}}{\G} (\tindeux - \todeux) /
\chiadeuxtot $} and Eq.(\ref{controloxi2f2ko}) it follows that
{\small $\left| \D{{\Fdeux}^{(2)}}{\G} \right|_{eq,PPAR-/-} > \left|
\D{{\Fdeux}^{(2)}}{\G} \right|_{eq,WT}$}. \cqfd

{\small
\bibliography{Metab}
\bibliographystyle{alpha}
}

\end{document}